\documentclass[onecolumn, prd, aps, tightenlines, preprintnumbers, showpacs, nofootinbib, superscriptaddress, notitlepage]{revtex4-1}

\pdfoutput=1

\usepackage{float}
\usepackage{amsmath}
\usepackage{color}
\usepackage{graphicx}
\usepackage[dvipsnames]{xcolor}
\usepackage{url}
\usepackage{epsfig}
\usepackage[T1]{fontenc}
\usepackage{multirow}
\usepackage{physics} 
\usepackage{booktabs} 
\usepackage{array} 
\usepackage{paralist} 
\usepackage{grffile}
\usepackage{verbatim} 
\usepackage{subfig} 
\usepackage{amsmath,amsthm,amssymb,bm,amsfonts}
\usepackage{slashed}
\usepackage[utf8]{inputenc}
\usepackage{hyperref}
\allowdisplaybreaks[1]

\usepackage{color}
\usepackage[normalem]{ulem}

\def\beg{\begin{equation}}
\def\eeg{\end{equation}}
\def\bea{\begin{eqnarray}}
\def\eea{\end{eqnarray}}

\usepackage{multirow}
\usepackage[title]{appendix}

\def\Tr{{\rm tr}}

\newcommand{\slv}{\raise.15ex\hbox{$/$}\kern-.53em\hbox{$v$}}
\newcommand{\slnbar}{\raise.15ex\hbox{$/$}\kern-.53em\hbox{$\bar{n}$}}
\newcommand{\slF}{\raise.15ex\hbox{$/$}\kern-.53em\hbox{$F$}}
\newcommand{\sllbar}{\raise.15ex\hbox{$/$}\kern-.40em\hbox{$\bar{l}$}}
\newcommand{\slh}{\raise.15ex\hbox{$/$}\kern-.40em\hbox{$h$}}
\newcommand{\slP}{\raise.15ex\hbox{$/$}\kern-.53em\hbox{$P$}}
\newcommand{\slR}{\raise.15ex\hbox{$/$}\kern-.53em\hbox{$R$}}
\newcommand{\slz}{\raise.15ex\hbox{$/$}\kern-.53em\hbox{$Z$}}
\newcommand{\slzbar}{\raise.15ex\hbox{$/$}\kern-.53em\hbox{$\bar{Z}$}}
\newcommand{\slQ}{\raise.15ex\hbox{$/$}\kern-.53em\hbox{$Q$}}
\newcommand{\slK}{\raise.15ex\hbox{$/$}\kern-.53em\hbox{$K$}}
\newcommand{\slkbar}{\raise.15ex\hbox{$/$}\kern-.53em\hbox{$\bar{k}$}}
\newcommand{\slkone}{\raise.15ex\hbox{$/$}\kern-.53em\hbox{$k_1$}}
\newcommand{\slpone}{\raise.15ex\hbox{$/$}\kern-.53em\hbox{$p_1$}}
\newcommand{\slpbarone}{\raise.15ex\hbox{$/$}\kern-.53em\hbox{$\bar{p}_1$}}
\newcommand{\slptwo}{\raise.15ex\hbox{$/$}\kern-.53em\hbox{$p_2$}}
\newcommand{\slpbartwo}{\raise.15ex\hbox{$/$}\kern-.53em\hbox{$\bar{p}_2$}}
\newcommand{\slqone}{\raise.15ex\hbox{$/$}\kern-.53em\hbox{$q_1$}}
\newcommand{\slD}{\raise.15ex\hbox{$/$}\kern-.53em\hbox{$\!D$}}
\newcommand{\slC}{\raise.15ex\hbox{$/$}\kern-.53em\hbox{$C$}}
\newcommand{\slA}{\raise.15ex\hbox{$/$}\kern-.73em\hbox{$A$}}
\newcommand{\slSigma}{\raise.15ex\hbox{$/$}\kern-.53em\hbox{$\Sigma$}}
\newcommand{\slpartial}{\raise.15ex\hbox{$/$}\kern-.53em\hbox{$\partial$}}
\newcommand{\slcalP}{\raise.15ex\hbox{$/$}\kern-.63em\hbox{$\cal P$}}
\newcommand{\sleps}{\raise.15ex\hbox{$/$}\kern-.53em\hbox{$\epsilon$}}
\newcommand{\slepsbar}{\raise.15ex\hbox{$/$}\kern-.53em\hbox{$\overline{\epsilon}$}}
\newcommand{\slepsstar}{\raise.15ex\hbox{$/$}\kern-.53em\hbox{$\epsilon$}^\star}
\newcommand{\slS}{\raise.15ex\hbox{$/$}\kern-.73em\hbox{$S$}}

\newcommand{\bb}{\mathbf}

\newcommand{\be}{\boldsymbol{\epsilon}}
\newcommand{\bk}{\mathbf{k}}
\newcommand{\bp}{\mathbf{p}}
\newcommand{\bq}{\mathbf{q}}
\newcommand{\bx}{\mathbf{x}}
\newcommand{\by}{\mathbf{y}}
\newcommand{\slp}{\slashed{p}}
\newcommand{\slq}{\slashed{q}}
\newcommand{\slk}{\slashed{k}}
\newcommand{\sll}{\slashed{l}}
\newcommand{\sln}{\slashed{n}}
\newcommand{\slep}{\slashed{\epsilon}}

\newcommand{\ubar}{\bar{u}}

\newcommand{\barq}{\bar{q}}
\newcommand{\de}{\cdot\boldsymbol{\epsilon}}
\newcommand{\des}{\cdot \boldsymbol{\epsilon}^*}

\newcommand{\mM}{\mathcal{M}}
\newcommand{\dfour}[1]{\frac{\dd^4 #1}{(2\pi)^4}}
\newcommand{\dthree}[1]{\frac{\dd^3 #1}{(2\pi)^3}}
\newcommand{\dtwo}[1]{\frac{\dd^2 #1}{(2\pi)^2}}

\newcommand{\p}{\prime}

\newcommand{\zho}{z_{h_1}}
\newcommand{\zht}{z_{h_2}}

\newcommand{\Delt}[2]{\left[\frac{1}{\bx_{3 #1}^2} + \frac{1}{\bx_{3#2}^2} - \frac{\bx_{#1 #2}^2}{\bx_{3 #1}^2 \bx_{3 #2}^2}\right]}

\newcommand{\td}{\tilde{\Delta}}

\begin{document}
\title{ One-Loop Corrections to Dihadron Production in DIS at Small $x$}

\author{Filip Bergabo}
\email{fbergabo@gradcenter.cuny.edu}
\affiliation{Department of Natural Sciences, Baruch College, CUNY, 17 Lexington Avenue, New York, NY 10010, USA}
\affiliation{City University of New York Graduate Center, 365 Fifth Avenue, New York, NY 10016, USA}

\author{Jamal Jalilian-Marian}
\email{jamal.jalilian-marian@baruch.cuny.edu}
\affiliation{Department of Natural Sciences, Baruch College, CUNY, 17 Lexington Avenue, New York, NY 10010, USA}
\affiliation{City University of New York Graduate Center, 365 Fifth Avenue, New York, NY 10016, USA}


\begin{abstract}
We calculate the one-loop corrections to dihadron production in Deep Inelastic Scattering (DIS) at small $x$ using the Color Glass Condensate formalism. We show that all UV and soft singularities cancel while the collinear divergences are absorbed into quark and anti quark-hadron fragmentation functions. Rapidity divergences lead to JIMWLK evolution of dipoles and quadrupoles describing multiple-scatterings of the quark anti-quark dipole on the target proton/nucleus. The resulting cross section is finite and can be used for phenomenological studies of dihadron angular correlations at small $x$ in a future Electron-Ion Collider (EIC).
\end{abstract}

\maketitle



\section{Introduction}\label{sec:intro}

Since the experimental observation of the fast rise of parton (especially gluon) distribution functions at HERA~\cite{Aaron:2009kv}, the phenomenon of gluon saturation~\cite{Gribov:1984tu,Mueller:1985wy} at small $x$ has been an active field of study. The Color Glass Condensate 
formalism~\cite{Iancu:2003xm,Jalilian-Marian:2005ccm,Weigert:2005us,Morreale:2021pnn} is an effective theory of QCD at small $x$ which provides a robust platform that can be used to investigate saturation dynamics. 
Nevertheless and despite the intense ongoing theoretical and experimental efforts, clear and unambiguous evidence for gluon saturation remains elusive. There is hope that the proposed Electron-Ion Collider (EIC)~\cite{Accardi:2012qut} will be able to unambiguously discover gluon saturation and to establish the kinematic region in which it is 
applicable. Perhaps the most promising process in which to discover gluon saturation is two-particle angular correlations which has been extensively studied using Leading Order (LO) expressions obtained in the Color Glass Condensate 
formalism~\cite{Kovner:2001vi,JalilianMarian:2004da,Jalilian-Marian:2005qbq, Marquet:2007vb,Albacete:2010pg,Stasto:2011ru,Lappi:2012nh,Jalilian-Marian:2012wwi,Jalilian-Marian:2011tvq, Zheng:2014vka,Stasto:2018rci,Albacete:2018ruq,Mantysaari:2019hkq,Hatta:2020bgy,Jia:2019qbl,Dominguez:2011wm,Metz:2011wb,Dominguez:2011br,Iancu:2013dta,Iancu:2021rup,Altinoluk:2015dpi,Hatta:2016dxp,Dumitru:2015gaa,Kotko:2015ura,Marquet:2016cgx,vanHameren:2016ftb,Marquet:2017xwy,Dumitru:2018kuw,Mantysaari:2019csc,Salazar:2019ncp,Boussarie:2021ybe,Kotko:2017oxg,Hagiwara:2017fye,Klein:2019qfb,Hatta:2021jcd,Iancu:2020mos,Caucal:2021ent,Kolbe:2020tlq,Gelis:2002fw,Altinoluk:2019fui,Boussarie:2019ero,Boussarie:2016ogo,Boussarie:2014lxa,Dumitru:2010ak}. There are already strong hints for the presence of gluon saturation dynamics in the observed disappearance of away side hadrons in inclusive dihadron production in deuteron (proton)-gold collisions at RHIC~\cite{Aschenauer:2016our,Braidot:2010ig,Adare:2011sc}. Nevertheless the EIC will measure dihadrons with much more accuracy and with much better control over the range of target $x$ contributing to the process. Therefore it is essential to go beyond a LO calculation in order to achieve a reasonable quantitative accuracy. Toward this end we calculate the one-loop corrections to inclusive dihadron production in DIS here. As expected there are several divergences that appear when going beyond a tree level calculation. We show that all such  divergences either cancel or can be absorbed into evolution (scale dependence) of physical quantities.  Our conclusion on factorization/cancellation of all singularities are in agreement with recent studies of one-loop corrections to dijet production in DIS~\cite{Caucal:2021ent,Taels:2022tza}, nevertheless our treatment of collinear divergences and their absorption into evolution of hadron fragmentation functions is new. Our final results are completely finite and can be used to investigate gluon saturation dynamics in inclusive dihadron production and angular correlations in DIS at small $x$. 

In the small $x$ limit scattering of a virtual photon on a target hadron or nucleus can be understood as a two-step process; first the virtual photon splits into a quark anti-quark pair (a dipole), which then scatters from the target hadron or nucleus. The total virtual photon-target scattering probability is expressed as a convolution of the probability for a photon to split into a quark at transverse position $\bx_1$ and an anti-quark at position $\bx_2$, with the probability for this dipole to scatter from the target. The leading order double inclusive production cross section can be written as

\bea
\frac{\dd \sigma^{\gamma^*A \to q\bar{q} X}}{\dd^2 \bb{p}\, \dd^2 \bb{q} \, \dd y_1 \, \dd y_2} &=& \frac{ e^2 Q^2(z_1z_2)^2 N_c}{(2\pi)^7} \delta(1-z_1-z_2)\int \dd^8 \bx \left[S_{122^\prime 1^\prime} - S_{12} - S_{1^\prime 2^\prime} + 1\right] \nonumber \\
&& e^{i\bb{p}\cdot(\bb{x}_1^\prime - \bb{x}_1)} e^{i\bb{q}\cdot(\bb{x}_2^\prime - \bb{x}_2)} 
\bigg[4z_1z_2K_0(|\bb{x}_{12}|Q_1)K_0(|\bb{x}_{1^\prime 2^\prime}|Q_1) + \nonumber \\
&&  (z_1^2 + z_2^2) \,
\frac{ \bb{x}_{12}\cdot \bb{x}_{1^\prime 2^\prime}}{|\bb{x}_{12}| |\bb{x}_{1^\prime 2^\prime}|} \, 
K_1(|\bb{x}_{12}|Q_1)K_1(|\bb{x}_{1^\prime 2^\prime}|Q_1) 
\bigg] .\label{LOdsig}
\eea
where the first and second terms above correspond to the contribution of the longitudinal and transverse polarizations of the virtual photon. Here $l^\mu$ is the momentum of the virtual photon with $l^2 = -Q^2$. We set the transverse momentum of the photon to zero without any loss of generality. Also $p^\mu$ ($q^\mu$) is the momentum of the outgoing quark (anti-quark) and $z_1$ ($z_2$) is its longitudinal momentum fraction relative to the photon. $\bx_1$ ($\bx_2$) is the transverse coordinate of the quark (anti-quark), and primed coordinates are used in the conjugate amplitude. Note that we write the differential variables on the left side using quark and anti-quark rapidities $y_1$ and $y_2$ which are related to their momentum fractions $z_1$ and $z_2$ via $\dd y_i = \dd z_i / z_i$. The electric charge of the quark is written as $e$, and in principle one needs to sum over quark flavors. We also use the following shorthand notations:

\begin{align}
Q_i = Q\sqrt{z_i(1-z_i)}, \,\,\,\,\,\, \bx_{ij} = \bx_i - \bx_j,\,\,\,\,\,\, \dd^8 \bx = \dd^2 \bx_1 \, \dd^2 \bx_2\, \dd^2 \bx_{1^\p} \, \dd^2 \bx_{2^\p}.
\end{align}

All the dynamics of the strong interactions and gluon saturation are contained in the dipoles $S_{ij}$ and quadrupoles $S_{ijkl}$, normalized correlation functions of two and four Wilson lines

\begin{align}
S_{ij} = \frac{1}{N_c} \Tr\left\langle V_i V_j^\dag \right\rangle, \,\,\,\,\,\,\,\,\, S_{ijkl} = \frac{1}{N_c}\Tr\left\langle V_i V_j^\dag V_k V_l^\dag\right\rangle, \label{dipquad}
\end{align}

\noindent where the index $i$ refers to the transverse coordinate $\bb{x}_i$ and the following notation is used for Wilson lines, 

\begin{align}
V_i &= \hat{P}\exp\left( ig \int \dd x^+ A^-(x^+,\bb{x}_i)\right).
\end{align}

The Wilson lines efficiently resum the multiple scatterings of the quark and anti-quark from the target hadron or nucleus. The angle brackets in Eq. \ref{dipquad} signify \emph{color averaging}. It is important to keep in mind that as this is a classical result the cross section has no non-trivial $x$ (or rapidity/energy) dependence. It is also easy to check that if one integrates over the phase space of the quark and anti-quark one recovers the standard expressions for the virtual photon-target total cross section at small $x$. 

\section{One-loop corrections}

\begin{figure}[H]
\centering
\includegraphics[width=70mm]{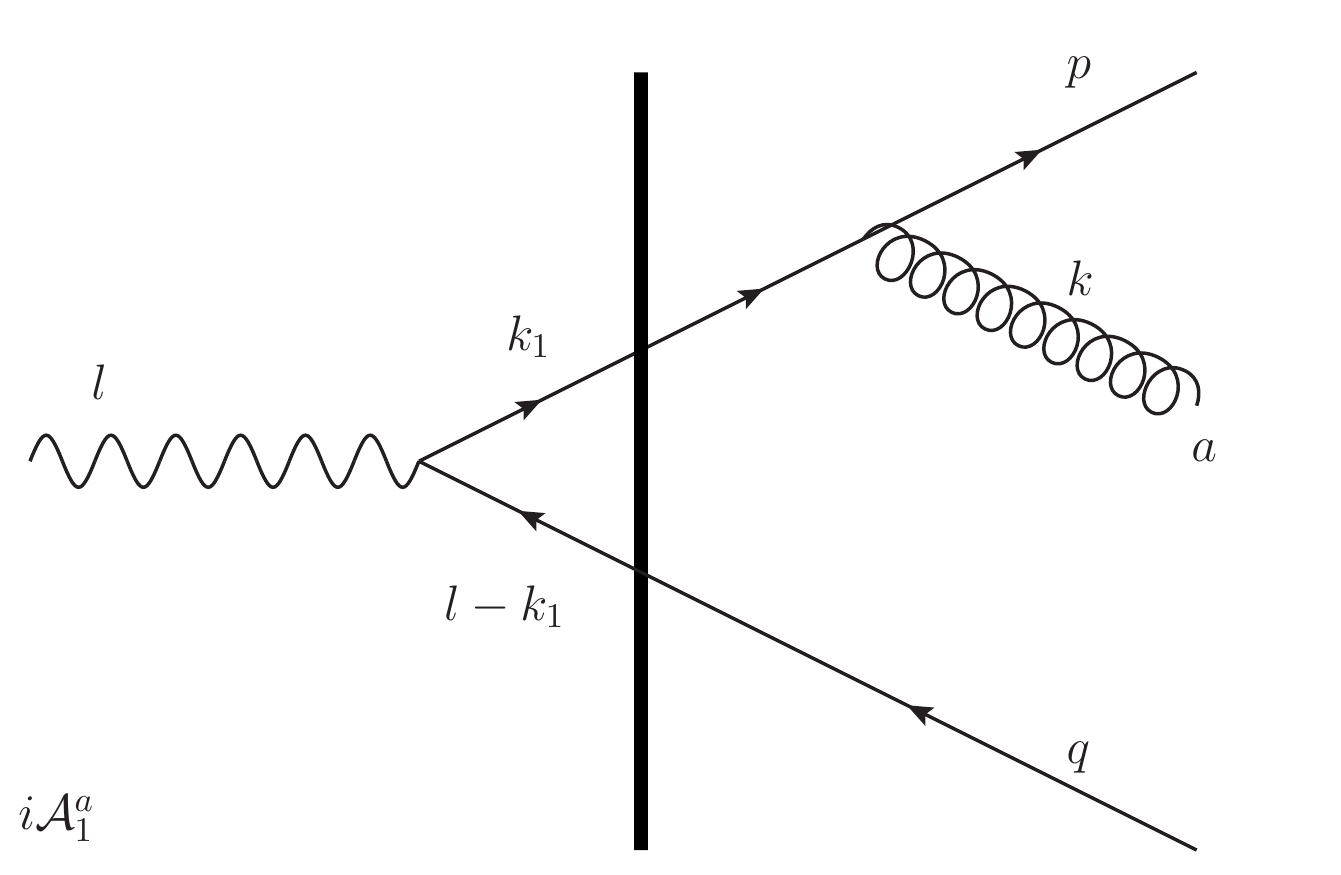}\includegraphics[width=70mm]{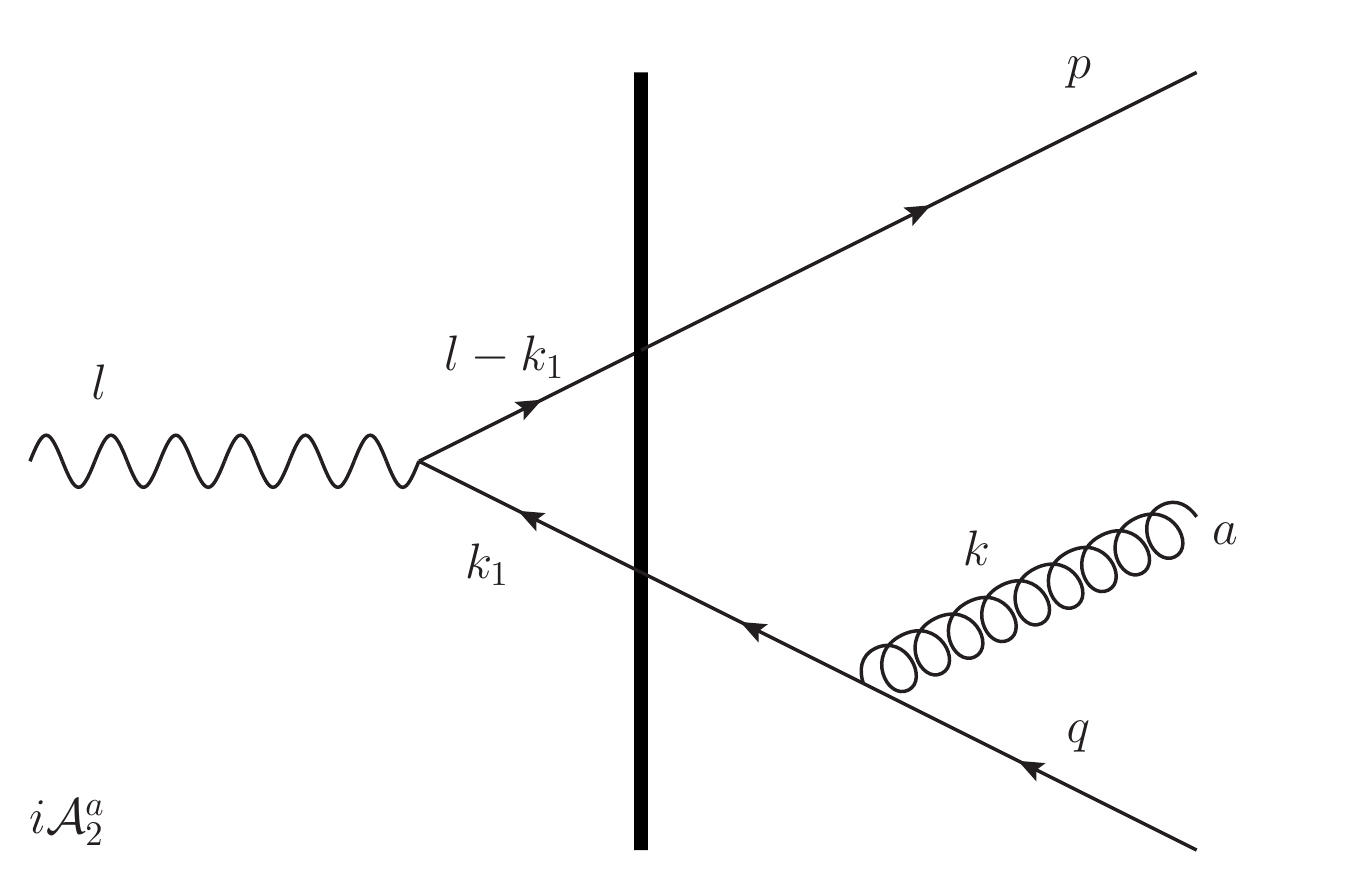}\\ \includegraphics[width=70mm]{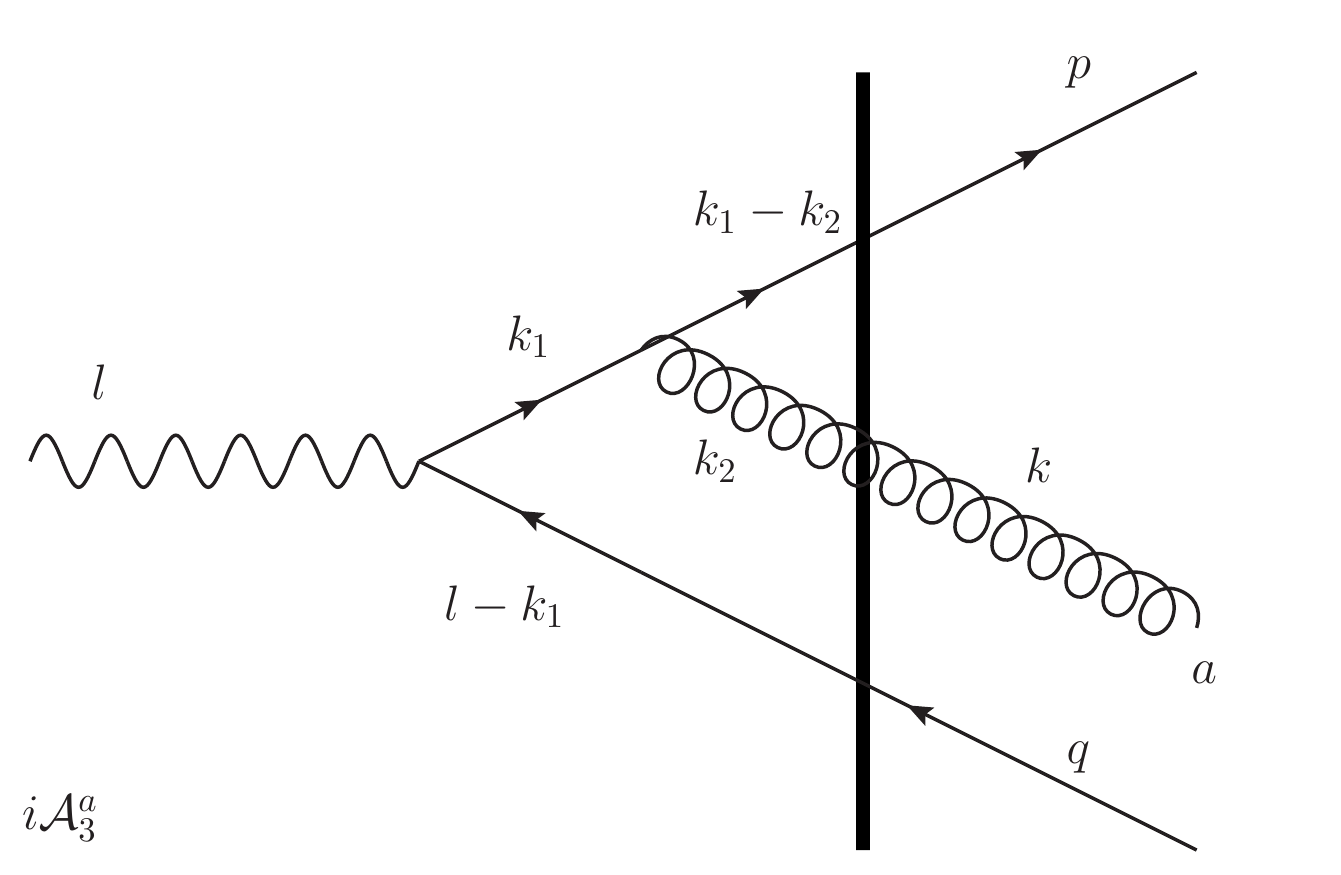}\includegraphics[width=70mm]{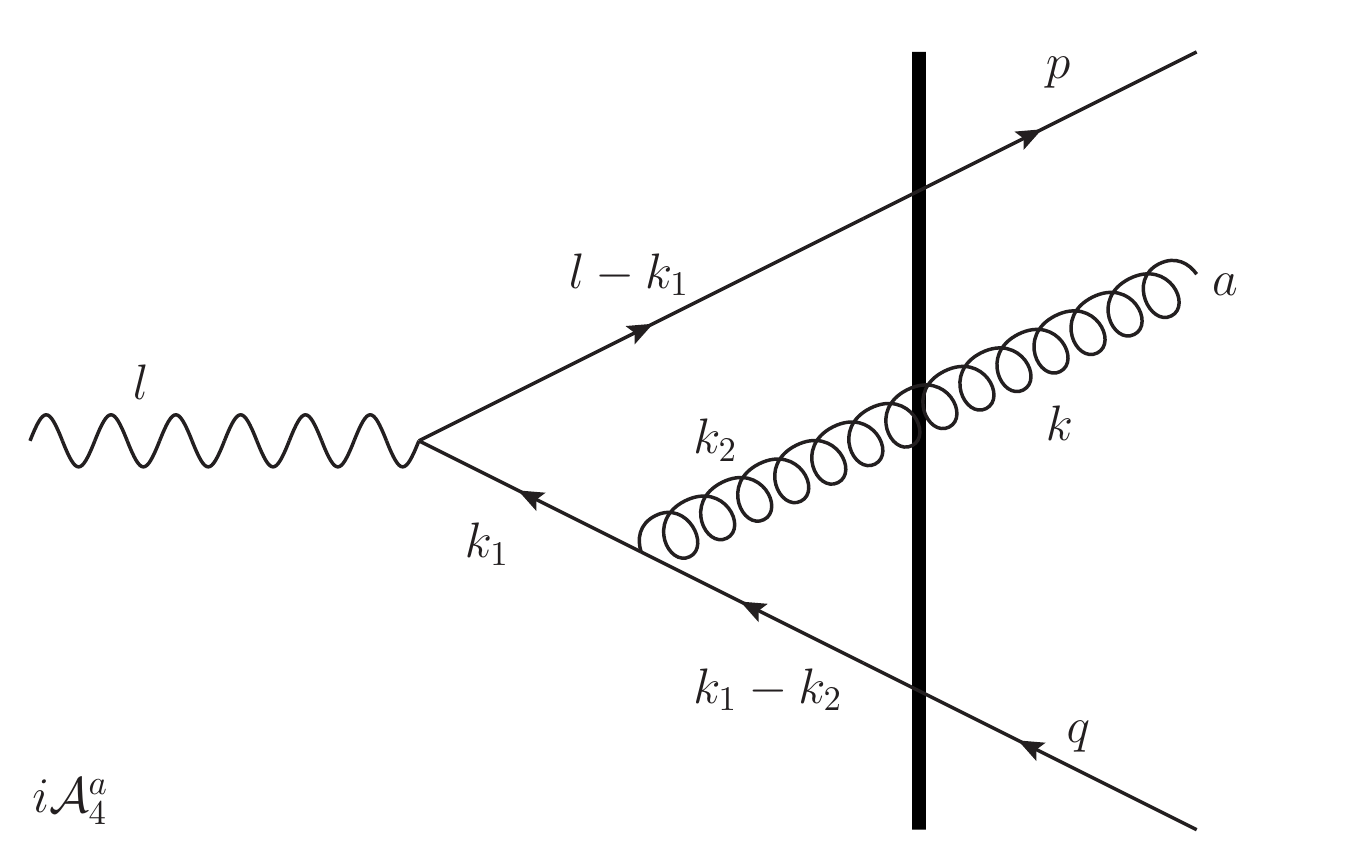}
\caption{The real corrections $i\mathcal{A}_1^a, ..., i\mathcal{A}_4^a$. The arrows on Fermion lines indicate Fermion number flow, all momenta flow to the right. The thick solid line indicates interaction with the target.}\label{fig:realdiags}
\end{figure}

One-loop corrections to the LO cross section above involve radiation of a gluon from either the quark or the anti-quark. The radiated gluon can either be real (radiated in the amplitude and absorbed in the complex conjugate amplitude) or virtual (radiated and absorbed in either the amplitude or the complex conjugate amplitude). The real contributions were already computed in~\cite{Ayala:2016lhd,Ayala:2017rmh}, here we will repeat the calculation and reproduce the previous results. The virtual 
corrections have also been computed, but for a different process, namely inclusive dijets~\cite{Caucal:2021ent,Taels:2022tza} rather than dihadrons. Furthermore we use spinor helicity methods to evaluate the Dirac Algebra which leads to a tremendous simplification of the calculations. The real corrections are shown in Fig. (\ref{fig:realdiags}) and are given by~\footnote{We define $i\mathcal{M}_i$ via $i\mathcal{A}_i = 2\pi \delta(l^+-p^+-q^+) i\mathcal{M}_i$.} for quark initiated radiation,

\bea
i\mM_1^a &=& -ieg \int \frac{\dd^3 k_1}{(2\pi)^3} \dd^4 \bx \frac{16 (l^+)^2 N_1 \, t^a V_1 V_2^\dag }{ k_1^2 (k_1-l)^2} e^{i\bk_1\cdot(\bx_1-\bx_2)}e^{-i(\bp+\bk)\cdot\bx_1}e^{-i\bq\cdot\bx_2}, \,\,\,\,\, k_1^+ = p^++k^+, \nonumber \\
i\mM_3^a &=& -eg \int \frac{\dd^3 k_1}{(2\pi)^3} \frac{\dd^3 k_2}{(2\pi)^3}\dd^6\bx \frac{ 16 (l^+)^2 N_3 \, (2k^+) V_1 t^b V_2^\dag U_3^{ab}e^{i\bk_1\cdot(\bx_1-\bx_2)}e^{i\bk_2\cdot(\bx_3-\bx_1)}e^{-i(\bp\cdot\bx_1+\bq\cdot\bx_2+ \bk\cdot\bx_3)}}{k_1^2 k_2^2 (k_1-l)^2 (k_1-k_2)^2} ,\nonumber  \\ 
k_1^+ &=& p^+ + k^+, \,\,\,\,\,\, k_2^+ = k^+. \nonumber \\
\eea
Here we use $\dthree{k} = \frac{\dd k^-}{(2\pi)} \frac{\dd^2 \bk}{(2\pi)^2}$ and the Dirac numerators $N_i$ are defined 
as

\bea
N_1 &\equiv&\frac{1}{16(l^+)^2} \frac{\ubar(p)\slep(k)^* (\slp+\slk)\sln\slk_1\slep(l)(\slk_1-\sll)\sln v(q)}{(p+k)^2}.\nonumber \\
N_3 &\equiv& \frac{1}{16(l^+)^2} \ubar(p)\sln (\slk_1-\slk_2) \gamma^\mu \slk_1 \slep(l) (\slk_1-\sll)\sln v(q) d_{\mu \nu}(k_2) \epsilon^\nu(k)^*.
\eea

\noindent We define the gauge vector $n^\mu$ in $(+,-,\perp)$ notation as $(0,1,\mathbf{0})$. The anti-quark diagrams $i\mM_2$ and $i\mM_4$ are related to the quark diagrams via swapping quark and anti-quark momenta  (i.e.: $\bp \leftrightarrow \bq$ and $z_1 \leftrightarrow z_2$). Note that when a gluon line multiply  scatters from the target one gets a Wilson line in the adjoint representation $U_{j}^{ab}$.

\begin{align}
U_j^{ab}=& \hat{P} \exp\left( ig \int \dd x^+ A_c^-(x^+,\bx_j) T^c\right)_{ab}, \,\,\,\,\,\,\,\,\,\,\,\,\,\,\, \text{where} \,\,\,\,\,\,\,\,T^{c}_{ab} = if^{abc}.
\end{align}

\begin{figure}[H]
\centering
\includegraphics[width=60mm]{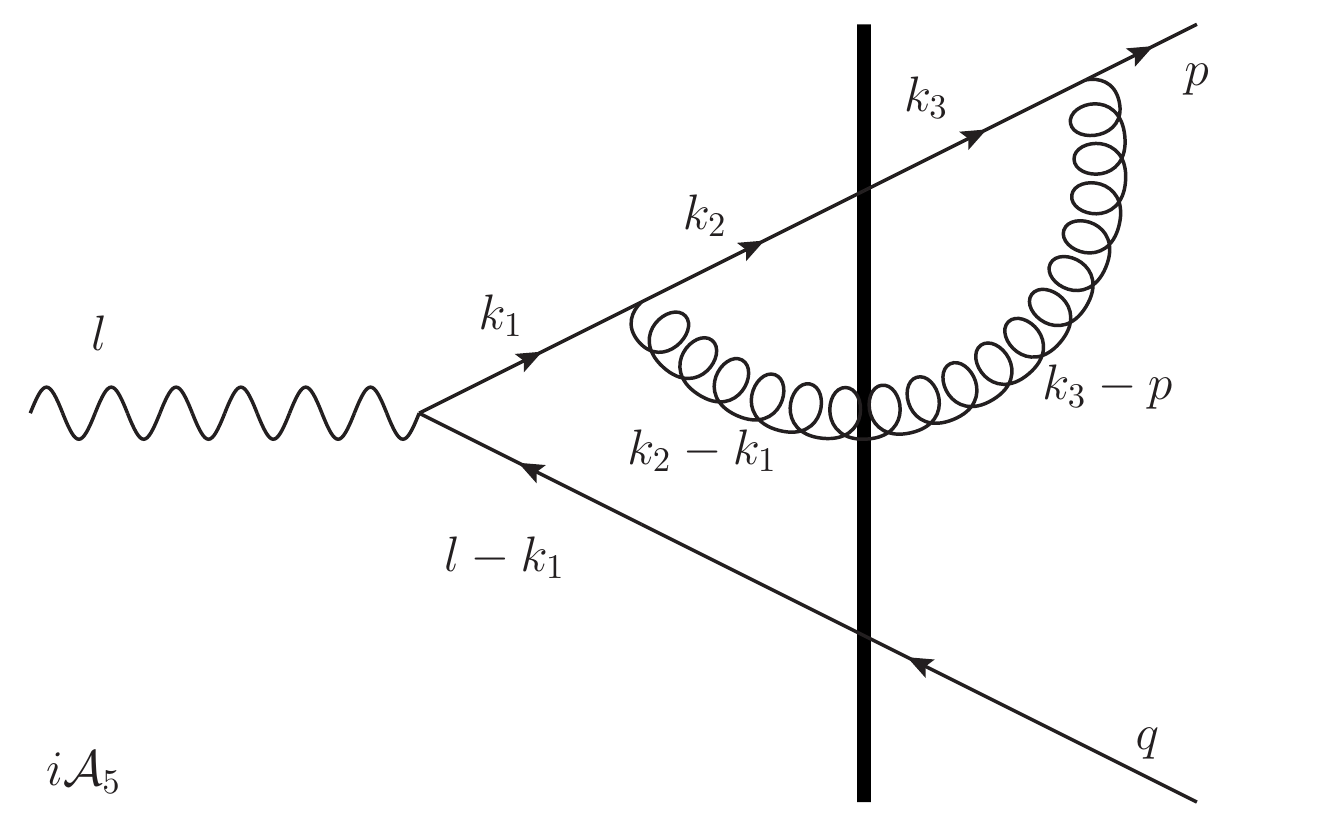}\includegraphics[width=60mm]{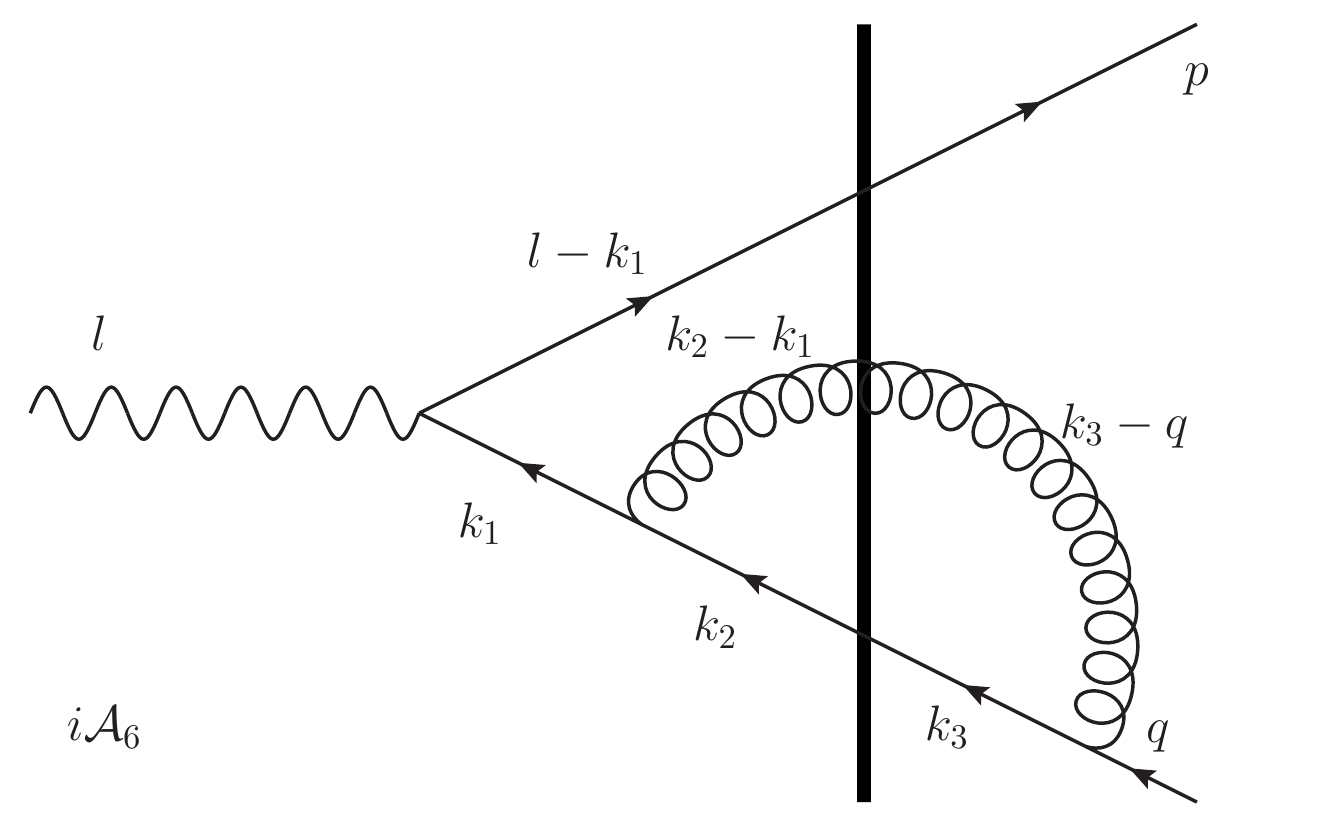}\\ \includegraphics[width=60mm]{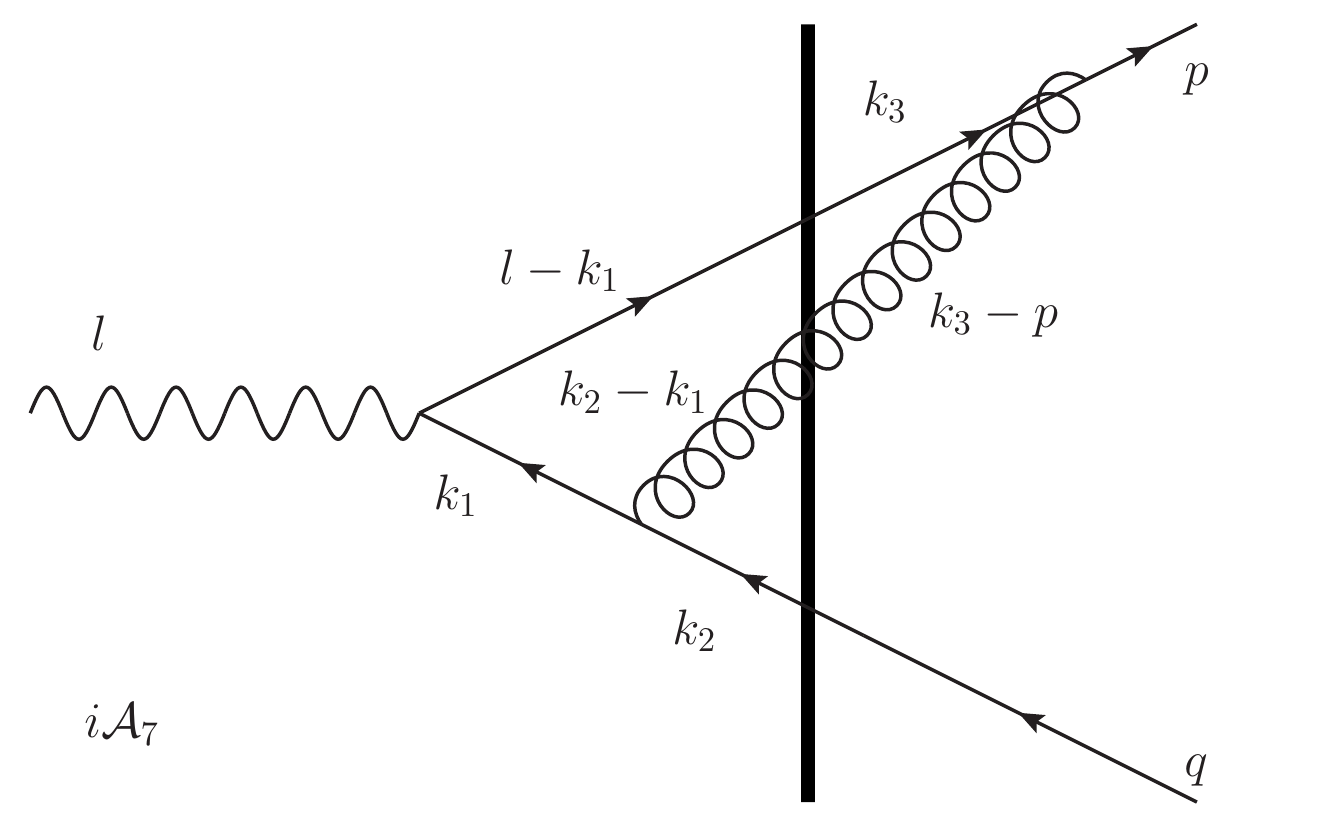}\includegraphics[width=60mm]{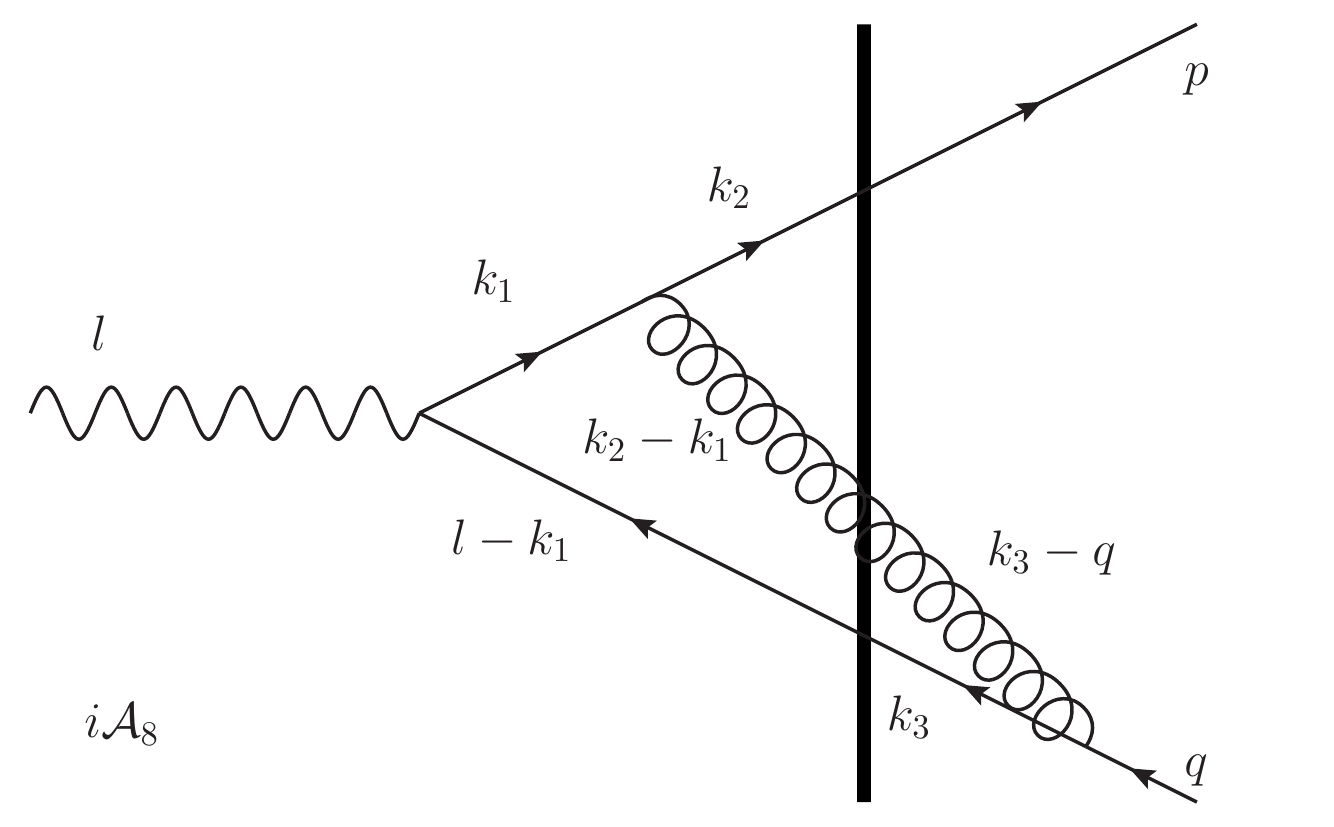}\\ \includegraphics[width=60mm]{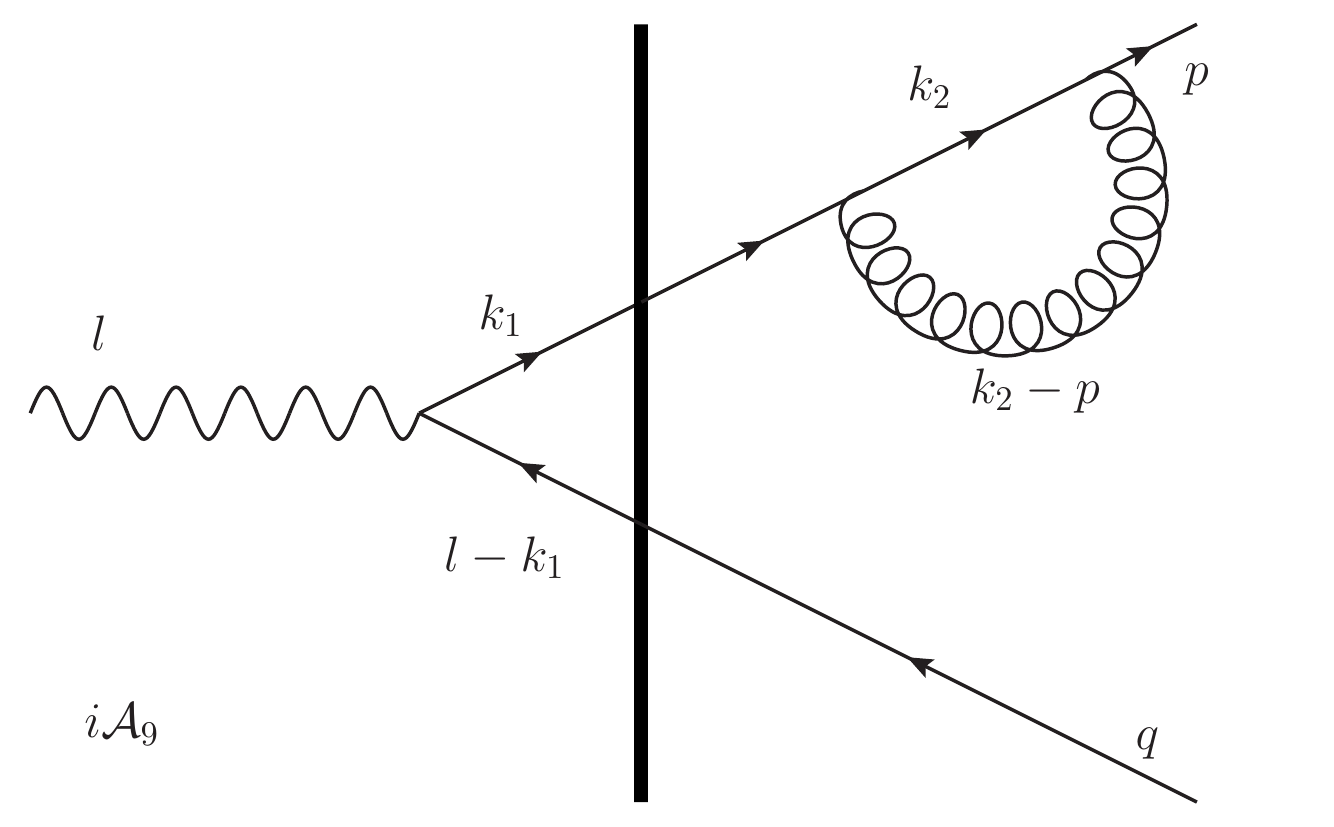}\includegraphics[width=60mm]{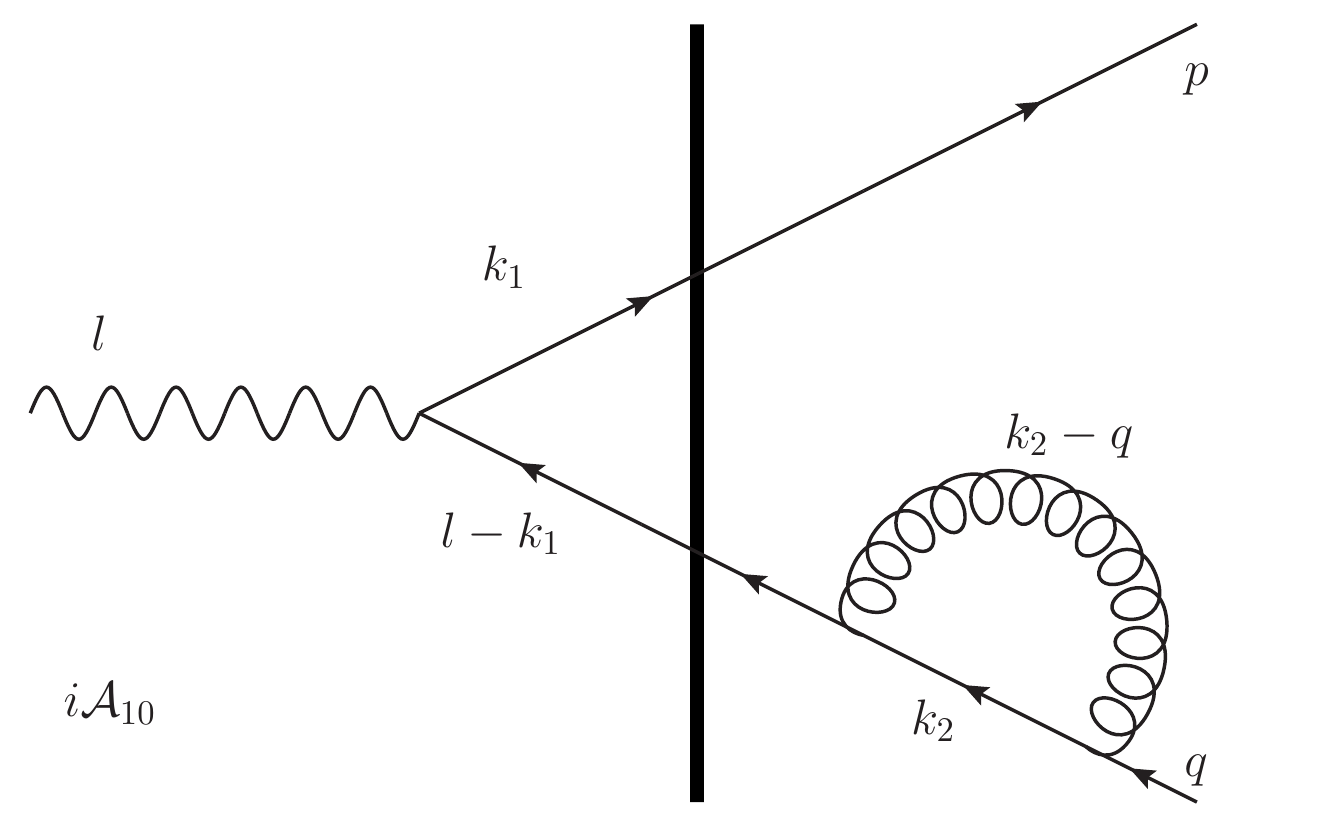}\\ \includegraphics[width=60mm]{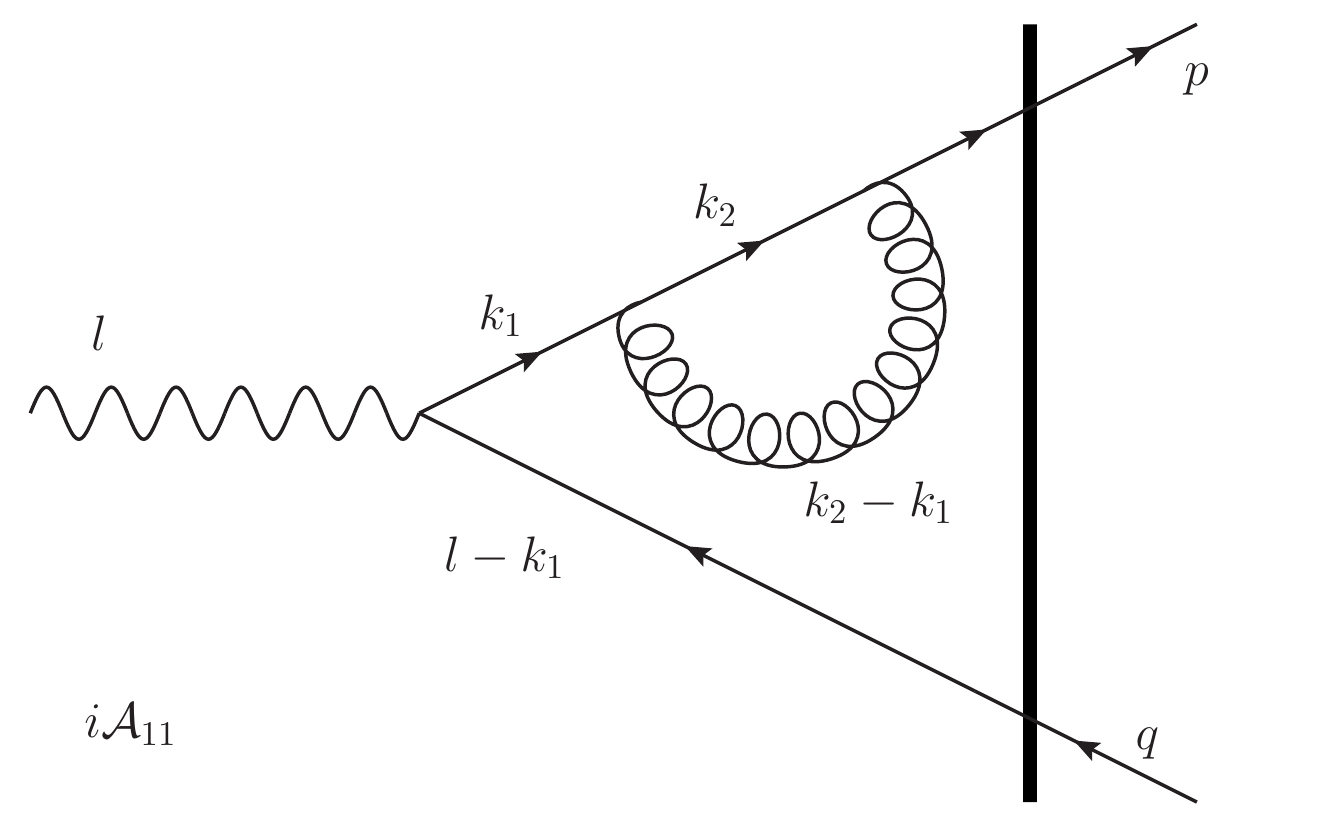}\includegraphics[width=60mm]{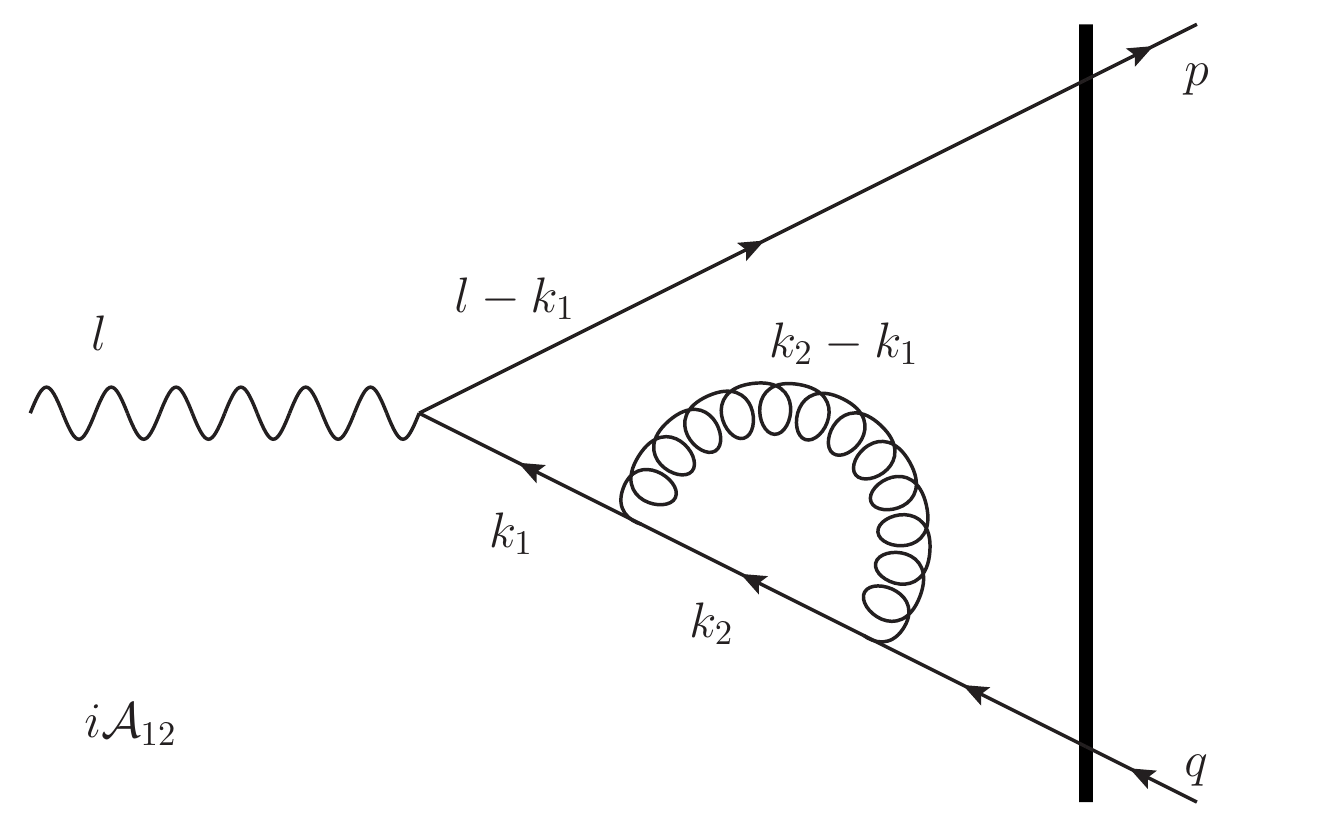}\\ \includegraphics[width=60mm]{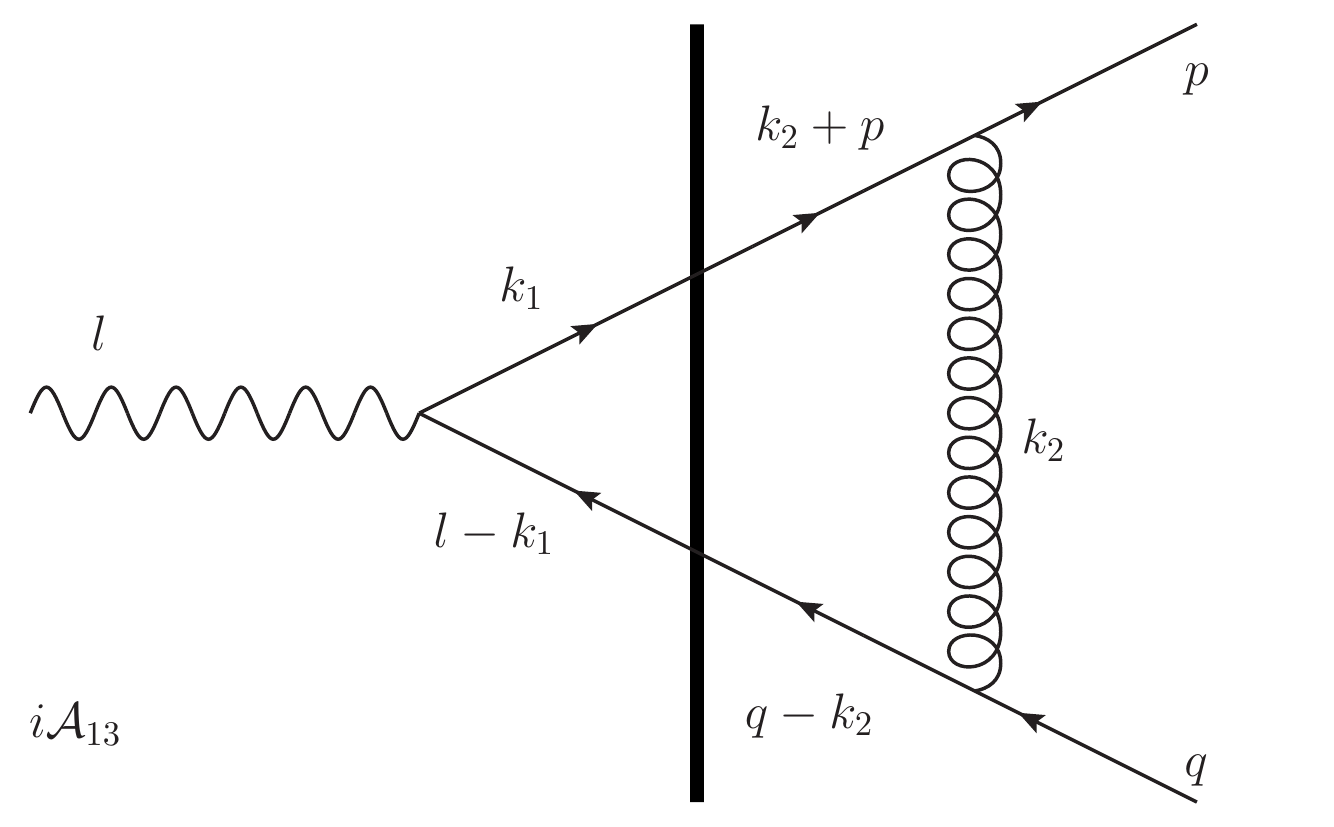}\includegraphics[width=60mm]{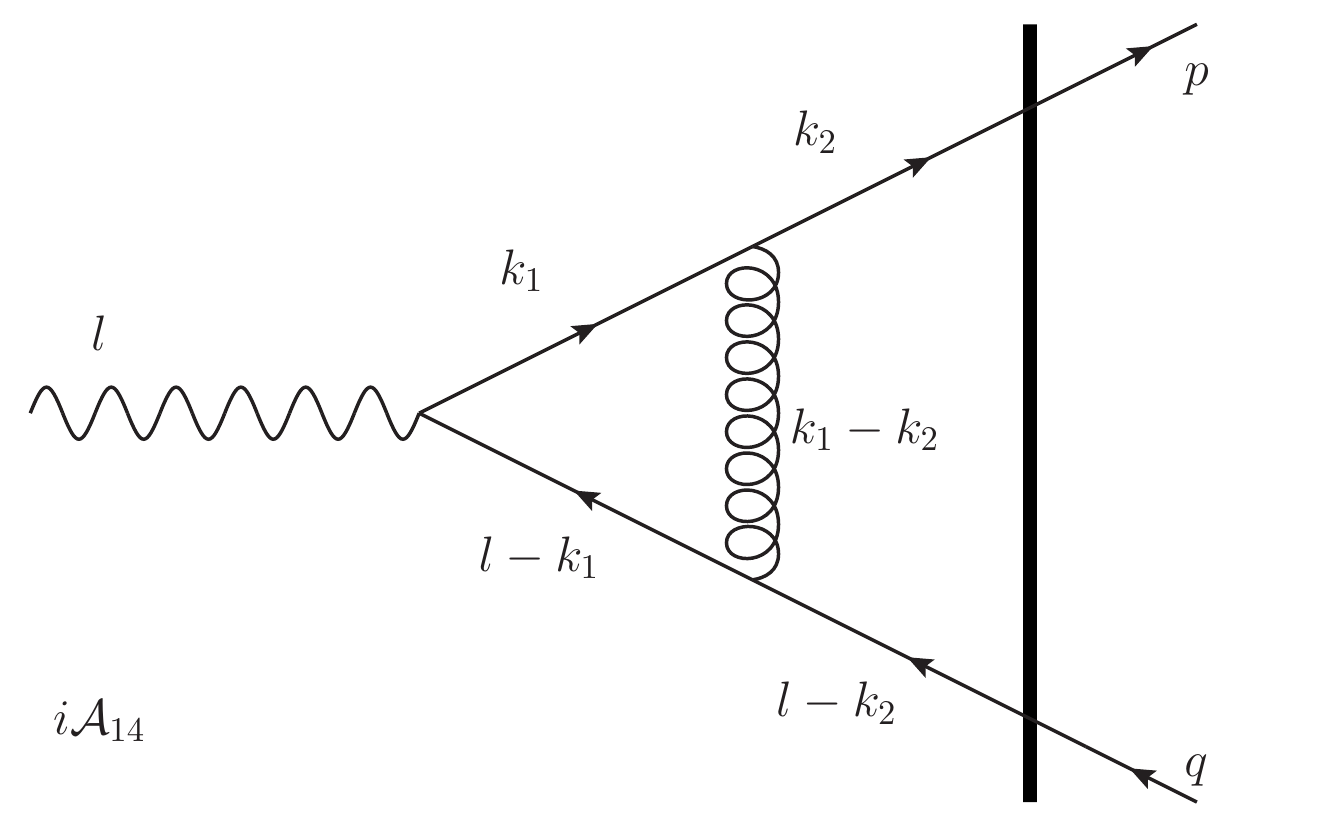}
\caption{The ten virtual NLO diagrams $i\mathcal{A}_5, ..., i\mathcal{A}_{14}$. The arrows on fermion lines indicate fermion number flow, all momenta flow to the right, \textit{except} for gluon momenta. The thick solid line indicates interaction with the target.}\label{virtualdiags}
\end{figure}

Virtual diagrams are shown in figure \ref{virtualdiags}, and the amplitudes can be written as follows,

\begin{align}
i\mathcal{M}_5 &= ieg^2  \int \frac{ \dd^3 k_1}{(2\pi)^3} \frac{ \dd^3 k_2}{(2\pi)^3} \frac{ \dd^4 k_3 }{(2\pi)^4}  \dd^6 \bx \frac{N_5 e^{i\bk_1\cdot(\bx_3-\bx_2)}e^{i\bk_2\cdot(\bx_1-\bx_3)}e^{i\bk_3\cdot(\bx_3-\bx_1)}e^{-i\bp\cdot\bx_3}e^{-i\bq\cdot\bx_2}}{k_1^2k_2^2k_3^2(k_1-l)^2(k_2-k_1)^2(k_3-p)^2}   2(k_3^+ - p^+)\nonumber \\ 
& t^a \left[ \theta(k_3^+)V_1 - \theta(-k_3^+) V^\dag_1\right] t^b V^\dag_2 \left[ \theta(k_3^+ - p^+)U^{ba}_3 - \theta(p^+ - k_3^+)U_3^{\dag ba}\right], \,\,\,\,\, k_2^+ = k_3^+, \,\,\,\,\,\, k_1^+ = p^+. \label{M5}\\
i\mathcal{M}_{7} &= ieg^2 \int \frac{\dd^3 k_1}{(2\pi)^3} \frac{\dd^3 k_2}{(2\pi)^3}\frac{\dd^4 k_3}{(2\pi)^4} \dd^6 \bb{x} \frac{N_{7} e^{i\bb{k}_3\cdot(\bb{x}_3-\bb{x}_1)} e^{i\bb{k}_2\cdot(\bb{x}_2-\bb{x}_3)}e^{i\bb{k}_1\cdot(\bb{x}_3-\bb{x}_1)}e^{-i\bb{p}\cdot \bb{x}_3} e^{-i\bb{q}\cdot\bb{x}_2}}{k_1^2k_2^2k_3^2(l-k_1)^2(k_1-k_2)^2(k_3-p)^2} 2(k_3^+-p^+) \nonumber \\ & t^a\Big[\theta(k_3^+) V_1 - \theta(-k_3^+)V_1^\dag\Big] t^b V_2^\dag \Big[\theta(k_3^+ - p^+)U_3^{ba} - \theta(p^+-k_3^+)U_3^{\dag ba}\Big], \,\,\,\,\,\, k_2^+ = q^+, \,\,\,\, k_1^+ = l^+ - k_3^+. \\
i\mM_9 &= \frac{ -eg^2}{p^2} \int \dthree{k_1} \dfour{k_2} \dd^4 \bx \frac{N_9 \, t^a t^a V_1 V_2^\dag \, e^{i\bk_1\cdot(\bx_1-\bx_2)} e^{-i(\bp\cdot\bx_1 + \bq\cdot\bx_2)}}{k_1^2 k_2^2 (k_1-l)^2 (k_2-p)^2}, \,\,\,\,\,\,\, k_1^+ = p^+.\\
i\mM_{11} &= -eg^2 \int \dthree{k_1} \dfour{k_2} \dd^4 \bx  \frac{N_{11} \, V_1 t^a t^a V_2^\dag \, e^{i\bk_1\cdot(\bx_1-\bx_2)}e^{-i(\bp\cdot\bx_1 + \bq\cdot\bx_2)}}{k_1^4 k_2^2 (k_1-l)^2 (k_1-k_2)^2}, \,\,\,\,\,\, k_1^+ = p^+.\\
i\mM_{13} &= eg^2 \int \dthree{k_1}\dfour{k_2} \dd^4 \bx \frac{N_{13}e^{i\bk_1\cdot\bx_{12}}e^{i\bk_2\cdot\bx_{21}}e^{-i(\bp\cdot\bx_1+\bq\cdot\bx_2)}}{k_1^2 k_2^2(k_1-l)^2(k_2+p)^2(k_2-q)^2} \nonumber \\
&\times t^a \Big[\theta(k_2^++p^+)V_1 -\theta(-k_2^+ - p^+)V_1^\dag\Big]\Big[\theta(k_2^+-q^+)V_2 - \theta(q^+-k_2^+)V_2^\dag\Big]t^a, \,\,\,\,\,\, k_1^+ = k_2^+ + p^+. \\
i\mathcal{M}_{14} &= -eg^2 \int \frac{ \dd^4 k_1}{(2\pi)^4}\frac{ \dd^3 k_2}{(2\pi)^3} \dd^4 \bx \frac{ N_{14}\, V_1 t^at^a V_2^\dag e^{i\bk_2\cdot(\bx_1 - \bx_2)} e^{-i(\bp\cdot \bx_1 + \bq\cdot \bx_2)}}{k_1^2 k_2^2 (k_1-l)^2  (k_2-l)^2 (k_1-k_2)^2}, \,\,\,\,\, k_2^+ = p^+. \label{M14}
\end{align}
with the virtual Dirac numerators defined as
\begin{align}
N_5 &\equiv \bar{u}(p) \gamma^\mu \slashed{k}_3 \slashed{n} \slashed{k}_2 \gamma^\nu \slashed{k}_1 \slashed{\epsilon}(l) (\slashed{k}_1 - \slashed{l})\slashed{n}v(q) d_{\nu \sigma}(k_2-k_1)d_{\mu}^{\, \sigma}(k_3-p). \\
N_{7} &\equiv \bar{u}(p)\gamma^\mu \slashed{k}_3 \slashed{n} (\slashed{l}-\slashed{k}_1) \slashed{\epsilon}(l)\slashed{k}_1 \gamma^\nu \slashed{k}_2 \slashed{n} v(q)d_{\nu \sigma}(k_2-k_1)d_\mu^\sigma (k_3-p). \\
N_9 & \equiv \ubar(p) \gamma^\mu (\slp-\slk_2) \gamma^\nu \slp \sln \slk_1 \slep(l) (\slk_1-\sll)\sln v(q) d_{\mu \nu}(k_2). \\
N_{11} &\equiv  \ubar(p) \sln \slk_1 \gamma^\mu \slk_2 \gamma^\nu \slk_1 \slep(l) (\slk_1-\sll)\sln v(q) d_{\mu \nu}(k_1-k_2).\\
N_{13} &\equiv \ubar(p) \gamma^\mu (\slk_2+\slp)\sln\slk_1\slep(l)(\slk_1-\sll)\sln(\slk_2-\slq)\gamma^\nu v(q)d_{\mu \nu}(k_2). \\
N_{14} &\equiv \bar{u}(p)\slashed{n}\slashed{k}_2\gamma^\mu \slashed{k}_1 \slashed{\epsilon}(l) (\slashed{k}_1 - \slashed{l})\gamma^\nu (\slashed{k}_2 - \slashed{l}) \slashed{n} v(q) d_{\mu \nu}(k_1 - k_2).
\end{align}

\noindent Here again the anti-quark diagrams ($i\mM_6, i\mM_8, i\mM_{10}, i\mM_{12}$) are related to the corresponding quark diagrams via swapping quark and anti-quark momenta. The tensor $d_{\mu \nu}$ is defined as follows.

\begin{align}
d_{\mu \nu}(p) = -g_{\mu \nu} + \frac{ p_\mu n_\nu + p_\nu n_\mu}{n \cdot p}.
\end{align}

\noindent In this study, we focus on the contribution from longitudinally polarized photons and we compute the numerators using the spinor helicity formalism. The necessary real numerators are in table \ref{numtable} and the virtual numerators are in Eq. \ref{N5} - \ref{N14}.

\begin{table}[H]
\centering
\begin{tabular}{| c
 | c | l |}
\hline
Numerator & $\lambda_\gamma ; \lambda_q ,\lambda_g$ & $N_i^{\lambda_\gamma ; \lambda_q, \lambda_g}$ \\[2pt]
\hline
\multirow{6}{1em}{\vspace{9ex}$N_1$} & $L;+,+$ & $-Q (z_1z_2)^{3/2}(1-z_2)\frac{[(z_1 \bk-z_3\bp)\de]}{(z_1\bk-z_3\bp)^2}$ \\[5pt] 
& $L;+,-$ & $-Q (z_2)^{3/2}\sqrt{z_1}  (1-z_2)^2\frac{[(z_1\bk-z_3\bp)\de]}{(z_1\bk-z_3\bp)^2}$ \\[5pt] 
\hline
\multirow{6}{1em}{\vspace{9ex}$N_2$} & $L;+,+$ & $ Q (z_1)^{3/2}\sqrt{z_2}(1-z_1)^2 \frac{[(z_2\bk-z_3\bq)\de]}{(z_2\bk-z_3\bq)^2}$ \\ [5pt] 
& $L;+,-$ & $  Q(z_1z_2)^{3/2}(1-z_1) \frac{[(z_2\bk-z_3\bq)\des]}{(z_2\bk-z_3\bq)^2}$ \\ [5pt] 
\hline
\multirow{6}{1em}{\vspace{9ex}$N_3$} & $L;+,+$ & $ Q(z_1z_2)^{3/2}(1-z_2) \left(\frac{\bk_2\de}{z_3}-\frac{\bk_1\de}{1-z_2}\right)$ \\ [5pt] 
& $L;+,-$ &  $Q(z_2)^{3/2}\sqrt{z_1}(1-z_2)^2 \left( \frac{ \mathbf{k}_2 \cdot \boldsymbol{\epsilon}^*}{z_3} - \frac{ \mathbf{k}_1 \cdot \boldsymbol{\epsilon}^*}{1-z_2}\right)$ \\ [5pt] 
\hline
\multirow{6}{1em}{\vspace{9ex}$N_4$} & $L;+,+$ & $- Q  (z_1)^{3/2} \sqrt{z_2}(1-z_1)^2\left( \frac{ \mathbf{k}_2 \cdot \boldsymbol{\epsilon}}{z_3} - \frac{ \mathbf{k}_1 \cdot \boldsymbol{\epsilon}}{1-z_1}\right) $\\[5pt] 
& $L;+,-$ & $-Q(z_1z_2)^{3/2}(1-z_1)\left( \frac{ \mathbf{k}_2 \cdot \boldsymbol{\epsilon}^*}{z_3} - \frac{ \mathbf{k}_1 \cdot \boldsymbol{\epsilon}^*}{1-z_1}\right)$ \\ [5pt] 
\hline
\end{tabular}
\caption{The minimal set of real numerators $N_1$ to $N_4$ in momentum fraction notation. Complex conjugation results in the numerator with all helicities flipped (while leaving longitudinal helicities unchanged), and any numerator where $\lambda_q = \lambda_{\bar{q}}$ is zero. We have also defined $z_3$ to be the gluon's momentum fraction relative to the photon ($z_3 = k^+/l^+$).}
\label{numtable}
\end{table}

\noindent We also need the Dirac numerators for the virtual diagrams, they are

\begin{align}
N_5^{L;+} =& \frac{2^5 (l^+)^2 Q  (z_1 z_2)^{3/2}}{(z_1-z_3)^2} \Bigg\{ z_1^2\left[\left(\bk_3-\frac{z_3}{z_1}\bp\right)\de\right]\left[\left(\bk_2-\frac{z_3}{z_1}\bk_1\right)\de^*\right] + z_3^2\left[\left(\bk_3-\frac{z_3}{z_1}\bp\right)\de^*\right]\left[\left(\bk_2-\frac{z_3}{z_1}\bk_1\right)\de\right]\Bigg\}. \label{N5}\\
N_{7}^{L;+} =& \frac{2^5 (l^+)^2 Q z_3 \sqrt{z_1z_2}}{(z_1-z_3)^2}\Bigg\{ z_1 z_2 \left[\left( \bk_3 - \frac{z_3}{z_1} \bp\right)\de\right] \Big[\Big( z_2 \bk_1-(1-z_3)\bk_2 \Big)\de^*\Big]\nonumber \\ & + z_3(1-z_3)\left[\left( \bk_3 - \frac{z_3}{z_1} \bp\right)\de^*\right] \Big[\Big( z_2 \bk_1-(1-z_3)\bk_2 \Big)\de \Big]\Bigg].\\
N_9^{L;+}=& 2^4 Q (l^+)^2 (z_1 z_2)^{3/2}\left[k_2^2 +\frac{(2z_1-z)}{z} (k_2-p)^2 - \frac{[z_1^2+(z_1-z)^2]}{z_1 z} p^2\right].\\
N_{11}^{L;+} =& 2^4 (l^+)^2 Q (z_1 z_2)^{3/2} \left[ \frac{\left[ (k_2^+)^2 + (p^+)^2\right]}{p^+(k_2^+-p^+)} k_1^2 + \frac{(p^++k_2^+)}{(p^+-k_2^+)} k_2^2 + (k_1-k_2)^2\right].\\
N_{13}^{L;+} = &2^5 Q (l^+)^2 (z_1+z_3)(z_2-z_3)\sqrt{z_1 z_2}\Bigg[ z_1z_3\left(\frac{\bp\de}{z_1}-\frac{\bq\de}{z_2}\right)\left(\frac{\bp\des}{z_1}-\frac{\bk_2\des}{z_3}\right) +z_3^2\left(\frac{\bk_2\de}{z_3}-\frac{\bq\de}{z_2}\right)\left(\frac{\bp\des}{z_1}-\frac{\bk_2\des}{z_3}\right) \nonumber \\
& -z_2z_3\left(\frac{\bk_2\de}{z_3}-\frac{\bq\de}{z_2}\right)\left(\frac{\bp\des}{z_1}-\frac{\bq\des}{z_2}\right) - p\cdot q -\frac{(z_1+z_3)}{2z_3}(k_2-q)^2 + \frac{(z_2-z_3)}{2z_3}(k_2+p)^2\Bigg].\\
N_{14}^{L;+} =& \frac{2^5 Q (l^+)^2 (z_1z_2)^{3/2} z_3}{(z_3-z_1)} \Bigg[ 2l^+(1-z_3)(k_1^- - k_2^-) + \frac{(z_3 - z_2)}{z_3} (\bk_1 \cdot \be)(\bk_1 \cdot \be^*) + \frac{(z_1-z_3)}{z_2} (\bk_2\cdot \be)(\bk_1\cdot \be^*) \nonumber \\
& + \frac{(1-z_3)(z_1-z_3)}{z_1z_3} (\bk_1 \cdot \be)(\bk_2 \cdot \be^*) + \frac{(z_2 - z_3)(1-z_3)}{z_1 z_2} (\bk_2 \cdot \be)(\bk_2 \cdot \be^*)\Bigg].\label{N14}
\end{align}

\noindent In all these virtual numerators, $z_3$ is the momentum fraction for the remaining internal $+$ momentum (see Eq. \ref{M5} - \ref{M14}). We have also defined the transverse vector $\be$ as

\begin{align}
\be = \frac{1}{\sqrt{2}}(1,i).
\end{align}




\section{Results}\label{Production cross section}

To calculate the $\mathcal{O}(\alpha_s)$ corrections to the production cross section we need to multiply the helicity amplitudes with the corresponding conjugate amplitudes (real amplitudes need to be multiplied with each other, virtual amplitudes need to be multiplied by the leading order amplitude $i\mathcal{M}$), include the phase space factors $\dd \Phi^{n}$, and divide by the incident flux $\mathcal{F}$ which we'll take to be $2l^+$. We also relabel and shift (where necessary) on the remaining $z_3$ integral in each expression so that the gluon's momentum fraction is always labeled $z$. We'll write the real corrections as $\sigma_{i\times j}$ for $i,j=1,...,4$ and the virtual corrections as $\sigma_i$ for $i=5,...,14$. 

\begin{align}
\dd \sigma^L_{\text{NLO}} = \sum_{i,j=1}^4 \dd \sigma^L_{i\times j} +2\,\text{Re} \sum_{i = 5}^{14} \dd \sigma_{i}^L. \label{breakdown}
\end{align}

\begin{align}
\dd \sigma^{ L}_{i\times j} = \frac{1}{\mathcal{F}} \int_{z,\bk} \left[ (i\mM_i^a)(i\mM_j^a)^{* L}\right] \dd \Phi^{(3)}, \,\,\,\,\,\, \dd \sigma^{L}_i = \frac{1}{\mathcal{F}} \left[ (i\mM_i)(i\mM)^{*,L}\right]\dd \Phi^{(2)}.
\end{align}

\begin{align}
\dd \Phi^{(3)} = 2l^+ \frac{ \dd^2 \bp \, \dd^2 \bq\, \dd^2 \bk \, \dd y_1 \, \dd y_2 \, \dd z}{(2\pi)^8 (4l^+)^2z} \delta(1-z_1-z_2-z), \,\,\,\,\,\,\, \dd \Phi^{(2)} = 2l^+ \frac{ \dd^2 \bp \, \dd^2 \bq\, \dd y_1 \, \dd y_2}{2(2\pi)^5 (2l^+)^2} \delta(1-z_1-z_2). 
\end{align}

\noindent Here the $L$ signifies that we are including contributions only from longitudinally polarized photons, and imply that we have summed over all outgoing polarizations. We note that all real corrections come with a $\delta(1-z_1-z_2-z)$ whereas all virtual corrections come with a $\delta(1-z_1-z_2)$. We omit these delta functions here for brevity and restore them at the end. In many cases, it is easiest to write the results in coordinate space with the radiation kernel $\Delta^{(3)}_{ij}$ defined as follows.

\begin{align}
\Delta^{(3)}_{ij} = \frac{\bx_{3i}\cdot\bx_{3j}}{\bx_{3i}^2\bx_{3j}^2}.
\end{align}

\noindent The real corrections are:

\begin{align}
\frac{\dd \sigma^{L}_{1\times 1}}{\dd^2 \bp\, \dd^2 \bq\, \dd y_1 \, \dd y_2} = &\frac{2 e^2 g^2  Q^2 N_c^2 z_2^3 (1-z_2)^2(z_1^2+(1-z_2)^2)}{(2\pi)^{10} z_1}\int \frac{\dd z}{z}  \int \dd^{10} x \,K_0(|\bx_{12}|Q_2)K_0(|\bx_{1^\prime 2^\prime}|Q_2)\Delta^{(3)}_{11^\p} \nonumber \\ 
&\left[ S_{122^\p 1^\p} - S_{12} - S_{1^\p 2^\p} + 1\right] e^{i\bp\cdot(\bx_1^\prime - \bx_1)}e^{i\bq\cdot(\bx_2^\prime - \bx_2)} e^{i\frac{z}{z_1}\bp\cdot(\bx_1^\p-\bx_1)} . \label{Real11} \\
\frac{\dd \sigma^{L}_{2\times 2}}{\dd^2 \bp\, \dd^2 \bq\, \dd y_1 \, \dd y_2} = & \frac{2 e^2 g^2  Q^2 N_c^2 z_1^3  (1-z_1)^2(z_2^2+(1-z_1)^2)}{(2\pi)^{10}  z_2}\int \frac{\dd z}{z}\int \dd^{10} \bx \, K_0(|\bx_{12}|Q_1) K_0(|\bx_{1^\p 2^\p}|Q_1) \Delta^{(3)}_{22^\p}\nonumber \\
& [S_{122^\p 1^\p} - S_{12} - S_{1^\p 2^\p} + 1]e^{i\bq\cdot(\bx_2^\p - \bx_2)} e^{i\bp\cdot(\bx_1^\p - \bx_1)}e^{i\frac{z}{z_2}\bq\cdot(\bx_2^\p-\bx_2)} . \label{Real22} \\
\frac{\dd \sigma^{L}_{1\times 2}}{\dd^2 \bp\, \dd^2 \bq\, \dd y_1 \, \dd y_2} = & \frac{-2e^2 g^2  Q^2 N_c^2z_1 z_2 (1-z_1) (1-z_2)(z_1(1-z_1)+z_2(1-z_2))}{(2\pi)^{10} }\int \frac{\dd z}{z}\int \dd^{10} \bx \,K_0(|\bx_{12}|Q_2)\nonumber \\ 
&K_0(|\bx_{1^\p2^\p}| Q_1)\Delta^{(3)}_{12^\p}[S_{12}S_{1^\p2^\p}-S_{12}-S_{1^\p2^\p}+1] e^{i\bp\cdot(\bx_1^\p-\bx_1)}e^{i\bq\cdot(\bx_2^\p-\bx_2)}e^{i\frac{z}{z_1}\bp\cdot(\bx_3-\bx_1)}e^{i\frac{z}{z_2}\bq\cdot(\bx_2^\p-\bx_3)}.\label{Real12}\\
\frac{\dd \sigma^{L}_{3\times 3}}{\dd^2 \bp\, \dd^2 \bq\, \dd y_1 \, \dd y_2}=& \frac{2e^2 g^2 Q^2 N_c^2 z_1 z_2^3(z_1^2+(1-z_2)^2)}{(2\pi)^{10}}\int \frac{\dd z}{z} \int \dd^{10}\bx\, K_0(Q X) K_0(Q X^\p) \Delta^{(3)}_{11^\p} \nonumber \\
&[S_{11^\p} S_{22^\p} - S_{13}S_{23} - S_{1^\p 3} S_{2^\p 3}+1] e^{i\bp\cdot(\bx_1^\p-\bx_1)}e^{i\bq\cdot(\bx_2^\p-\bx_2)}.\\
\frac{\dd \sigma^{L}_{4\times 4}}{\dd^2 \bp\, \dd^2 \bq\, \dd y_1 \, \dd y_2} =& \frac{2e^2 g^2 Q^2 N_c^2 z_1^3 z_2(z_2^2+(1-z_1)^2)}{(2\pi)^{10}}\int \frac{\dd z}{z} \int \dd^{10}\bx \,K_0(Q X) K_0(Q X^\p) \Delta^{(3)}_{22^\p}\nonumber \\
&[S_{11^\p} S_{22^\p} - S_{13}S_{23} - S_{1^\p 3} S_{2^\p 3}+1] e^{i\bp\cdot(\bx_1^\p-\bx_1)}e^{i\bq\cdot(\bx_2^\p-\bx_2)}.\\
\frac{\dd \sigma^{L}_{3\times 4}}{\dd^2 \bp\, \dd^2 \bq\, \dd y_1 \, \dd y_2} =& \frac{-2e^2 g^2  Q^2N_c^2 z_1^2 z_2^2(z_1(1-z_1)+z_2(1-z_2))}{(2\pi)^{10}}\int \frac{\dd z}{z} \int \dd^{10}\bx \, K_0(Q X) K_0(Q X^\p) \Delta^{(3)}_{12^\p} \nonumber \\
&[S_{11^\p} S_{22^\p} - S_{13}S_{23} - S_{1^\p 3} S_{2^\p 3}+1] e^{i\bp\cdot(\bx_1^\p-\bx_1)}e^{i\bq\cdot(\bx_2^\p-\bx_2)}.\\
\frac{\dd \sigma^{L}_{1\times 3}}{\dd^2 \bp\, \dd^2 \bq\, \dd y_1 \, \dd y_2}=& \frac{-2e^2 g^2 Q^2 N_c^2 z_2^3 (1-z_2)(z_1^2+(1-z_2)^2)}{(2\pi)^{10} }\int \frac{\dd z}{z} \int \dd^{10}x\, K_0(|\bx_{12}|Q_2)K_0(QX^\p) \Delta^{(3)}_{11^\p} \nonumber \\
& [S_{122^\p3} S_{1^\p3} - S_{1^\p 3} S_{2^\p 3} - S_{12} + 1] e^{i\bp\cdot(\bx_1^\p - \bx_1)}e^{i\bq\cdot(\bx_2^\p-\bx_2)} e^{i\frac{z}{z_1}\bp\cdot(\bx_3-\bx_1)}.\\
\frac{\dd \sigma^{L}_{1\times 4}}{\dd^2 \bp\, \dd^2 \bq\, \dd y_1 \, \dd y_2}=& \frac{2e^2 g^2 Q^2 N_c^2 z_1 z_2^2 (1-z_2)(z_1(1-z_1)+z_2(1-z_2))}{(2\pi)^{10}}\int \frac{\dd z}{z} \int \dd^{10}x  K_0(|\bx_{12}|Q_2)K_0(QX^\p) \Delta^{(3)}_{12^\p} \nonumber \\
&  [S_{122^\p3} S_{1^\p3} - S_{1^\p 3} S_{2^\p 3} - S_{12} + 1]  e^{i\bp\cdot(\bx_1^\p - \bx_1)}e^{i\bq\cdot(\bx_2^\p-\bx_2)}e^{i\frac{z}{z_1}\bp\cdot(\bx_3-\bx_1)}.\\
\frac{\dd \sigma^{L}_{2\times 3}}{\dd^2 \bp\, \dd^2 \bq\, \dd y_1 \, \dd y_2} =& \frac{2e^2 g^2 Q^2 N_c^2  z_1^2 z_2 (1-z_1)(z_1(1-z_1)+z_2(1-z_2))}{(2\pi)^{10}}\int \frac{\dd z}{z} \int \dd^{10}x K_0(|\bx_{12}|Q_1) K_0(QX^\p) \Delta^{(3)}_{21^\p} \nonumber \\
&[S_{1231^\p} S_{2^\p 3} - S_{1^\p 3} S_{2^\p 3} - S_{12} + 1] e^{i\bp\cdot(\bx_1^\p-\bx_1)}e^{i\bq\cdot(\bx_2^\p-\bx_2)}e^{i\frac{z}{z_2}\bq\cdot(\bx_3-\bx_2)}.\\ 
\frac{\dd \sigma^{L}_{2\times 4}}{\dd^2 \bp\, \dd^2 \bq\, \dd y_1 \, \dd y_2}=& \frac{-2e^2 g^2 Q^2 N_c^2  z_1^3  (1-z_1)(z_2^2+(1-z_1)^2)}{(2\pi)^{10}}\int \frac{\dd z}{z} \int \dd^{10}x  K_0(|\bx_{12}|Q_1) K_0(QX^\p) \Delta^{(3)}_{22^\p}\nonumber \\
&[S_{1231^\p} S_{2^\p 3} - S_{1^\p 3} S_{2^\p 3} - S_{12} + 1]  e^{i\bp\cdot(\bx_1^\p-\bx_1)}e^{i\bq\cdot(\bx_2^\p-\bx_2)}e^{i\frac{z}{z_2}\bq\cdot(\bx_3-\bx_2)}. \label{Real24}
\end{align}

\noindent This concludes the real corrections, next we have the virtual corrections:

\begin{align}
\frac{\dd \sigma^{L}_5}{\dd^2 \bp \, \dd^2 \bq \, \dd y_1 \, \dd y_2} = & \frac{2e^2g^2Q^2 N_c^2z_2^3 z_1}{(2\pi)^{10} } \int_0^{z_1} \frac{\dd z}{z} \dd^{10} \bx \,[S_{322^\p 1^\p} S_{13} - S_{13}S_{23} - S_{1^\p 2^\p} + 1](z_1^2+(z_1-z)^2) \nonumber \\
&\frac{K_0(QX_5)K_0(|\bx_{1^\p 2^\p}|Q_1)}{\bx_{31}^2} e^{i\bp\cdot(\bx_1^\p-\bx_1)}e^{i\bq\cdot(\bx_2^\p-\bx_2)} e^{-i\frac{z}{z_1} \bp\cdot(\bx_3-\bx_1)}. \label{dsig5}\\
\frac{\dd \sigma^{L}_6}{\dd^2 \bp \, \dd^2 \bq \, \dd y_1 \, \dd y_2} = &\frac{2e^2g^2Q^2N_c^2z_1^3 z_2}{(2\pi)^{10}} \int_0^{z_2} \frac{\dd z}{z} \dd^{10} \bx [S_{132^\p 1^\p}S_{23} - S_{13}S_{23} - S_{1^\p 2^\p}+1](z_2^2+(z_2-z)^2) \nonumber \\
&\frac{K_0(QX_6)K_0(|\bx_{1^\p 2^\p}|Q_1)}{\bx_{32}^2}e^{i\bp\cdot(\bx_1^\p-\bx_1)}e^{i\bq\cdot(\bx_2^\p-\bx_2)}e^{-i\frac{z}{z_2}\bq\cdot(\bx_3-\bx_2)}. \label{dsig6}\\
\frac{\dd \sigma^{L}_{7}}{\dd^2 \bp \, \dd^2 \bq \, \dd y_1 \, \dd y_2} = & \frac{-2e^2g^2Q^2 N_c^2 z_2^2 z_1}{(2\pi)^{10}} \int_0^{z_1}\frac{\dd z \,(z_1- z)}{z} \dd^{10} \bx\,  [S_{322^\p 1^\p} S_{13} - S_{13}S_{23} - S_{1^\p 2^\p} + 1](z_1z_2+(z_1-z)(z_2+z)) \nonumber \\
& K_0(QX_5)  K_0(|\bx_{1^\p2^\p}|Q_1) \Delta^{(3)}_{12} e^{i\bp\cdot(\bx_1^\p-\bx_1)}e^{i\bq\cdot(\bx_2^\p-\bx_2)} e^{-i\frac{z}{z_1}\bp\cdot(\bx_3-\bx_1)}. \label{dsig7} \\
\frac{\dd \sigma^{L}_{8}}{\dd^2 \bp \, \dd^2 \bq \, \dd y_1 \, \dd y_2} = & \frac{-2e^2g^2 Q^2 N_c^2z_1^2 z_2}{(2\pi)^{10}} \int_0^{z_2} \frac{\dd z \, (z_2-z)}{z} \dd^{10}\bx\,  [S_{132^\p 1^\p}S_{23} - S_{13}S_{23} - S_{1^\p 2^\p}+1] (z_1z_2+(z_2-z)(z_1+z))\nonumber\\
&K_0(QX_6)K_0(|\bx_{1^\p2^\p}|Q_1) \Delta^{(3)}_{12} e^{i\bp\cdot(\bx_1^\p-\bx_1)}e^{i\bq\cdot(\bx_2^\p-\bx_2)}e^{-i\frac{z}{z_2}\bq\cdot(\bx_3-\bx_2)}.\label{dsig8}\\
\frac{\dd \sigma_9^{L}}{\dd^2 \bp \, \dd^2 \bq\, \dd y_1\, \dd y_2} = &\frac{-e^2 g^2 Q^2 N_c^2 (z_1 z_2)^3}{(2\pi)^8} \int \dd^8 \bx \big[S_{122^\p 1^\p} - S_{12} - S_{1^\p 2^\p} + 1\big] K_0(|\bx_{12}|Q_1) K_0(|\bx_{1^\p 2^\p}|Q_1) e^{i\bp\cdot(\bx_1^\p-\bx_1)}e^{i\bq\cdot(\bx_2^\p-\bx_2)} \nonumber \\
&\times \int_0^{z_1} \frac{\dd z}{z}\frac{ [z_1^2+(z_1-z)^2]}{z_1^2} \int \dtwo{\bk_2} \frac{1}{\left(\bk_2-\frac{z}{z_1}\bp\right)^2}.\label{dsig9} \\
\frac{\dd \sigma_{10}^{L}}{\dd^2 \bp \, \dd^2 \bq\, \dd y_1\, \dd y_2} = &\frac{-e^2 g^2 Q^2 N_c^2 (z_1 z_2)^3}{(2\pi)^8} \int \dd^8 \bx \big[S_{122^\p 1^\p} - S_{12} - S_{1^\p 2^\p} + 1\big] K_0(|\bx_{12}|Q_1) K_0(|\bx_{1^\p 2^\p}|Q_1) e^{i\bp\cdot(\bx_1^\p-\bx_1)}e^{i\bq\cdot(\bx_2^\p-\bx_2)} \nonumber \\
&\times \int_0^{z_2} \frac{\dd z}{z}\frac{ [z_2^2+(z_2-z)^2]}{z_2^2} \int \dtwo{\bk_2} \frac{1}{\left(\bk_2-\frac{z}{z_2}\bq\right)^2}.\label{dsig10}\\
\frac{\dd \sigma_{11}^{L}}{\dd^2 \bp\, \dd^2 \bq\, \dd y_1 \, \dd y_2} =& \frac{-2e^2g^2Q^2N_c^2(z_1 z_2)^3}{(2\pi)^7} \int \dd^8 \bx\big[S_{122^\p1^\p}-S_{12}-S_{1^\p 2^\p}+1\big] K_0(|\bx_{1^\p 2^\p}|Q_1) e^{i\bp\cdot(\bx_1^\p-\bx_1)}e^{i\bq\cdot(\bx_2^\p-\bx_2)} \nonumber\\
&\int_0^{z_1} \frac{\dd z}{z} \frac{[z_1^2+(z_1-z)^2]}{z_1^2} \int \dtwo{\bk_2} \int \dtwo{\bk_1}  \frac{e^{i\bk_1\cdot(\bx_1-\bx_2)}}{\big[\bk_1^2+Q_1^2\big]\left[ \left(\bk_2-\frac{z}{z_1}\bk_1\right)^2+\frac{z(z_1-z)}{z_1^2 z_2}\bk_1^2 +\frac{z}{z_1}(z_1-z)Q^2\right]}.\label{dsig11}\\
\frac{\dd \sigma_{12}^{L}}{\dd^2 \bp\, \dd^2 \bq\, \dd y_1 \, \dd y_2} =& \frac{-2e^2g^2Q^2N_c^2(z_1 z_2)^3}{(2\pi)^7} \int \dd^8 \bx\big[S_{122^\p1^\p}-S_{12}-S_{1^\p 2^\p}+1\big] K_0(|\bx_{1^\p 2^\p}|Q_1) e^{i\bp\cdot(\bx_1^\p-\bx_1)}e^{i\bq\cdot(\bx_2^\p-\bx_2)} \nonumber\\
&\int_0^{z_2} \frac{\dd z}{z} \frac{[z_2^2+(z_2-z)^2]}{z_2^2} \int \dtwo{\bk_2} \int \dtwo{\bk_1}  \frac{e^{i\bk_1\cdot(\bx_1-\bx_2)}}{\big[\bk_1^2+Q_1^2\big]\left[ \left(\bk_2-\frac{z}{z_2}\bk_1\right)^2+\frac{z(z_2-z)}{z_2^2 z_1}\bk_1^2 +\frac{z}{z_2}(z_2-z)Q^2\right]}.\label{dsig12}
\end{align}

\begin{align}
& \frac{\dd \sigma_{13(1)}^{L}}{\dd^2\bp\,\dd^2\bq\,\dd y_1\,\dd y_2} =  \frac{-e^2g^2 Q^2 N_c^2 (z_1z_2)^2}{(2\pi)^8}\int_{0}^{z_2}\frac{ \dd z}{z} (z_1+z)(z_2-z) \int \dd^8\bx \,K_0\left(|\bx_{12}|Q\sqrt{(z_1+z)(z_2-z)}\right)K_0(|\bx_{1^\p 2^\p}|Q_1)\nonumber \\
& e^{i\bp\cdot(\bx_{1^\p }-\bx_1)}e^{i\bq\cdot(\bx_{2^\p }-\bx_2)}[S_{12}S_{1^\p 2^\p}-S_{12}-S_{1^\p 2^\p} + 1]\nonumber \\
& \int \dtwo{\bk}\Bigg[ \frac{-\frac{z(z_2-z)[z_1z_2+(z_1+z)(z_2-z)]}{z_1^2 z_2^3} (z_2\bp-z_1\bq)^2}{\left[(\bk+(z_2-z)\bp-(z_1+z)\bq)^2-\frac{(z_1+z)(z_2-z)}{z_1z_2}(z_2\bp-z_1\bq)^2 -i\epsilon\right] \left(\bk-\frac{z}{z_2}\bq\right)^2} \nonumber \\
&+ \frac{\frac{z[z_1(z_1+z)+z_2(z_2-z)]}{z_1 z_2}}{\left[(\bk+(z_2-z)\bp-(z_1+z)\bq)^2-\frac{(z_1+z)(z_2-z)}{z_1z_2}(z_2\bp-z_1\bq)^2 -i\epsilon\right]} -\frac{\frac{[z_1 z_2+(z_1+z)(z_2-z)]}{z_1z_2}}{ \left(\bk-\frac{z}{z_2}\bq\right)^2}\Bigg] e^{i\bk\cdot(\bx_2 -\bx_1)}.\label{dsig131} \\
& \frac{\dd \sigma_{13(2)}^{L}}{\dd^2\bp\,\dd^2\bq\,\dd y_1\,\dd y_2} =  \frac{-e^2g^2 Q^2 N_c^2 (z_1z_2)^2}{(2\pi)^8}\int_{0}^{z_1}\frac{ \dd z}{z} (z_2+z)(z_1-z) \int \dd^8\bx \,K_0\left(|\bx_{12}|Q\sqrt{(z_2+z)(z_1-z)}\right)K_0(|\bx_{1^\p 2^\p}|Q_1)\nonumber \\
& e^{i\bp\cdot(\bx_{1^\p }-\bx_1)}e^{i\bq\cdot(\bx_{2^\p }-\bx_2)}[S_{12}S_{1^\p 2^\p}-S_{12}-S_{1^\p 2^\p} + 1]\nonumber \\
& \int \dtwo{\bk}\Bigg[ \frac{-\frac{z(z_1-z)[z_1z_2+(z_2+z)(z_1-z)]}{z_2^2 z_1^3} (z_2\bp-z_1\bq)^2}{\left[(\bk+(z_1-z)\bq-(z_2+z)\bp)^2-\frac{(z_2+z)(z_1-z)}{z_1z_2}(z_2\bp-z_1\bq)^2 +i\epsilon\right] \left(\bk-\frac{z}{z_1}\bp\right)^2} \nonumber \\
&+ \frac{\frac{z[z_2(z_2+z)+z_1(z_1-z)]}{z_1 z_2}}{\left[(\bk+(z_1-z)\bq-(z_2+z)\bp)^2-\frac{(z_2+z)(z_1-z)}{z_1z_2}(z_2\bp-z_1\bq)^2 +i\epsilon\right]} -\frac{\frac{[z_1 z_2+(z_2+z)(z_1-z)]}{z_1z_2}}{ \left(\bk-\frac{z}{z_1}\bp\right)^2}\Bigg] e^{i\bk\cdot(\bx_1-\bx_2)}.\label{dsig132}
\end{align}

\noindent Here we have kept the $i\epsilon$ in the long denominators, since it proves useful for evaluating the $\bk$ integral.

\begin{align}
\frac{\dd \sigma_{14(1)}^{ L}}{\dd^2 \bp\, \dd^2 \bq \, \dd y_1 \, \dd y_2} =& \frac{e^2 g^2 Q^2 N_c^2 (z_1 z_2)^2}{(2\pi)^7} \int \dd^8 \bx [ S_{122^\p 1^\p} - S_{12} - S_{1^\p 2^\p} + 1] K_0(|\bx_{1^\p 2^\p}|Q_1) e^{i\bp\cdot(\bx_1^\p - \bx_1)} e^{i\bq\cdot(\bx_2^\p - \bx_2)} \nonumber \\
& \int_0^{z_1} \frac{\dd z}{z} \dtwo{\bk_1}\dtwo{\bk_2} e^{i\bk_2\cdot(\bx_1-\bx_2)} \Bigg[ \frac{(z_1 z_2 +(z_1-z)(z_2+z))}{\Big[\bk_2^2+Q_1^2\Big] \left[\left(\bk_1-\frac{z_1-z}{z_1}\bk_2\right)^2+\frac{z(z_1-z)}{z_2 z_1^2} \bk_2^2 + \frac{z}{z_1}(z_1-z)Q^2\right]} \nonumber \\
& + \frac{\frac{(z_1-z)(z_2+z)}{z_1 z_2} (z_1 z_2+(z_1-z)(z_2+z))}{\Big[\bk_1^2 + (z_1-z)(z_2+z)Q^2\Big] \left[\left(\bk_1-\frac{z_1-z}{z_1}\bk_2\right)^2+\frac{z(z_1-z)}{z_2 z_1^2} \bk_2^2 + \frac{z}{z_1}(z_1-z)Q^2\right]} \nonumber\\
&- \frac{\frac{z(z_1-z)}{z_1} (z_1 z_2 +(z_1-z)(z_2+z))Q^2}{\Big[\bk_1^2 + (z_1-z)(z_2+z)Q^2\Big]\Big[\bk_2^2 + Q_1^2\Big] \left[\left(\bk_1-\frac{z_1-z}{z_1}\bk_2\right)^2+\frac{z(z_1-z)}{z_2 z_1^2} \bk_2^2 + \frac{z}{z_1}(z_1-z)Q^2\right]} \nonumber \\
&- \frac{z (z_1(z_1-z)+z_2(z_2+z))}{\Big[\bk_1^2+(z_1-z)(z_2+z)Q^2\Big]\Big[\bk_2^2+Q_1^2\Big]}  \Bigg]. \label{dsig141} \\
\frac{\dd \sigma_{14(2)}^{ L}}{\dd^2 \bp\, \dd^2 \bq \, \dd y_1 \, \dd y_2} =& \frac{e^2 g^2 Q^2 N_c^2 (z_1 z_2)^2}{(2\pi)^7} \int \dd^8 \bx [ S_{122^\p 1^\p} - S_{12} - S_{1^\p 2^\p} + 1] K_0(|\bx_{1^\p 2^\p}|Q_1) e^{i\bp\cdot(\bx_1^\p - \bx_1)} e^{i\bq\cdot(\bx_2^\p - \bx_2)} \nonumber \\
& \int_0^{z_2} \frac{\dd z}{z} \dtwo{\bk_1}\dtwo{\bk_2} e^{i\bk_2\cdot(\bx_1-\bx_2)} \Bigg[ \frac{(z_1 z_2 +(z_1+z)(z_2-z))}{\Big[\bk_2^2+Q_1^2\Big] \left[\left(\bk_1-\frac{z_2-z}{z_2}\bk_2\right)^2+\frac{z(z_2-z)}{z_1 z_2^2} \bk_2^2 + \frac{z}{z_2}(z_2-z)Q^2\right]} \nonumber \\
& + \frac{\frac{(z_1+z)(z_2-z)}{z_1 z_2} (z_1 z_2+(z_1+z)(z_2-z))}{\Big[\bk_1^2 + (z_1+z)(z_2-z)Q^2\Big] \left[\left(\bk_1-\frac{z_2-z}{z_2}\bk_2\right)^2+\frac{z(z_2-z)}{z_1 z_2^2} \bk_2^2 + \frac{z}{z_2}(z_2-z)Q^2\right]} \nonumber\\
&- \frac{\frac{z(z_2-z)}{z_2} (z_1 z_2 +(z_1+z)(z_2-z))Q^2}{\Big[\bk_1^2 + (z_1+z)(z_2-z)Q^2\Big]\Big[\bk_2^2 + Q_1^2\Big] \left[\left(\bk_1-\frac{z_2-z}{z_2}\bk_2\right)^2+\frac{z(z_2-z)}{z_1 z_2^2} \bk_2^2 + \frac{z}{z_2}(z_2-z)Q^2\right]} \nonumber \\
&- \frac{z (z_2(z_2-z)+z_1(z_1+z))}{\Big[\bk_1^2+(z_1+z)(z_2-z)Q^2\Big]\Big[\bk_2^2+Q_1^2\Big]}  \Bigg]. \label{dsig142}
\end{align}

\noindent These expressions constitute the full result for the one-loop corrections to inclusive quark anti-quark 
production cross section. We have written these results all in terms of the dipole and quadrupole functions defined in Eq.~(\ref{dipquad}) in the large $N_c$ limit and ignored all subleading $N_c$ terms. We have also used the following notation for the coordinate dependence of some of the Bessel functions:

\begin{align}
X &= \sqrt{z_1 z_2 \bx_{12}^2 + z_1 z \bx_{13}^2 + z_2 z \bx_{23}^2},\nonumber \\
X_5 &= \sqrt{z_2(z_1-z)\bx_{12}^2 + z(z_1-z) \bx_{13}^2 + z_2 z\, \bx_{23}^2}, \nonumber \\
X_6 &= \sqrt{z_1(z_2-z)\bx_{12}^2 + z_1 z\,\bx_{13}^2 + z(z_2-z)\bx_{23}^2}.
\end{align}

\noindent Note that when $z \to 0$ these all become $|\bx_{12}|\sqrt{z_1 z_2}$. The primed version $X^\prime$ that appears in some real corrections is the same as $X$ above but with $\bx_1,\bx_2 \to \bx_1^\p,\bx_2^\p$. 




\section{Divergences}
The expressions for the one-loop corrections to quark anti-quark inclusive production cross section given above are formal, i.e. they contain divergences that need to be regulated and/or eliminated before one obtains a meaningful result. There are $4$ categories of divergences:

\noindent $\bullet$ Ultraviolet (UV) divergences when loop momentum $\bk \rightarrow \infty$ or equivalently in coordinate space, when the transverse coordinate of the radiated gluon approaches the transverse coordinate $\bx_{i }$ of either quark or anti-quark when integrated, i.e.  $\bx_{3 } \rightarrow \bx_{i }$ such that $|\bx_{3 } - \bx_{i }| \rightarrow 0$. We will show that the real corrections are UV finite and that all UV divergences in the virtual corrections cancel against each other so that the virtual corrections are also UV finite by themselves.

\noindent $\bullet$ Soft divergences when $k^\mu \rightarrow 0$, which in this context corresponds to {\it both} transverse momentum in the loop $\bk$ {\it and} the radiated gluon momentum fraction $z$ go to zero simultaneously, $\bk , z \rightarrow 0$. Both the real and virtual corrections contain soft divergences, however all soft divergences cancel between real and virtual corrections.  

\noindent $\bullet$ Collinear divergences when the radiated gluon momentum becomes parallel to either quark or anti-quark momentum, i.e. when the angle $\theta$ between the radiated gluon $3$-momentum and its parent quark (anti-quark) $3$-momentum goes to zero at {\it finite} $\bk$ and $z$ (note that finite in this context means that $\bk$ is neither zero nor infinite). These appear, for example as 
${1 \over (p +k)^2} \sim {1 \over (z_1 \bk - z \bp)^2} \sim \frac{1}{1-\cos\theta}$ for radiation from a quark and similarly for radiation from a anti-quark ($z_1, \bp \rightarrow z_2, \bq$). These collinear divergences are absorbed into quark-hadron (anti quark-hadron) fragmentation functions which makes the fragmentation functions scale dependent, i.e. they evolve with the renormalization scale $\mu^2$.

\noindent $\bullet$ Rapidity divergences when the momentum fraction $z$ of the gluon goes to zero while the transverse momentum $\bk$ of the gluon remains finite. These rapidity divergences are absorbed into ($x$) renormalization of the dipoles and quadrupoles making them energy/rapidity dependent and evolving according to the BK and JIMWLK evolution equations~\cite{Balitsky:1995ub,Kovchegov:1999yj,Jalilian-Marian:1997qno,Jalilian-Marian:1997jhx,Jalilian-Marian:1997ubg,Kovner:2000pt,Iancu:2000hn,Ferreiro:2001qy}.  

Here we give more details of the divergences and their cancellation/absorption into renormalized quantities. The easiest way to see cancellations of divergences is via power counting which is what we will use in this section. We will also provide results using dimensional regularization in the $\overline{\text{MS}}$ scheme which is the more common method. Using dimensional regularization raises issues when using spinor helicity methods as one changes the number of spacetime dimensions from $4$ (where spinor helicity methods are formulated) to $4 - 2\epsilon$. There are various ways of dealing with this issue, here we will use the FDH (Four Dimensional Helicity) scheme which treats both internal and external states in $4$-dimensions. As expected the finite terms after regulating divergences will depend on the regularization scheme.
 

	\subsection{UV divergences} \label{uv}
	
For the real corrections (Eq. \ref{Real11}-\ref{Real24}) all momentum integrations have been performed and results are given in terms of a two-dimensional coordinate space integration $\dd^2 \bx_3$ and an integration over momentum fraction $z$ (relative to the LO result in Eq. \ref{LOdsig}). In this context UV divergences manifest themselves as poles when
$\bx_3 \rightarrow \bx_1$ or 	$\bx_3 \rightarrow \bx_2$, i.e. when the transverse coordinate of the radiated gluon $\bx_3$ approaches that of the quark $\bx_1$ or the anti-quark $\bx_2$. We focus on the quark case and consider the limit $\bx_3 \rightarrow \bx_1$. The case of radiation from the anti-quark is identical.

We note that the only possible UV singularity in our expressions is in the gluon radiation kernel 
$\Delta_{i j}^{(3)}$ which appears in all real corrections. Writing it as
\begin{align}
\Delta^{(3)}_{ij} = \frac{\bx_{3i}\cdot\bx_{3j}}{\bx_{3i}^2 \bx_{3j}^2} = \frac12 \left[ \frac{1}{\bx_{3i}^2} + \frac{1}{\bx_{3j}^2} - \frac{{\bx_{ij}^2}}{\bx_{3i}^2 \bx_{3j}^2}\right],
\end{align}
we see that the integral of this kernel over $\bx_3$ is finite as $\bx_3 \rightarrow \bx_i$ due to cancellations between the first and third terms. Similarly, as $\bx_3 \rightarrow \bx_j$ there is a cancellation between the second and third terms. The other factors appearing in the real corrections are all non-singular in the UV limit. We therefore conclude that there are no UV divergences in the real corrections and that {\it all real corrections are UV finite}.

For the virtual corrections, several of the diagrams are UV divergent. However when combined they all cancel each other in the UV limit, specifically 
\begin{align}
&\left[\dd \sigma_{5}+\dd \sigma_{11}\right]_{\text{UV}} = 0, \nonumber \\
&\left[\dd \sigma_{6}+\dd \sigma_{12} \right]_{\text{UV}} = 0, \nonumber \\
&\left[\dd \sigma_{9}+\dd\sigma_{10}+\dd\sigma_{14(1)}+\dd\sigma_{14(2)}\right]_{\text{UV}}  = 0. \label{uvcancels}
\end{align}

\noindent The remaining virtual diagrams (7,8,13) are all UV finite. Therefore, \textit{all UV divergences are canceled} when adding up all the relevant pieces of the differential cross section. The UV-finiteness of the production cross section holds for any value of $z$. We provide more details in Appendix \ref{UVapp}.


\subsection{Soft Divergences}\label{soft}
Soft divergences appear when all components of a loop momentum go to zero, i.e. 
$k^\mu \rightarrow 0$ which in this context means $\bk$ and $z$ {\it both} 
go to zero (in coordinate space this becomes $\bx_3 \to \infty$ and $z \to 0$). Soft divergences in the real corrections generically 
appear either as 
\begin{equation}
\int {\dd^2 \bk \over \bk^2}\, e^{i \bk\cdot (\bx_{i} - \bx_{j })}\,\,\,\,\,\,\,\, \text{or} \,\,\,\,\,\,\,\,\int {\dd^2 \bk \over \bk^2}\label{radkern}
\end{equation}
in the virtual corrections (after possibly shifting the transverse momentum).  
In coordinate space soft divergences are contained in
\begin{equation}
\int \dd^2 \bx_{3} \, \Delta^{(3)}_{i j} = 
{1 \over 2} \, \int \dd^2 \bx_{3} \, 
\left[{1 \over (\bx_{3 } - \bx_{i })^2} + {1 \over (\bx_{3 } - \bx_{j })^2} - 
{(\bx_{i } - \bx_{j })^2  \over (\bx_{3 } - \bx_{i })^2 \, (\bx_{3 }- \bx_{j })^2}\right]
\end{equation}
from which it is clear that the first two terms in $\Delta^{(3)}_{i j}$ are
divergent when $\bx_{3 } \rightarrow \infty$. Adding all the singular parts 
of the production cross section we get the following cancellations between the various terms,
\begin{align}
&\left[\dd \sigma_{1\times 1} + 2\,\dd \sigma_{9} \right]_{\text{soft}}= 0, \nonumber \\
&\left[\dd \sigma_{2\times 2} + 2\,\dd \sigma_{10}\right]_{\text{soft}} = 0, \nonumber \\
&\left[\dd \sigma_{1\times 2} + \dd\sigma_{13(1)}+\dd\sigma_{13(2)} \right]_{\text{soft}}= 0, \nonumber \\
&\left[\dd \sigma_{3\times 3} + \dd \sigma_{4\times 4} + 2\,\dd \sigma_{3\times 4}\right]_{\text{soft}} = 0,\nonumber\\
&\left[\dd \sigma_{1\times 3} + \dd \sigma_{1\times 4} \right]_{\text{soft}}= 0,\nonumber\\
&\left[\dd \sigma_{2\times 3} + \dd \sigma_{2\times 4}\right]_{\text{soft}} = 0,\nonumber\\
&\left[\dd \sigma_{5}+\dd\sigma_{7} \right]_{\text{soft}}= 0, \nonumber \\
&\left[\dd \sigma_{6}+\dd\sigma_{8}\right]_{\text{soft}} = 0, \nonumber \\
&\left[\dd \sigma_{11} + \dd \sigma_{14(1)}\right]_{\text{soft}} = 0,  \nonumber \\
&\left[\dd \sigma_{12} + \dd \sigma_{14(2)}\right]_{\text{soft}} = 0 .
\end{align}

The necessary relative factors of $2$ in these expressions can be seen from Eq. \ref{breakdown}. These cancellations are easiest to see when the cross section is written in coordinate space in the soft $z$ limit (Appendix \ref{softzcoord}). It is worth noting that there are singular terms $\sim {1\over z} \, \log ({1\over z})$ which can be seen in dimensional regularization of the remaining integrals
which cancel between $\dd \sigma_{11,\text{soft}}$ and $\dd \sigma_{14(1),\text{soft}}$ 
and similarly between $\dd \sigma_{12,\text{soft}}$ and $\dd \sigma_{14(2),\text{soft}}$ (this is shown in Appendix \ref{dimreg}).
Thus there are no soft singularities (both $\bk, z \rightarrow 0$) left after adding 
all the real and virtual terms. 


\subsection{Rapidity Divergences }
From eqs. (\ref{Real11} - \ref{dsig142}) it is clear that the production 
cross section contains terms which are singular in the
$\int {\dd z\over z}$ integral as $z \rightarrow 0$. This happens at finite $\bk$ (or $\bx_3$) 
so that rapidity divergences are different from the soft divergences that have been 
shown to cancel among various terms. Following \cite{Caucal:2021ent} we  
introduce a longitudinal momentum fraction factorization scale $z_f$ and divide the $z$ integration into 
two regions: $z > z_f$ and $z < z_f$ (here we write the upper limit as $1$, in our expressions the upper limit is always either $z_1$ or $z_2$). 
\begin{align}
\int_0^1 \frac{\dd z}{z} f (z) =  \left\{\int_0^{z_f} \frac{\dd z}{z}  + 
\int_{z_f}^1 \frac{\dd z}{z} \, \right\} \, f (z).
\end{align}

\noindent In the region where $z > z_f$ there 
are formally no divergences left anymore (there are still collinear divergences in this 
region but they will be absorbed into fragmentation functions - this is the topic of the next section) so the cross 
section is completely finite. On the other hand in the second region where $z < z_f$ 
we encounter the familiar rapidity divergence $y = \log (1/z)$ which we expect will be absorbed into 
JIMWLK evolution of the quadrupoles and dipoles. We now focus on the 
$z \rightarrow 0$ limit of our expressions while $\bk$ (or equivalently $\bx_3$) 
is kept finite (see Appendix \ref{softzcoord}). 

\begin{align}
&\frac{ \dd \sigma_{\text{NLO}}^{L}}{\dd^2 \bp \, \dd^2 \bq\, \dd y_1 \, \dd y_2} = \frac{2e^2 g^2 Q^2 N_c^2 (z_1 z_2)^3}{(2\pi)^{10}} \int_0^{z_f} \frac{\dd z}{z} \int \dd^{10}\bx\,K_0(|\bx_{12}|Q_1)K_0(|\bx_{1^\p 2^\p}|Q_1) e^{i\bp\cdot\bx_{1^\p 1}}e^{i\bq\cdot\bx_{2^\p 2}} \delta(1-z_1-z_2) \nonumber \\
&\Bigg\{ [S_{122^\p1^\p}-S_{12}-S_{1^\p 2^\p} +1]\left( \frac{3}{\bx_{31}^2} + \frac{3}{\bx_{32}^2} +\frac{1}{\bx_{31^\p}^2} +\frac{1}{\bx_{32^\p}^2} -\frac{8}{\bx_3^2} - \td_{11^\p}-\td_{22^\p}-2\td_{12}\right) \nonumber \\
&+[S_{12}S_{1^\p 2^\p}-S_{12}-S_{1^\p 2^\p}+1] \left(\frac{2}{\bx_{32}^2} - \frac{2}{\bx_{32^\p}^2} +2\td_{12^\p} -2\td_{12}\right)\nonumber \\
&+[S_{11^\p}S_{22^\p}-S_{13}S_{23}-S_{1^\p 3}S_{2^\p 3}+1]\left( -\frac{1}{\bx_{31}^2} +\frac{1}{\bx_{31^\p}^2} +\frac{1}{\bx_{32}^2} -\frac{1}{\bx_{32^\p}^2} -\td_{11^\p} - \td_{22^\p} + 2\td_{12^\p}\right) \nonumber \\
&+[S_{122^\p 3} S_{1^\p 3}-S_{1^\p 3}S_{2^\p 3} - S_{12}+1]\left(\frac{2}{\bx_{32^\p}^2}-\frac{2}{\bx_{31^\p}^2} +2\td_{11^\p}-2\td_{12^\p}\right)\nonumber \\
&+[S_{1231^\p}S_{2^\p 3}-S_{1^\p 3}S_{2^\p 3}-S_{12}+1]\left(\frac{2}{\bx_{31^\p}^2}-\frac{2}{\bx_{32^\p}^2} +2\td_{22^\p}-2\td_{21^\p}\right)\nonumber \\
&+[S_{322^\p 1^\p}S_{13}-S_{13}S_{23}-S_{1^\p 2^\p}+1]\left(\frac{2}{\bx_{31}^2}-\frac{2}{\bx_{32}^2}+2\td_{12}\right) \nonumber \\
&+[S_{132^\p 1^\p}S_{23}-S_{13}S_{23}-S_{1^\p 2^\p}+1]\left(\frac{2}{\bx_{32}^2}-\frac{2}{\bx_{31}^2}+2\td_{12}\right)\Bigg\}. \label{1strap}
\end{align}

\noindent Here we have defined

\begin{align}
\td_{ij} = \frac{\bx_{ij}^2}{\bx_{3i}^2 \bx_{3j}^2}.
\end{align}

\noindent Inside the curly brackets, the first line comes from adding $\sigma_{1\times 1}+\sigma_{2\times 2}+2\sigma_9+2\sigma_{10}+2\sigma_{11}+2\sigma_{12}+2\sigma_{14(1)}+2\sigma_{14(2)}$. The second line is $2\sigma_{1\times 2}+2\sigma_{13(1)}+2\sigma_{13(2)}$. The third line is $\sigma_{3\times 3}+\sigma_{4\times 4}+2\sigma_{3\times 4}$. Fourth line is $2\sigma_{1\times 3}+2\sigma_{1\times 4}$. Fifth line is $2\sigma_{2\times 3}+2\sigma_{2\times 4}$. Sixth line is $2\sigma_5 + 2\sigma_7$, and the last line is $2\sigma_6+2\sigma_8$. Notice that cross terms have been doubled since we need to add their complex conjugate. To simplify the expression we use a combination of shifting $\bx_3$ where possible and using symmetry between primed and unprimed coordinates. We find that everything can be simplified down to the following expression.

\begin{align}
&\frac{ \dd \sigma_{\text{NLO}}^{L}}{\dd^2 \bp \, \dd^2 \bq\, \dd y_1 \, \dd y_2} = 
\frac{2e^2 g^2 Q^2 N_c^2 (z_1 z_2)^3}{(2\pi)^{10}} \, \delta(1-z_1-z_2) 
\int_0^{z_f} \frac{\dd z}{z} \int \dd^{10}\bx\,K_0(|\bx_{12}|Q_1)K_0(|\bx_{1^\p 2^\p}|Q_1)
\nonumber \\
& e^{i\bp\cdot\bx_{1^\p 1}}e^{i\bq\cdot\bx_{2^\p 2}} 
\Bigg\{\left(\td_{12} + \td_{22^\p} - \td_{12^\p}\right)S_{132^\p 1^\p}S_{23} + 
\left(\td_{1^\p 2^\p} + \td_{22^\p} - \td_{2 1^\p}\right)S_{1^\p32 1}S_{2^\p3} 
\nonumber \\
&+ 
\left(\td_{12} + \td_{11^\p}-\td_{21^\p}\right)S_{322^\p 1^\p}S_{13} 
+ \left(\td_{1^\p 2^\p} + \td_{11^\p}-\td_{12^\p}\right)S_{32^\p2 1}S_{1^\p 3}
\nonumber \\
& - \left(  \td_{11^\p}+\td_{22^\p}+\td_{12}+\td_{1^\p 2^\p}\right) S_{122^\p 1^\p} -\left(\td_{12} + \td_{1^\p 2^\p}-\td_{12^\p} -\td_{21^\p} \right) S_{12}S_{1^\p 2^\p}  \nonumber \\
& -\left(\td_{11^\p} + \td_{22^\p} -\td_{12^\p} - \td_{21^\p} \ \right)S_{11^\p}S_{22^\p}  - 2\td_{12}\left(S_{13}S_{23} - S_{12}\right) - 2\td_{1^\p 2^\p}\left(S_{1^\p 3}S_{2^\p 3} - S_{1^\p 2^\p}\right)\Bigg\}.
\end{align}

\noindent Here the first seven terms inside the curly bracket give the JIMWLK evolution of the quadrupole 
$S_{122^\p 1^\p}$ with the factorization scale $z_f$ (See Eq. $5$ in \cite{Jalilian-Marian:2011jrd}). The last two terms correspond to the rapidity evolution of the dipoles $S_{12}$ and $S_{1^\p 2^\p}$  as given by the BK equation~\cite{Balitsky:1995ub,Kovchegov:1999yj}. This demonstrates that the rapidity divergences can be absorbed into rapidity evolution of the LO cross section. The contribution of the $z$-integration above the factorization scale $z_f$ is finite and constitutes part of the NLO corrections to the Leading Order cross section.


\subsection{Collinear divergences}
	\subsubsection{Leading Order dihadron production}

We now focus on the collinear divergences present in some of the one-loop corrections in (Eq. \ref{Real11} - \ref{dsig142}) and show that they can absorbed into evolution of quark-hadron and anti quark-hadron fragmentation functions. As this is completely new we will provide the full details of the calculation. We start by writing the inclusive dihadron production cross section in terms of the partonic cross section as

\begin{align}
\frac{\dd \sigma^{\gamma^* A \to h_1 h_2 X}}{\dd^2 \bp_h \, \dd^2 \bq_h \, \dd y_1 \, \dd y_2}=\int_0^1 \dd z_{h_1} \int_0^1 \dd z_{h_2}\, \frac{1}{z_{h_1}^2 z_{h_2}^2}\frac{ \dd \sigma^{\gamma^* A \to q\bar{q}X}}{\dd^2 \bp \, \dd^2 \bq \, \dd y_1 \, \dd y_2} D_{h_1/q}(z_{h_1}) D_{h_2/\bar{q}}(z_{h_2}). \label{hadronparton}
\end{align}
where $\bp, \bq$ are the quark and anti-quark transverse momenta while $\bp_h, \bq_h$ are the transverse momenta of the two produced hadrons. As partons and hadrons are assumed to be massless their rapidities are the same so that $y_1 \equiv y_p = y_{h}$ (and similarly for the anti-quark). 
Hadronization is assumed to be describable in terms of a non-perturbative quark (or anti-quark)-hadron fragmentation function $D_{h/q} (z_{h_1})$ which is then perturbatively renormalized (becomes $\mu^2$ dependent) after absorption of the collinear divergences. The momentum fractions  $ z_{h_1},  z_{h_2}$ are defined via 
$p_{h_1}^\mu = z_{h_1}p^\mu$ and $q_{h_2}^\mu = z_{h_2}q^\mu$ so that  $0 < z_{h_i} < 1$. Using our Leading Order result from Eq. (\ref{LOdsig}) (only the longitudinal part) we can write this as

\begin{align}
\frac{\dd \sigma_{LO}^{\gamma^* A \to h_1 h_2 X}}{\dd^2 \bp_h \, \dd^2 \bq_h \, \dd y_1 \, \dd y_2}=&\int_0^1 \dd \zho \int_0^1 \dd \zht \frac{4e^2 Q^2(z_1z_2)^3 N_c}{(2\pi)^7 (\zho \zht)^2} \int \dd^8 \bx \left[S_{122^\p1^\p}-S_{12}-S_{1^\p 2^\p}+1\right] 
\nonumber \\
&
K_0(|\bx_{12}|Q_1)K_0(|\bx_{1^\p 2^\p}|Q_1)\,  e^{i\bp\cdot(\bx_1^\p-\bx_1)}\, e^{i\bq\cdot(\bx_2^\p-\bx_2)} \nonumber \\
& D_{h_1/q}(\zho) D_{h_2/\barq}(\zht)\, \delta(1-z_1-z_2),
\end{align}

\noindent where $z_1$ and $z_2$ are related to hadronic momentum fractions by 

\begin{align}
z_1 = \frac{p_h^+}{z_{h_1}l^+}, \,\,\,\,\,\, z_2 = \frac{q_h^+}{z_{h_2} l^+}. \label{zrel}
\end{align}

\noindent Anticipating factorization of the one-loop corrections and to make our expression more compact we define the  common factor $H(\bp,\bq,\bx,z_1)$ as

\begin{align}
H(\bp,\bq,\bx,z_1) \equiv  \left[S_{122^\p 1^\p}-S_{12}-S_{1^\p 2^\p} + 1\right] 
K_0 \left(|\bx_{12}|Q_1\right) \, K_0 \left(|\bx_{1^\p2^\p}|Q_1\right) \,
e^{i\bp\cdot(\bx_1^\p-\bx_1)}e^{i\bq\cdot(\bx_2^\p-\bx_2)},\label{Hdef}
\end{align}

\noindent with $Q_1^2 \equiv z_1(1-z_1)\, Q^2$ so that the common factor implicitly depends on $Q^2$. The Leading Order cross section can be then expressed in the following compact form

\begin{align}
\frac{\dd \sigma_{LO}^{\gamma^* A \to h_1 h_2 X}}
{\dd^2 \bp_h \, \dd^2 \bq_h \, \dd y_1 \, \dd y_2}=
&\int_0^1 \dd \zho \int_0^1 \dd \zht \frac{4e^2 Q^2(z_1 z_2)^3 N_c}{(2\pi)^7 (\zho \zht)^2}\int \dd^8 \bx\, H\left(\bp,\bq,\bx,z_2\right) \nonumber \\
&\delta\left(1-z_1-z_2\right) D_{h_1/q}(\zho) D_{h_2/\barq}(\zht).
\label{eq:LO-hadronic}
\end{align}


\subsubsection{The virtual correction}

First we'll focus on the fragmentation of the quark into a hadron $h_1$. Treatment of anti quark fragmentation will be identical. Schematically, the fragmentation function receives corrections from the following diagrams:

\begin{figure}[H]
\centering
\includegraphics[width=120mm]{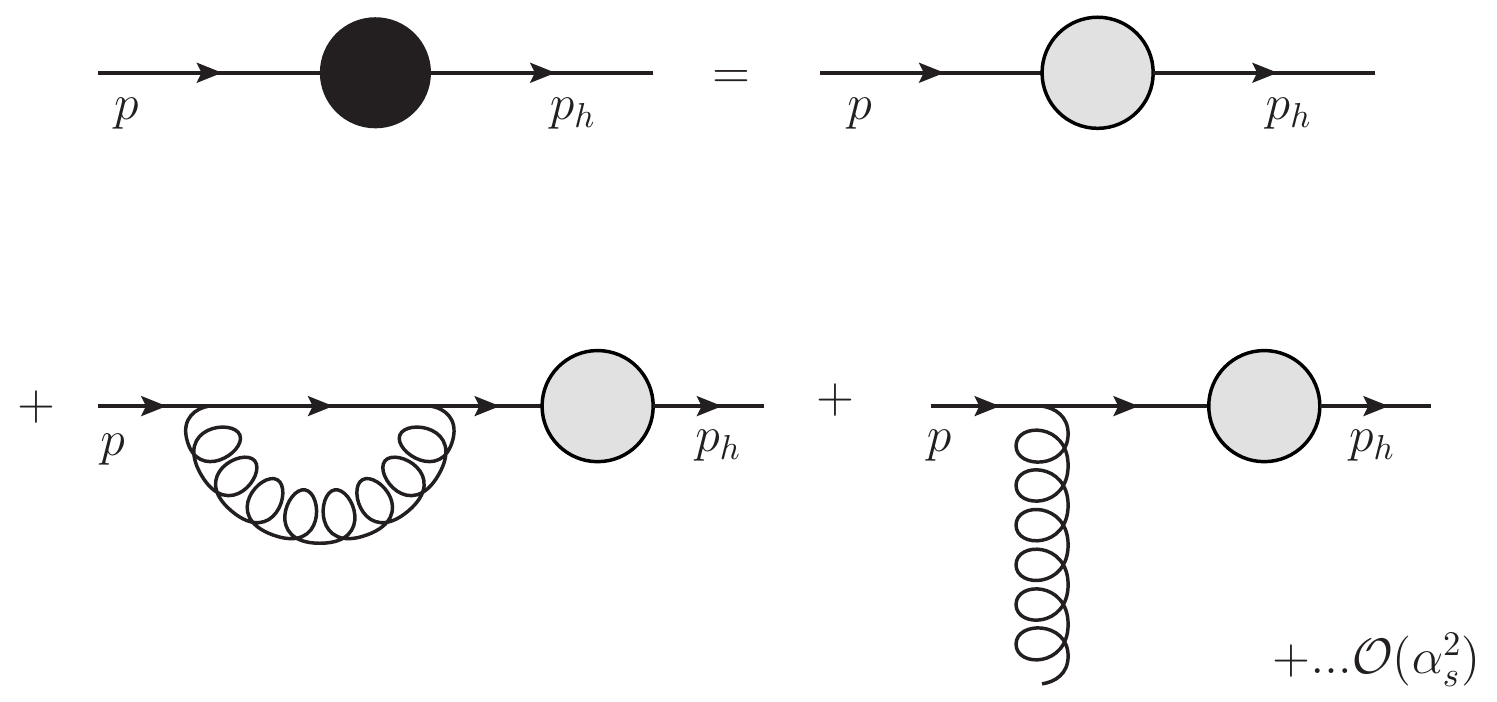}
\caption{The full fragmentation function $D_{h_1/q}(\zho,\mu^2)$ (solid black circle) can be written as the bare fragmentation function (faint gray circle) plus the $\mathcal{O}(\alpha_s)$ corrections shown. These need to be added at the cross section level, and we'll keep terms only up to $\mathcal{O}(\alpha_s)$.}\label{phfig}
\end{figure}

\noindent Here in Fig. \ref{phfig} the first diagram after the equals sign corresponds to the leading order hadronic cross section computed in the previous subsection. The two $\mathcal{O}(\alpha_s)$ corrections correspond to our virtual correction $\sigma_9$ and the real correction $\sigma_{1\times 1}$. So, let's now compute the hadronic cross section contribution from the virtual correction $ \sigma_9$ (combining Eq. \ref{dsig9} with Eq. \ref{hadronparton}).

\begin{align}
&\frac{\dd \sigma_{9}^{\gamma^* A\to h_1 h_2 X}}{\dd^2 \bp_h \, \dd^2 \bq_h \,\dd y_1 \, \dd y_2} = - \int_0^1 \dd \zho \int_0^1 \dd \zht  \frac{4e^2  Q^2 (z_1z_2)^3N_c}{(2\pi)^7(z_{h_1}z_{h_2})^2}\int \dd^8 \bx\, H(\bp,\bq,\bx,z_2) D^0_{h_1/q}(\zho)D^0_{h_2/\barq}(\zht)\nonumber  \\
&\times \alpha_s  C_F\int_0^{z_1} \frac{\dd z}{z} \left[ \frac{z_1^2+(z_1-z)^2}{z_1^2}\right] \int \dtwo{\bk} \frac{1}{\left(\bk-\frac{z}{z_1}\bp\right)^2} \delta(1-z_1-z_2).
\end{align}

\noindent Here we relaxed the large $N_c$ approximation taking one factor of $N_c \to 2C_F$, and have also used  $g^2 \to 4 \pi \alpha_s$. Let's define a new variable for the $z$ integration

\begin{align}
\xi = \frac{z_1-z}{z_1}.
\end{align}

\noindent This defines $\xi$ to be the fraction of longitudinal momentum that the quark carries after radiating the gluon. In terms of this, we get

\begin{align}
&\frac{\dd \sigma_{9}^{\gamma^* A\to h_1 h_2 X}}{\dd^2 \bp_h \, \dd^2 \bq_h \,\dd y_1 \, \dd y_2} = - \int_0^1 \dd \zho \int_0^1 \dd \zht  \frac{4e^2  Q^2 (z_1z_2)^3N_c}{(2\pi)^7(z_{h_1}z_{h_2})^2}\int \dd^8 \bx\, H(\bp,\bq,\bx,z_2)  
D^0_{h_2/\barq}(\zht)D^0_{h_1/q}(\zho)\nonumber  \\
&\times \alpha_s \, C_F\, \int^{1}_{0}\dd \xi \, \frac{(1+\xi^2)}{(1-\xi)}\, 
 \int \dtwo{\bk} \frac{1}{\left(\bk-(1-\xi)\bp\right)^2} \delta(1-z_1-z_2).\label{virtfrag}
\end{align}

\noindent This can now be added to the leading order hadronic cross section. Note that we actually need to add $2\sigma_9$ because it appears twice in the formula since it's a cross term (Eq. \ref{breakdown}).


\subsubsection{The real correction}

Looking at the relevant diagrams in Fig. \ref{phfig}, the last one we need for the $\mathcal{O}(\alpha_s)$ corrections to the quark fragmentation function is the real correction $\sigma_{1\times 1}$. 

\begin{align}
&\frac{\dd \sigma^{L}_{1\times 1}}{\dd^2 \bp\, \dd^2 \bq\, \dd y_1 \, \dd y_2} = \frac{2 e^2 g^2  Q^2 N_c^2 z_2^3 (1-z_2)^2(z_1^2+(1-z_2)^2)}{(2\pi)^{8} z_1} \int \dd^{8} x \,K_0(|\bx_{12}|Q_2)K_0(|\bx_{1^\prime 2^\prime}|Q_2) \nonumber \\
&\left[ S_{122^\p 1^\p} - S_{12} - S_{1^\p 2^\p} + 1\right]  e^{i\bp\cdot(\bx_1^\prime - \bx_1)}e^{i\bq\cdot(\bx_2^\prime - \bx_2)}\int \frac{\dd z}{z}  \int \dtwo{\bk} \frac{e^{i\bk\cdot(\bx_1^\p-\bx_1)}}{\left(\bk-\frac{z}{z_1}\bp\right)^2} \delta(1-z_1-z_2-z). 
\end{align}

\noindent We have rewritten the integral over the radiation kernel (in Eq. \ref{Real11}) in momentum space using

\begin{align}
e^{i\frac{z}{z_1}\bp\cdot(\bx_1^\p-\bx_1)} \int \dd^2 \bx_3 \Delta^{(3)}_{11^\p} = \int \dd^2 \bk \frac{e^{i\bk\cdot(\bx_1^\p-\bx_1)}}{\left(\bk-\frac{z}{z_1}\bp\right)^2}.
\end{align}

\noindent Looking at Eq. \ref{hadronparton} we see that this correction contributes the following expression to the hadronic cross section.

\begin{align}
\frac{\dd \sigma^{\gamma^* A\to h_1 h_2 X}_{1\times 1}}{\dd^2 \bp_h\, \dd^2 \bq_h \, \dd y_1\, \dd y_2} =& \int_0^1 \dd \zho \int_0^1 \dd\zht \frac{2 e^2 g^2  Q^2 N_c^2 z_2^3 (1-z_2)^2(z_1^2+(1-z_2)^2)}{(2\pi)^{8} z_1 (\zho \zht)^2} \int \dd^{8} x \,K_0(|\bx_{12}|Q_2)\nonumber \\ 
&K_0(|\bx_{1^\prime 2^\prime}|Q_2)\left[ S_{122^\p 1^\p} - S_{12} - S_{1^\p 2^\p} + 1\right]  e^{i\bp\cdot(\bx_1^\prime - \bx_1)}e^{i\bq\cdot(\bx_2^\prime - \bx_2)}\nonumber \\
&\int \frac{\dd z}{z}  \int \dtwo{\bk} \frac{e^{i\bk\cdot(\bx_1^\p-\bx_1)}}{\left(\bk-\frac{z}{z_1}\bp\right)^2} \delta(1-z_1-z_2-z) D^0_{h_1/q}(\zho) D^0_{h_2/\barq}(\zht).
\end{align}

\noindent Now we'll write this using $H(\bp,\bq,\bx,z_2)$ defined in Eq. \ref{Hdef} and again relax the large $N_c$ approximation taking one factor of $N_c$ to $2C_F$, and also write $g^2 = 4\pi \alpha_s$.

\begin{align}
&\frac{\dd \sigma^{\gamma^* A\to h_1 h_2 X}_{1\times 1}}{\dd^2 \bp_h\, \dd^2 \bq_h \, \dd y_1\, \dd y_2} = \int_0^1 \dd \zho \int_0^1 \dd\zht \frac{8 e^2   Q^2 N_c(z_1 z_2)^3}{(2\pi)^{7}  (\zho \zht)^2}\int \dd^8\bx\,H(\bp,\bq,\bx,z_2)D^0_{h_1/q}(\zho) D^0_{h_2/\barq}(\zht)  \nonumber \\
&\times \alpha_s C_F\int \frac{\dd z}{z} \frac{(z_1+z)^2(z_1^2+(z_1+z)^2)}{z_1^4} \delta(1-z_1-z_2-z)\int \dtwo{\bk} \frac{e^{i\bk\cdot(\bx_1^\p-\bx_1)}}{\left(\bk-\frac{z}{z_1}\bp\right)^2}  .
\end{align}

\noindent We factored out $(z_1z_2)^3$ to match the leading order expression and used the delta function to write $1-z_2$ as $z_1+z$. We then perform a change of variables in the $z$ integral by defining 

\begin{align}
\xi = \frac{z_1}{z_1+z}.
\end{align}

\noindent This $\xi_1$ is then the fraction of the outgoing quark's momentum relative to the parent quark that emitted the gluon (i.e.: the same definition as we used for the virtual correction). 

\begin{align}
&\frac{\dd \sigma^{\gamma^* A\to h_1 h_2 X}_{1\times 1}}{\dd^2 \bp_h\, \dd^2 \bq_h \, \dd y_1\, \dd y_2} = \int_0^1 \dd \zho \int_0^1 \dd\zht \frac{8 e^2   Q^2 N_c(z_1 z_2)^3}{(2\pi)^{7}  (\zho \zht)^2}\int \dd^8 \bx\,H(\bp,\bq,\bx,z_2) D^0_{h_2/\barq}(\zht)  \nonumber \\
&\times \alpha_s C_F\int \frac{\dd \xi}{\xi^5}\frac{(1+\xi^2)}{(1-\xi)} \delta(1-z_2-z_1/\xi) D^0_{h_1/q}(\zho)
\, 
\int \dtwo{\bk} \frac{e^{i\bk\cdot(\bx_1^\p-\bx_1)}}{\left(\bk-\frac{(1-\xi)}{\xi}\bp\right)^2} .
\end{align}

\noindent Now, $z_1$ is no longer an external variable, as remarked earlier it can be written in terms of $\zho$ using Eq. \ref{zrel}. So one can then write $z_1$ in terms of $\zho$, 

\begin{align}
&\frac{\dd \sigma^{\gamma^* A\to h_1 h_2 X}_{1\times 1}}{\dd^2 \bp_h\, \dd^2 \bq_h \, \dd y_1\, \dd y_2} = \int_0^1 \dd \zho \int_0^1 \dd\zht \frac{8 e^2 Q^2 N_c (p_h^+)^3 z_2^3}{(2\pi)^{7} (l^+)^3 \zho^5 \zht^2}\int \dd^8 \bx\,H(\bp,\bq,\bx,z_2)D^0_{h_2/\barq}(\zht)  \nonumber \\
&\times \alpha_s C_F\int \frac{\dd \xi}{\xi^5}\frac{(1+\xi^2)}{(1-\xi)} \delta\left(1-z_2-\frac{p_h^+}{\zho l^+ \xi}\right) D^0_{h_1/q}(\zho) \, \int \dtwo{\bk} \frac{e^{i\bk\cdot(\bx_1^\p-\bx_1)}}{\left(\bk-\frac{(1-\xi)}{\xi}\bp\right)^2} .
\end{align}

\noindent Now that the $\zho$ dependence is all explicit we make a substitution in the $\zho$ integral, defining the new variable

\begin{align}
\zho^\p = \zho \xi, \,\,\,\,\,\, \dd \zho^\p = \xi \dd \zho.
\end{align}

\noindent In order to complete this substitution, we need to take $\zho \to \frac{\zho^\p}{\xi}$ everywhere. This also means that the bounds on the $\zho^\p$ integral go from 0 to $\xi$. We'll write this using a step function $\theta(\xi-\zho^\p)$ so we can keep the explicit bounds from $0$ to $1$.  

\begin{align}
&\frac{\dd \sigma^{\gamma^* A\to h_1 h_2 X}_{1\times 1}}{\dd^2 \bp_h\, \dd^2 \bq_h \, \dd y_1\, \dd y_2} = \int_0^1 \dd \zho^\p \int_0^1 \dd\zht \frac{8 e^2 Q^2 N_c (p_h^+)^3 z_2^3}{(2\pi)^{7} (l^+)^3 \zho^{\p 5} \zht^2}\int \dd^8 \bx\,H(\bp,\bq,\bx,z_2) D^0_{h_2/\barq}(\zht)  \nonumber \\
&\times \alpha_s C_F\int \frac{\dd \xi}{\xi}\frac{(1+\xi^2)}{(1-\xi)} D^0_{h_1/q}\left(\frac{\zho^\p}{\xi}\right)\theta(\xi-\zho^\p)\delta\left(1-z_2-\frac{p_h^+}{\zho^\p l^+}\right) \, \, 
\int \dtwo{\bk} \frac{e^{i\bk\cdot(\bx_1^\p-\bx_1)}}{\left(\bk-\frac{(1-\xi)}{\xi}\bp\right)^2} .
\end{align}

\noindent  We can now write things back in terms of $z_1$ to get rid of $p_h^+$ and $l^+$ to match the leading order and virtual correction. We can also use the step function to set the bounds on the $\xi$ integral. Finally, we remove the primes from $\zho^\p$ since it's an integration variable.

\begin{align}
&\frac{\dd \sigma^{\gamma^* A\to h_1 h_2 X}_{1\times 1}}{\dd^2 \bp_h\, \dd^2 \bq_h \, \dd y_1\, \dd y_2} = \int_0^1 \dd \zho \int_0^1 \dd\zht \frac{8 e^2 Q^2 N_c (z_1 z_2)^3}{(2\pi)^{7}  (\zho \zht)^2}\int \dd^8 \bx\,H(\bp,\bq,\bx,z_2) D^0_{h_2/\barq}(\zht)  \nonumber \\
&\times\delta\left(1-z_1-z_2\right) \alpha_s C_F\int_{\zho}^1 \frac{\dd \xi}{\xi}\frac{(1+\xi^2)}{(1-\xi)} D^0_{h_1/q}\left(\frac{\zho}{\xi}\right) \, \, 
\int \dtwo{\bk} \frac{e^{i\bk\cdot(\bx_1^\p-\bx_1)}}{\left(\bk-\frac{(1-\xi)}{\xi}\bp\right)^2} .\label{realfrag}
\end{align}



\subsubsection{Evolution of the fragmentation function}

Now we would like to add up the three terms in Fig. \ref{phfig}. These terms are Eq. \ref{eq:LO-hadronic}, \ref{virtfrag}, and \ref{realfrag}. Note that we also need to double Eq. \ref{virtfrag} since it's a cross term. Adding up these three gives the following,

\begin{align}
&\int_0^1 \dd \zho \int_0^1 \dd \zht \frac{4e^2 Q^2 (z_1z_2)^3 N_c}{(2\pi)^7 (\zho \zht)^2} \int \dd^8 \bx \, H(\bp,\bq,\bx,z_2) D^0_{h_2/\barq}(\zht) \delta(1-z_1-z_2) \nonumber \\
&\times \Bigg[ D^0_{h_1/q}(\zho) +2\alpha_s C_F \Bigg\{ \int_{\zho}^1 \frac{\dd \xi}{\xi} \frac{(1+\xi^2)}{(1-\xi)} D^0_{h_1/q}\left(\frac{\zho}{\xi}\right) \int \dtwo{\bk} \frac{e^{i\frac{\bk}{\xi}\cdot(\bx_1^\p-\bx_1)}}{\left(\bk-(1-\xi)\bp\right)^2} \nonumber \\
&- D^0_{h_1/q}(\zho) \int_0^1 \dd \xi \frac{(1+\xi^2)}{(1-\xi)} \int \dtwo{\bk} \frac{1}{(\bk-(1-\xi)\bp)^2}\Bigg\}\Bigg].
\end{align}

\noindent Now we turn our attention to the two transverse momentum integrals. The one with the exponential comes from the real correction $\sigma_{1\times 1}$, and contains both a soft divergence, $\bk \to 0$ and $z \to 0$ (equivalently $\xi \to 1$) at the same time, and a collinear divergence as $\bk \to (1-\xi)\bp$. The integral without an exponential (from the virtual correction $\sigma_9$) contains a soft divergence, a collinear divergence, and a UV divergence as $|\bk| \to \infty$. We showed in section \ref{soft} that the soft divergences cancel between these two terms. In section \ref{uv} we also showed that the UV divergence in the virtual correction is canceled with other terms in the cross section. Therefore, the only divergence left to consider here is the collinear divergence in both integrals.

To see how regulating these divergences 
leads to renormalization of the quark-hadron fragmentation function it is perhaps easiest to demonstrate this using an
explicit cutoff where the lower limit on $|\bk|$ is $\Lambda$ and the upper limit is $\mu$. 

\begin{align}
&\int_0^1 \dd \zho \int_0^1 \dd \zht \frac{4e^2 Q^2 (z_1z_2)^3 N_c}{(2\pi)^7 (\zho \zht)^2} \int \dd^8 \bx \, H(\bp,\bq,\bx,z_2) D^0_{h_2/\barq}(\zht) \delta(1-z_1-z_2) \nonumber \\
&\times \Bigg[ D^0_{h_1/q}(\zho) +\frac{\alpha_s C_F}{2\pi} \Bigg\{ \int_{\zho}^1 \frac{\dd \xi}{\xi} \frac{(1+\xi^2)}{(1-\xi)} D^0_{h_1/q}\left(\frac{\zho}{\xi}\right)-  \int_0^1 \dd \xi \frac{(1+\xi^2)}{(1-\xi)}D^0_{h_1/q}(\zho) \Bigg\}\log\left(\frac{\mu^2}{\Lambda^2}\right)\Bigg].\label{beforePqq}
\end{align}

\noindent We would like to combine the terms inside the curly brackets, to do so we'll define the quark-quark splitting function $P_{qq}(\xi)$:

\begin{align}
P_{qq}(\xi) \equiv C_F\left[ \frac{(1+\xi^2)}{(1-\xi)_+} +\frac32 \delta(1-\xi)\right], \label{Pqq}
\end{align}

\noindent where the $+$ distribution is defined inside an integral via

\begin{align}
\int_z^1 \dd\xi \frac{f(\xi)}{(1-\xi)_+} \equiv \int_z^1 \frac{\dd \xi}{(1-\xi)} (f(\xi)-f(1)) +f(1)\log(1-z).\label{plusdist}
\end{align}

\noindent This allows us to write Eq. \ref{beforePqq} as

\begin{align}
&\int_0^1 \dd \zho \int_0^1 \dd \zht \frac{4e^2 Q^2 (z_1z_2)^3 N_c}{(2\pi)^7 (\zho \zht)^2} \int \dd^8 \bx \, H(\bp,\bq,\bx,z_2) D^0_{h_2/\barq}(\zht) \delta(1-z_1-z_2) \nonumber \\
&\times \Bigg[ D^0_{h_1/q}(\zho) +\frac{\alpha_s }{2\pi} \Bigg\{ \int_{\zho}^1 \frac{\dd \xi}{\xi} P_{qq}(\xi) D^0_{h_1/q}\left(\frac{\zho}{\xi}\right) \Bigg\}\log\left(\frac{\mu^2}{\Lambda^2}\right)\Bigg].\label{afterPqq}
\end{align}

\noindent Here we have used the following result:

\begin{align}
\int_{\zho}^1 \frac{\dd \xi}{\xi} P_{qq}(\xi) D^0_{h_1/q}\left(\frac{\zho}{\xi}\right) =  \int_{\zho}^1 \frac{\dd \xi}{\xi} \frac{(1+\xi^2)}{(1-\xi)} D^0_{h_1/q}\left(\frac{\zho}{\xi}\right)-  \int_0^1 \dd \xi \frac{(1+\xi^2)}{(1-\xi)}D^0_{h_1/q}(\zho) . \label{proofid}
\end{align}

\noindent To prove this, one first expands the $P_{qq}$ function using Eq. \ref{Pqq}, then further expands the term with the plus distribution according to Eq. \ref{plusdist}. One of the terms is then exactly the first term on the right side of Eq. \ref{proofid}. In the remaining term with a $\xi$ integral one separates into two terms using $\int_{\zho}^1 = \int_0^1 - \int_0^{\zho}$. One can then use the fact that 

\begin{align}
\frac{2}{1-\xi} = \frac{(1+\xi^2)}{(1-\xi)} +1 +\xi.
\end{align}

\noindent The first term here paired with the $\int_0^1$ integrals then gives exactly the second term on the right side of Eq. \ref{proofid}. In all the other terms, the integral over $\xi$ can be performed and these terms all cancel in the full expression.

Next, we can also combine with the leading order term (first term in the square brackets) by including a $\xi$ integral that evaluates to $1$:

\begin{align}
&\int_0^1 \dd \zho \int_0^1 \dd \zht \frac{4e^2 Q^2 (z_1z_2)^3 N_c}{(2\pi)^7 (\zho \zht)^2} \int \dd^8 \bx \, H(\bp,\bq,\bx,z_2) D^0_{h_2/\barq}(\zht) \delta(1-z_1-z_2) \nonumber \\
&\times\int_{\zho}^1 \frac{\dd \xi}{\xi} D^0_{h_1/q}\left(\frac{\zho}{\xi}\right) \Bigg[ \delta(1-\xi) +\frac{\alpha_s }{2\pi}P_{qq}(\xi)\log\left(\frac{\mu^2}{\Lambda^2}\right)\Bigg].
\end{align}

\noindent Then, we define the DGLAP evolved quark-hadron fragmentation function $D_{h_1/q}(\zho,\mu^2)$ as

\begin{align}
D_{h_1/q}(\zho,\mu^2)= \int_{\zho}^1 \frac{\dd \xi}{\xi} D^0_{h_1/q}\left(\frac{\zho}{\xi}\right) \Bigg[ \delta(1-\xi) +\frac{\alpha_s }{2\pi}P_{qq}(\xi)\log\left(\frac{\mu^2}{\Lambda^2}\right)\Bigg],
\end{align}

\noindent in terms of which our expression becomes

\begin{align}
\frac{\dd \sigma^{\gamma^*A\to h_1 h_2 X}_{LO+9+1\times 1}}{\dd^2 \bp_h \, \dd^2 \bq_h \, \dd y_1 \, \dd y_2} =& \int_0^1 \dd \zho \int_0^1 \dd \zht \frac{4e^2 Q^2 (z_1z_2)^3 N_c}{(2\pi)^7 (\zho \zht)^2} \int \dd^8 \bx \, H(\bp,\bq,\bx,z_2) \nonumber \\
& D^0_{h_2/\barq}(\zht)D_{h_1/q}(\zho,\mu^2) \delta(1-z_1-z_2) .
\end{align}

\noindent It must be noted that here we used an explicit cutoff $\mu$ to regulate these collinear divergences. However the most common way to handle these divergences is via dimensional regularization (dim reg). This is done in the next subsection.

\subsubsection{Evolution of the Fragmentation Function in Dimensional Regularization}

As most parametrizations of the fragmentation functions are defined in the $\overline{\text{MS}}$ scheme it is important to relate our result with that given in the $\overline{\text{MS}}$ scheme and using dimensional regularization as this would affect the finite pieces. The real integral can be regulated without complication via dimensional regularization.

\begin{align}
& \int \dtwo{\bk} \frac{e^{i\bk\cdot(\bx_1^\p-\bx_1)}}{\left(\bk-\frac{(1-\xi)}{\xi}\bp\right)^2} \to \mu^{2-d} \int \frac{\dd^d \bk}{(2\pi)^d}  \frac{e^{i\bk\cdot (\bx_1^\p - \bx_1)}}{\left(\bk-\frac{(1-\xi)}{\xi}\bp\right)^2} \nonumber \\
&= \mu^{2-d} e^{i\frac{(1-\xi)}{\xi}\bp\cdot(\bx_1^\p - \bx_1)} \int \frac{\dd^d \bk}{(2\pi)^d} \frac{e^{i\bk\cdot(\bx_1^\p -\bx_1)}}{\bk^2} =  \frac{\mu^{2-d} e^{i\frac{(1-\xi)}{\xi}\bp\cdot(\bx_1^\p - \bx_1)}}{(2\pi)^d}\int \dd k \, k^{d-3} \int \dd \Omega_d e^{ik |\bx_1^\p -\bx_1|\cos\theta}.
\end{align}

\noindent The limit $d \to 2$ is safe in the angular integral.

\begin{align}
&\frac{\mu^{2-d} e^{i\frac{(1-\xi)}{\xi}\bp\cdot(\bx_1^\p - \bx_1)}}{(2\pi)^{d-1}}\int \dd k \, k^{d-3}J_0\left(k |\bx_1^\p - \bx_1|\right) \nonumber \\
&= \frac{\mu^{2-d} e^{i\frac{(1-\xi)}{\xi}\bp\cdot(\bx_1^\p - \bx_1)}}{(2\pi)^{d-1}} 2^{d-3} |\bx_1^\p-\bx_1|^{2-d} \frac{\Gamma[d/2-1]}{\Gamma[2-d/2]}, \,\,\,\,\,\, 2<d < 7/2, \nonumber \\
&=  \frac{ e^{i\frac{(1-\xi)}{\xi}\bp\cdot(\bx_1^\p - \bx_1)}}{2\pi} \left[ \frac{1}{\epsilon} -\log\left(\pi e^{\gamma_E}\mu |\bx_1^\p-\bx_1|\right)\right] + \mathcal{O}(\epsilon),\,\,\,\,\, \epsilon = d-2 > 0. \label{realdr}
\end{align}

\noindent This completes our result for regulating the real integral via dimensional regularization. Next we look at the virtual integral which, after a shift, becomes

\begin{align}
\int \dtwo{\bk}\frac{1}{\bk^2}.
\end{align}

\noindent This divergent integral is zero in dimensional regularization~\cite{Leibbrandt:1975dj,Leibbrandt:1987qv}. This can be understood as the result of the collinear (here looks soft due to the shift) divergence that exactly cancels the ultraviolet divergence in dimensional regularization. One can see this by separating the integration into two terms with a cutoff and doing each separately\cite{Chirilli:2011km,Chirilli:2012jd,Caucal:2021ent}. Now, it's important to note that in our case we have already used the UV divergence to cancel other UV divergences in the virtual corrections (see section \ref{uv}). Thus we need to be careful with this integral. Following~\cite{Caucal:2021ent} let's separate into two regions,

\begin{align}
\int \dtwo{\bk} \frac{1}{\bk^2} \to \mu^{2-d} \int \frac{\dd^d \bk}{(2\pi)^d} \frac{1}{\bk^2} = \frac{\mu^{2-d} 2\pi^{d/2}}{(2\pi)^{d} \Gamma[d/2]} \left[ \int_0^\Lambda \dd k\, k^{d-3} + \int_\Lambda^\infty \dd k \, k^{d-3}\right].
\end{align}

\noindent The first term contains the infrared (collinear) divergence, and the second term contains the UV divergence. These integrals can both be done, which yields

\begin{align}
 \frac{\mu^{2-d} 2\pi^{d/2}}{(2\pi)^{d} \Gamma[d/2]} \left[   \frac{\Lambda^{-\epsilon_{UV}}}{\epsilon_{UV}} -\frac{\Lambda^{-\epsilon_{IR}}}{\epsilon_{IR}}\right].
\end{align}

\noindent Here $\epsilon = 2-d$ with  $\epsilon_{UV} > 0$ and $\epsilon_{IR} < 0$. If one formally sets $\epsilon_{UV} = \epsilon_{IR}$, then one finds that the expression is zero. For our purposes, let's write both expressions separately and expand each for small $\epsilon$:

\begin{align}
\left[\int \dtwo{\bk}\frac{1}{\bk^2}\right]_{IR} = \frac{1}{2\pi} \left[ -\frac{1}{\epsilon_{IR}}   -\frac12 \log\left(\frac{4\pi e^{-\gamma_E}\mu^2}{\Lambda^2}\right)\right] + \mathcal{O}(\epsilon).
\end{align}

\begin{align}
\left[\int \dtwo{\bk}\frac{1}{\bk^2}\right]_{UV} = \frac{1}{2\pi} \left[ \frac{1}{\epsilon_{UV}}  +\frac12 \log\left(\frac{4\pi e^{-\gamma_E} \mu^2}{\Lambda^2}\right)\right] + \mathcal{O}(\epsilon).
\end{align}

\noindent Notice that the finite terms cancel if one adds both of these together to form the full integral (which gives Eq. 5.45 in~\cite{Caucal:2021ent}). We would like for the finite term to look similar to the finite term in the result for the real integral (Eq. \ref{realdr}), so let's add a finite term to the IR piece and also subtract it from the UV piece:

\begin{align}
\left[\int \dtwo{\bk}\frac{1}{\bk^2}\right]_{IR} = \frac{1}{2\pi} \left[ -\frac{1}{\epsilon_{IR}}   -\frac12 \log\left(\frac{4\pi e^{-\gamma_E}\mu^2}{\Lambda^2}\right) + \log\left( \frac{2 e^{-\frac32 \gamma_E}}{\pi^{1/2} \Lambda |\bx_1^\p - \bx_1|}\right)\right] + \mathcal{O}(\epsilon).
\end{align}

\begin{align}
\left[\int \dtwo{\bk}\frac{1}{\bk^2}\right]_{UV} = \frac{1}{2\pi} \left[ \frac{1}{\epsilon_{UV}}  +\frac12 \log\left(\frac{4\pi e^{-\gamma_E} \mu^2}{\Lambda^2}\right)- \log\left( \frac{2 e^{-\frac32 \gamma_E}}{\pi^{1/2} \Lambda |\bx_1^\p - \bx_1|}\right)\right] + \mathcal{O}(\epsilon).
\end{align}

\noindent We have added and subtracted the same quantity from the full expression. Combining the logarithms in both terms gives

\begin{align}
\left[\int \dtwo{\bk}\frac{1}{\bk^2}\right]_{IR} = \frac{1}{2\pi} \left[ -\frac{1}{\epsilon_{IR}} -\log\left(e^{\gamma_E} \pi \mu |\bx_1^\p - \bx_1|\right)\right] + \mathcal{O}(\epsilon). \label{eir}
\end{align}

\begin{align}
\left[\int \dtwo{\bk}\frac{1}{\bk^2}\right]_{UV} = \frac{1}{2\pi} \left[ \frac{1}{\epsilon_{UV}}  + \log\left(e^{\gamma_E} \pi \mu |\bx_1^\p - \bx_1|\right)\right] + \mathcal{O}(\epsilon).\label{euv}
\end{align}

\noindent Then, since $\epsilon_{IR}$ is negative, we can label it as $-\epsilon$ with $\epsilon > 0$ to match the real case. This addition and subtraction of a finite piece can also be thought of as choosing a particular value for the cutoff $\Lambda$. So, Eq. \ref{eir} will contribute to the fragmentation function, while Eq. \ref{euv} will contribute to the rest of the NLO corrections. The divergence in Eq. \ref{eir} is collinear and will be absorbed into the bare quark hadron fragmentation function to make it scale dependent. The divergence in Eq. \ref{euv} is canceled with the other UV divergences in virtual diagrams as before. Finally we note that the real integral (Eq. \ref{realdr}) has an overall exponential which we set to $1$. This can be motivated by noticing that it does not affect the nature of the collinear singularity and that splitting function favors $\xi \rightarrow 1$ (soft radiation). 

Using the results of our transverse integrals in the expression for the hadronic cross section we get

\begin{align}
&\int_0^1 \dd \zho \int_0^1 \dd \zht \frac{4e^2 Q^2 (z_1z_2)^3 N_c}{(2\pi)^7 (\zho \zht)^2} \int \dd^8 \bx \, H(\bp,\bq,\bx,z_2) D^0_{h_2/\barq}(\zht) \delta(1-z_1-z_2) \nonumber \\
&\times \Bigg[ D^0_{h_1/q}(\zho) +\frac{\alpha_s C_F}{\pi} \Bigg\{ \int_{\zho}^1 \frac{\dd \xi}{\xi} \frac{(1+\xi^2)}{(1-\xi)} D^0_{h_1/q}\left(\frac{\zho}{\xi}\right) \nonumber \\
&- D^0_{h_1/q}(\zho) \int_0^1 \dd \xi \frac{(1+\xi^2)}{(1-\xi)}\Bigg\} \left( \frac{1}{\epsilon} -\log\left(\pi e^{\gamma_E}\mu |\bx_1^\p-\bx_1|\right)\right)\Bigg].
\end{align}

\noindent Next we again introduce the quark-quark splitting function as in the previous subsection (Eq. \ref{Pqq}). In terms of this, the full expression becomes

\begin{align}
&\int_0^1 \dd \zho \int_0^1 \dd \zht \frac{4e^2 Q^2 (z_1z_2)^3 N_c}{(2\pi)^7 (\zho \zht)^2} \int \dd^8 \bx \, H(\bp,\bq,\bx,z_2) D^0_{h_2/\barq}(\zht) \delta(1-z_1-z_2) \nonumber \\
&\times\int_{\zho}^1 \frac{\dd \xi}{\xi} D^0_{h_1/q}\left(\frac{\zho}{\xi}\right) \Bigg[ \delta(1-\xi) +\frac{\alpha_s }{\pi}P_{qq}(\xi) \left( \frac{1}{\epsilon} -\log\left(\pi e^{\gamma_E}\mu |\bx_1^\p-\bx_1|\right)\right)\Bigg].
\end{align}

\noindent Then, we define the DGLAP evolved quark-hadron fragmentation function $D_{h_1/q}(\zho,\mu^2)$ in dimensional regularization as

\begin{align}
D_{h_1/q}(\zho,\mu^2)= \int_{\zho}^1 \frac{\dd \xi}{\xi} D^0_{h_1/q}\left(\frac{\zho}{\xi}\right) \Bigg[ \delta(1-\xi) +\frac{\alpha_s }{\pi}P_{qq}(\xi) \left( \frac{1}{\epsilon} -\log\left(\pi e^{\gamma_E}\mu |\bx_1^\p-\bx_1|\right)\right)\Bigg],
\end{align}

\noindent in terms of which our expression becomes

\begin{align}
\frac{\dd \sigma^{\gamma^*A\to h_1 h_2 X}_{LO+9+1\times 1}}{\dd^2 \bp_h \, \dd^2 \bq_h \, \dd y_1 \, \dd y_2} =& \int_0^1 \dd \zho \int_0^1 \dd \zht \frac{4e^2 Q^2 (z_1z_2)^3 N_c}{(2\pi)^7 (\zho \zht)^2} \int \dd^8 \bx \, H(\bp,\bq,\bx,z_2) \nonumber \\
& D^0_{h_2/\barq}(\zht)D_{h_1/q}(\zho,\mu^2) \delta(1-z_1-z_2),
\end{align}

\noindent where we recall that

\begin{align}
z_1 = \frac{p_h^+}{z_{h_1}l^+}, \,\,\,\,\,\, z_2 = \frac{q_h^+}{z_{h_2} l^+}. 
\end{align}

\noindent Finally, if one repeats the exact same procedure for the anti-quark line (i.e.: adding up $LO + 10 + 2\times2$), one gets

\begin{align}
\frac{\dd \sigma^{\gamma^*A\to h_1 h_2 X}_{LO+10+2\times 2}}{\dd^2 \bp_h \, \dd^2 \bq_h \, \dd y_1 \, \dd y_2} =& \int_0^1 \dd \zho \int_0^1 \dd \zht \frac{4e^2 Q^2 (z_1z_2)^3 N_c}{(2\pi)^7 (\zho \zht)^2} \int \dd^8 \bx \, H(\bp,\bq,\bx,z_1) \nonumber \\
&D^0_{h_1/q}(\zho)D_{h_2/\barq}(\zht,\mu^2) \delta(1-z_1-z_2).
\end{align}

\noindent This shows that absorption of collinear divergences in $\sigma_9, \sigma_{1\times 1}, \sigma_{10}$, 
and $\sigma_{2\times 2}$ into bare quark and anti quark-hadron fragmentation functions lead to evolution, i.e. scale dependence of these fragmentation functions.

Therefore our final result for the regulated dihadron production cross section can be written as the sum of several terms (Eq. \ref{convolutioneq}). The first term contains the $z$ integration region below $z_f$ where the leading order cross section is evolved with the BK/JIMWLK evolution equations. The second term includes the integration region $z > z_f$ where the leading order cross section multiplies the DGLAP evolved fragmentation functions for both quark and anti-quark. Finally the last term constitutes all the remaining contributions to the NLO cross section which is finite.

\begin{align}
\dd \sigma^{\gamma^* A \to h_1 h_2 X}= \dd \sigma_{LO}\otimes \text{JIMWLK} + \dd\sigma_{LO}\otimes D_{h_1/q}(\zho , \mu^2) D_{h_2/\barq}(\zht , \mu^2) + \dd\sigma_{NLO}^{\text{finite}}.\label{convolutioneq}
\end{align}

\noindent Here we imply the presence of the bare fragmentation functions in the first and last terms.

In summary, we have calculated the one-loop corrections to inclusive quark anti-quark production in DIS at small $x$ for longitudinal photons. We have shown that all divergences that appear at the one-loop level are either canceled or absorbed into JIMWLK evolution of dipoles and quadrupoles, and into DGLAP evolution of parton-hadron fragmentation functions. These results are well suited for further phenomenological studies of angular correlations of the dihadrons produced in DIS at small $x$~\cite{bergabo:2021woe}. A particularly interesting limit is the so-called
back to back limit where a suppression of the away side peak is observed experimentally. Here one will be sensitive to Sudakov radiation which will be significant~\cite{mueller:2013wwa,Taels:2022tza,Zheng:2014vka} and must be included in a full phenomenological study. One can also use our results here and integrate over the phase space of one of the outgoing particles, thus obtaining the single inclusive hadron production cross section in DIS at small $x$ at Next-to-Leading Order. This is under investigation and will be reported elsewhere~\cite{bergabo:SI}.



\paragraph{Acknowledgements:}We gratefully acknowledge support from the DOE Office of Nuclear Physics through Grant No. DE-SC0002307 and by PSC-CUNY through grant No. 63158-0051. We would like to thank T. Altinoluk, G. Beuf, R. Boussarie, P. Caucal, L. Dixon, Y. Kovchegov, C. Marquet, F. Salazar, M. Tevio, R. Venugopalan, W. Vogelsang and B. Xiao for helpful discussions. 

\begin{appendices}


\section{Cancellation of UV Divergences}\label{UVapp}

UV divergences appear in the corrections $\dd \sigma_{5}$, $\dd \sigma_6$, $\dd \sigma_9$, $\dd \sigma_{10}$, $\dd \sigma_{11}$, $\dd \sigma_{12}$, $\dd \sigma_{14(1)}$, and $\dd \sigma_{14(2)}$. In this appendix, we will show that these divergences all cancel between these corrections. First, we note that the UV divergence in $\dd\sigma_5$ appears in the limit $\bx_3 \to \bx_1$. If we take this limit in our result (Eq. \ref{dsig5}) everywhere except where it causes the divergence, we get the following.

\begin{align}
\frac{\dd \sigma_{5,UV}^{L}}{\dd^2 \bp \, \dd^2 \bq \, \dd y_1 \, \dd y_2} =&\frac{2e^2 g^2 Q^2 N_c^2 (z_1 z_2)^3}{(2\pi)^{10}} \int_0^{z_1} \frac{\dd z}{z} \left[\frac{z_1^2+(z_1-z)^2}{z_1^2}\right]\int \dd^8 \bx [S_{122^\p 1^\p}-S_{12}-S_{1^\p 2^\p}+1] \nonumber \\
&K_0(|\bx_{12}|Q_1)K_0(|\bx_{1^\p 2^\p}|Q_1) e^{i\bp\cdot(\bx_1^\p-\bx_1)}e^{i\bq\cdot(\bx_2^\p-\bx_2)}
\int_{\bx_3 \to \bx_1} \frac{\dd^2 \bx_3 }{\bx_{31}^2}.
\end{align}

\noindent Here we have used the notation $\int_{\bx_3 \to \bx_1}$ to signify that the limit has been taken inside the $\bx_3$ integral. The divergence as $\bx_3 \to \bx_1$ can equivalently be written as a divergent momentum integral as the transverse momentum goes to infinity: $\int_{\bk\to \infty} \frac{\dd ^2 \bk}{\bk^2}$ . Making this substitution in $\dd \sigma_5$, and similarly for the UV divergence as $\bx_3 \to \bx_2$ in $\dd \sigma_6$, and writing the UV divergent parts of all the other UV divergent diagrams, we have

\begin{align}
\frac{\dd \sigma_{5,UV}^{L}}{\dd^2 \bp \, \dd^2 \bq \, \dd y_1 \, \dd y_2} = &\frac{2e^2 g^2 Q^2 N_c^2 (z_1 z_2)^3}{(2\pi)^{10}} \int_0^{z_1} \frac{\dd z}{z} \left[\frac{z_1^2+(z_1-z)^2}{z_1^2}\right]\int \dd^8 \bx [S_{122^\p 1^\p}-S_{12}-S_{1^\p 2^\p}+1] \nonumber \\
&K_0(|\bx_{12}|Q_1)K_0(|\bx_{1^\p 2^\p}|Q_1) e^{i\bp\cdot(\bx_1^\p-\bx_1)}e^{i\bq\cdot(\bx_2^\p-\bx_2)} \int_{|\bk|\to \infty} \frac{\dd^2 \bk}{\bk^2}. \\
\frac{\dd \sigma_{6,UV}^{L}}{\dd^2 \bp \, \dd^2 \bq \, \dd y_1 \, \dd y_2} = &\frac{2e^2 g^2 Q^2 N_c^2 (z_1 z_2)^3}{(2\pi)^{10}} \int_0^{z_2} \frac{\dd z}{z} \left[\frac{z_2^2+(z_2-z)^2}{z_2^2}\right]\int \dd^8 \bx [S_{122^\p 1^\p}-S_{12}-S_{1^\p 2^\p}+1] \nonumber \\
&K_0(|\bx_{12}|Q_1)K_0(|\bx_{1^\p 2^\p}|Q_1) e^{i\bp\cdot(\bx_1^\p-\bx_1)}e^{i\bq\cdot(\bx_2^\p-\bx_2)} \int_{|\bk|\to\infty} \frac{\dd^2 \bk}{\bk^2}.\\
\frac{\dd \sigma_{9,UV}^{L}}{\dd^2 \bp \, \dd^2 \bq \, \dd y_1 \, \dd y_2} = &\frac{-e^2 g^2 Q^2 N_c^2 (z_1 z_2)^3}{(2\pi)^{10}} \int_0^{z_1} \frac{\dd z}{z} \left[\frac{z_1^2+(z_1-z)^2}{z_1^2}\right]\int \dd^8 \bx [S_{122^\p 1^\p}-S_{12}-S_{1^\p 2^\p}+1] \nonumber \\
&K_0(|\bx_{12}|Q_1)K_0(|\bx_{1^\p 2^\p}|Q_1) e^{i\bp\cdot(\bx_1^\p-\bx_1)}e^{i\bq\cdot(\bx_2^\p-\bx_2)} \int_{|\bk| \to \infty} \frac{\dd^2 \bk}{\bk^2}. \\
\frac{\dd \sigma_{10,UV}^{L}}{\dd^2 \bp \, \dd^2 \bq \, \dd y_1 \, \dd y_2} = &\frac{-e^2 g^2 Q^2 N_c^2 (z_1 z_2)^3}{(2\pi)^{10}} \int_0^{z_2} \frac{\dd z}{z} \left[\frac{z_2^2+(z_2-z)^2}{z_2^2}\right]\int \dd^8 \bx [S_{122^\p 1^\p}-S_{12}-S_{1^\p 2^\p}+1] \nonumber \\
&K_0(|\bx_{12}|Q_1)K_0(|\bx_{1^\p 2^\p}|Q_1) e^{i\bp\cdot(\bx_1^\p-\bx_1)}e^{i\bq\cdot(\bx_2^\p-\bx_2)} \int_{|\bk|\to\infty} \frac{\dd^2 \bk}{\bk^2}. \\
\frac{\dd \sigma_{11,UV}^{L}}{\dd^2 \bp \, \dd^2 \bq \, \dd y_1 \, \dd y_2} = &\frac{-2e^2 g^2 Q^2 N_c^2 (z_1z_2)^3}{(2\pi)^{10}} \int_0^{z_1} \frac{\dd z}{z} \left[ \frac{z_1^2+(z_1-z)^2}{z_1^2}\right] \int \dd^8 \bx [S_{122^\p 1^\p}-S_{12}-S_{1^\p 2^\p}+1] \nonumber \\
&K_0(|\bx_{12}|Q_1)K_0(|\bx_{1^\p 2^\p}|Q_1) e^{i\bp\cdot(\bx_1^\p-\bx_1)}e^{i\bq\cdot(\bx_2^\p-\bx_2)} \int_{|\bk|\to\infty} \frac{\dd^2 \bk}{\bk^2}. \\
\frac{\dd \sigma_{12,UV}^{L}}{\dd^2 \bp \, \dd^2 \bq \, \dd y_1 \, \dd y_2} = &\frac{-2e^2 g^2 Q^2 N_c^2 (z_1z_2)^3}{(2\pi)^{10}} \int_0^{z_2} \frac{\dd z}{z} \left[ \frac{z_2^2+(z_2-z)^2}{z_2^2}\right] \int \dd^8 \bx [S_{122^\p 1^\p}-S_{12}-S_{1^\p 2^\p}+1] \nonumber \\
&K_0(|\bx_{12}|Q_1)K_0(|\bx_{1^\p 2^\p}|Q_1) e^{i\bp\cdot(\bx_1^\p-\bx_1)}e^{i\bq\cdot(\bx_2^\p-\bx_2)} \int_{|\bk|\to\infty} \frac{\dd^2 \bk}{\bk^2}. \\
\frac{\dd \sigma_{14(1),UV}^{L}}{\dd^2 \bp \, \dd^2 \bq \, \dd y_1 \, \dd y_2} = &\frac{e^2 g^2 Q^2 N_c^2 (z_1 z_2)^3}{(2\pi)^{10}} \int_0^{z_1} \frac{\dd z}{z} \left[ 1+\frac{(z_1-z)(z_2+z)}{z_1 z_2} - z\frac{z_1(z_1-z)+z_2(z_2+z)}{z_1 z_2}\right] \nonumber \\
& \int \dd^8 \bx [S_{122^\p 1^\p}-S_{12}-S_{1^\p 2^\p}+1]K_0(|\bx_{12}|Q_1)K_0(|\bx_{1^\p 2^\p}|Q_1) e^{i\bp\cdot(\bx_1^\p-\bx_1)}e^{i\bq\cdot(\bx_2^\p-\bx_2)} \int_{|\bk|\to \infty} \frac{\dd^2 \bk}{\bk^2}. \\
\frac{\dd \sigma_{14(2),UV}^{L}}{\dd^2 \bp \, \dd^2 \bq \, \dd y_1 \, \dd y_2} = &\frac{e^2 g^2 Q^2 N_c^2 (z_1 z_2)^3}{(2\pi)^{10}} \int_0^{z_2} \frac{\dd z}{z} \left[ 1+\frac{(z_1+z)(z_2-z)}{z_1 z_2} - z\frac{z_1(z_1+z)+z_2(z_2-z)}{z_1 z_2}\right] \nonumber \\
& \int \dd^8 \bx [S_{122^\p 1^\p}-S_{12}-S_{1^\p 2^\p}+1]K_0(|\bx_{12}|Q_1)K_0(|\bx_{1^\p 2^\p}|Q_1) e^{i\bp\cdot(\bx_1^\p-\bx_1)}e^{i\bq\cdot(\bx_2^\p-\bx_2)} \int_{|\bk|\to \infty} \frac{\dd^2 \bk}{\bk^2}.
\end{align}

\noindent One can immediately verify from the above expressions that the UV divergent part of $\sigma_5$ is canceled by the UV divergent part of $\sigma_{11}$. Similarly $\sigma_{6,UV}$ cancels with $\sigma_{12,UV}$. The remaining terms can be written (after some simplification of the $z$ factors) as follows:

\begin{align}
\frac{\dd \sigma_{9+14(1),UV}^{L}}{\dd^2 \bp \, \dd^2 \bq \, \dd y_1 \, \dd y_2} = &\frac{e^2 g^2 Q^2 N_c^2 (z_1 z_2)^3}{(2\pi)^{10}} \int_0^{z_1} \dd z\, \left[ \frac{2(z_1-z)}{z_1} - \frac{z}{z_1^2}\right]  \int \dd^8 \bx [S_{122^\p 1^\p}-S_{12}-S_{1^\p 2^\p}+1]\nonumber \\
&K_0(|\bx_{12}|Q_1)K_0(|\bx_{1^\p 2^\p}|Q_1) e^{i\bp\cdot(\bx_1^\p-\bx_1)}e^{i\bq\cdot(\bx_2^\p-\bx_2)} \int_{|\bk|\to \infty} \frac{\dd^2 \bk}{\bk^2}.
\end{align}

\begin{align}
\frac{\dd \sigma_{10+14(2),UV}^{L}}{\dd^2 \bp \, \dd^2 \bq \, \dd y_1 \, \dd y_2} = &\frac{e^2 g^2 Q^2 N_c^2 (z_1 z_2)^3}{(2\pi)^{10}} \int_0^{z_2} \dd z\, \left[\frac{2(z_2-z)}{z_2} - \frac{z}{z_2^2}\right]\int \dd^8 \bx [S_{122^\p 1^\p}-S_{12}-S_{1^\p 2^\p}+1] \nonumber \\
& K_0(|\bx_{12}|Q_1)K_0(|\bx_{1^\p 2^\p}|Q_1) e^{i\bp\cdot(\bx_1^\p-\bx_1)}e^{i\bq\cdot(\bx_2^\p-\bx_2)} \int_{|\bk|\to \infty} \frac{\dd^2 \bk}{\bk^2}.
\end{align}

\noindent The integrals over $z$ in these two expressions can now be done, we find the following:

\begin{align}
\frac{\dd \sigma_{9+14(1),UV}^{L}}{\dd^2 \bp \, \dd^2 \bq \, \dd y_1 \, \dd y_2} = &\frac{e^2 g^2 Q^2 N_c^2 (z_1 z_2)^3}{(2\pi)^{10}} \left[z_1-\frac12 \right] \int \dd^8 \bx [S_{122^\p 1^\p}-S_{12}-S_{1^\p 2^\p}+1] \nonumber \\
&K_0(|\bx_{12}|Q_1)K_0(|\bx_{1^\p 2^\p}|Q_1) e^{i\bp\cdot(\bx_1^\p-\bx_1)}e^{i\bq\cdot(\bx_2^\p-\bx_2)} \int_{|\bk|\to \infty} \frac{\dd^2 \bk}{\bk^2}.
\end{align}

\begin{align}
\frac{\dd \sigma_{10+14(2),UV}^{L}}{\dd^2 \bp \, \dd^2 \bq \, \dd y_1 \, \dd y_2} = &\frac{e^2 g^2 Q^2 N_c^2 (z_1 z_2)^3}{(2\pi)^{10}} \left[z_2-\frac12\right] \int \dd^8 \bx [S_{122^\p 1^\p}-S_{12}-S_{1^\p 2^\p}+1] \nonumber \\
&K_0(|\bx_{12}|Q_1)K_0(|\bx_{1^\p 2^\p}|Q_1) e^{i\bp\cdot(\bx_1^\p-\bx_1)}e^{i\bq\cdot(\bx_2^\p-\bx_2)} \int_{|\bk|\to \infty} \frac{\dd^2 \bk}{\bk^2}.
\end{align}

\noindent Now, if we add up $\sigma_9,\sigma_{10},\sigma_{14(1)}$ and $\sigma_{14(2)}$ we see that the result is proportional to $z_1 + z_2 - 1$ which is zero by virtue of the delta function (not shown here). So, what we have found is precisely Eq. \ref{uvcancels}. Therefore, \textit{all UV divergences are canceled} when adding up all the relevant pieces of the differential cross section.


\section{Coordinate Space Results at Soft $z$}\label{softzcoord}

Here we have taken the limit $z \to 0$ in all our expressions everywhere except where it causes a divergence. The result for each expression has also been written in coordinate space where we have recalled the definition of the radiation kernel (Eq. \ref{radkern}). The expressions in this appendix are useful both for showing that soft divergences are canceled and that rapidity divergences can be combined to show evolution of the dipoles and quadrupoles according to the BK and JIMWLK equations.

\begin{align}
\frac{\dd \sigma^{L}_{1\times 1}}{\dd^2 \bp\, \dd^2 \bq\, \dd y_1 \, \dd y_2} = &\frac{2 e^2 g^2  Q^2 N_c^2(z_1z_2)^3}{(2\pi)^{10}}\int \frac{\dd z}{z}  \int \dd^{10} x \,K_0(|\bx_{12}|Q_1)K_0(|\bx_{1^\prime 2^\prime}|Q_1)\Delt{1}{1^\p} \nonumber \\ 
&\left[ S_{122^\p 1^\p} - S_{12} - S_{1^\p 2^\p} + 1\right] e^{i\bp\cdot(\bx_1^\prime - \bx_1)}e^{i\bq\cdot(\bx_2^\prime - \bx_2)}  \delta(1-z_1-z_2).  \\
\frac{\dd \sigma^{L}_{2\times 2}}{\dd^2 \bp\, \dd^2 \bq\, \dd y_1 \, \dd y_2} = & \frac{2 e^2 g^2  Q^2 N_c^2(z_1 z_2)^3}{(2\pi)^{10}  }\int \frac{\dd z}{z}\int \dd^{10} \bx \, K_0(|\bx_{12}|Q_1) K_0(|\bx_{1^\p 2^\p}|Q_1) \Delt{2}{2^\p}\nonumber \\
& [S_{122^\p 1^\p} - S_{12} - S_{1^\p 2^\p} + 1]e^{i\bq\cdot(\bx_2^\p - \bx_2)} e^{i\bp\cdot(\bx_1^\p - \bx_1)}\delta(1-z_1-z_2). \\
\frac{\dd \sigma^{L}_{1\times 2}}{\dd^2 \bp\, \dd^2 \bq\, \dd y_1 \, \dd y_2} = & \frac{-2e^2 g^2  Q^2 N_c^2(z_1 z_2)^3}{(2\pi)^{10} }\int \frac{\dd z}{z}\int \dd^{10} \bx \,K_0(|\bx_{12}|Q_1)K_0(|\bx_{1^\p2^\p}| Q_1)\Delt{1}{2^\p}\nonumber \\ 
&[S_{12}S_{1^\p2^\p}-S_{12}-S_{1^\p2^\p}+1] e^{i\bp\cdot(\bx_1^\p-\bx_1)}e^{i\bq\cdot(\bx_2^\p-\bx_2)}\delta(1-z_1-z_2).\\
\frac{\dd \sigma^{L}_{3\times 3}}{\dd^2 \bp\, \dd^2 \bq\, \dd y_1 \, \dd y_2}=& \frac{2e^2 g^2 Q^2 N_c^2 (z_1 z_2)^3}{(2\pi)^{10}}\int \frac{\dd z}{z} \int \dd^{10}\bx\, K_0(|\bx_{12}|Q_1) K_0(|\bx_{1^\p 2^\p}|Q_1) \Delt{1}{1^\p} \nonumber \\
&[S_{11^\p} S_{22^\p} - S_{13}S_{23} - S_{1^\p 3} S_{2^\p 3}+1] e^{i\bp\cdot(\bx_1^\p-\bx_1)}e^{i\bq\cdot(\bx_2^\p-\bx_2)}\delta(1-z_1-z_2).\\
\frac{\dd \sigma^{L}_{4\times 4}}{\dd^2 \bp\, \dd^2 \bq\, \dd y_1 \, \dd y_2} =& \frac{2e^2 g^2 Q^2 N_c^2  (z_1z_2)^3}{(2\pi)^{10}}\int \frac{\dd z}{z} \int \dd^{10}\bx \,K_0(|\bx_{12}|Q_1) K_0(|\bx_{1^\p 2^\p}|Q_1) \Delt{2}{2^\p}\nonumber \\
&[S_{11^\p} S_{22^\p} - S_{13}S_{23} - S_{1^\p 3} S_{2^\p 3}+1] e^{i\bp\cdot(\bx_1^\p-\bx_1)}e^{i\bq\cdot(\bx_2^\p-\bx_2)}\delta(1-z_1-z_2).\\
\frac{\dd \sigma^{L}_{3\times 4}}{\dd^2 \bp\, \dd^2 \bq\, \dd y_1 \, \dd y_2} =& \frac{-2e^2 g^2  Q^2N_c^2(z_1 z_2)^3}{(2\pi)^{10}}\int \frac{\dd z}{z} \int \dd^{10}\bx \, K_0(|\bx_{12}|Q_1) K_0(|\bx_{1^\p 2^\p}|Q_1) \Delt{1}{2^\p} \nonumber \\
&[S_{11^\p} S_{22^\p} - S_{13}S_{23} - S_{1^\p 3} S_{2^\p 3}+1] e^{i\bp\cdot(\bx_1^\p-\bx_1)}e^{i\bq\cdot(\bx_2^\p-\bx_2)}\delta(1-z_1-z_2).\\
\frac{\dd \sigma^{L}_{1\times 3}}{\dd^2 \bp\, \dd^2 \bq\, \dd y_1 \, \dd y_2}=& \frac{-2e^2 g^2 Q^2 N_c^2(z_1z_2)^3}{(2\pi)^{10} }\int \frac{\dd z}{z} \int \dd^{10}x\, K_0(|\bx_{12}|Q_2)K_0(|\bx_{1^\p 2^\p}|Q_1) \Delt{1}{1^\p} \nonumber \\
& [S_{122^\p3} S_{1^\p3} - S_{1^\p 3} S_{2^\p 3} - S_{12} + 1] e^{i\bp\cdot(\bx_1^\p - \bx_1)}e^{i\bq\cdot(\bx_2^\p-\bx_2)}\delta(1-z_1-z_2).\\
\frac{\dd \sigma^{L}_{1\times 4}}{\dd^2 \bp\, \dd^2 \bq\, \dd y_1 \, \dd y_2}=& \frac{2e^2 g^2 Q^2 N_c^2(z_1 z_2)^3}{(2\pi)^{10}}\int \frac{\dd z}{z} \int \dd^{10}x  K_0(|\bx_{12}|Q_2)K_0(|\bx_{1^\p 2^\p}|Q_1) \Delt{1}{2^\p} \nonumber \\
&  [S_{122^\p3} S_{1^\p3} - S_{1^\p 3} S_{2^\p 3} - S_{12} + 1]  e^{i\bp\cdot(\bx_1^\p - \bx_1)}e^{i\bq\cdot(\bx_2^\p-\bx_2)}\delta(1-z_1-z_2).\\
\frac{\dd \sigma^{L}_{2\times 3}}{\dd^2 \bp\, \dd^2 \bq\, \dd y_1 \, \dd y_2} =& \frac{2e^2 g^2 Q^2 N_c^2 (z_1 z_2)^3}{(2\pi)^{10}}\int \frac{\dd z}{z} \int \dd^{10}x K_0(|\bx_{12}|Q_1) K_0(|\bx_{1^\p 2^\p}|Q_1) \Delt{2}{1^\p} \nonumber \\
&[S_{1231^\p} S_{2^\p 3} - S_{1^\p 3} S_{2^\p 3} - S_{12} + 1] e^{i\bp\cdot(\bx_1^\p-\bx_1)}e^{i\bq\cdot(\bx_2^\p-\bx_2)}\delta(1-z_1-z_2).\\ 
\frac{\dd \sigma^{L}_{2\times 4}}{\dd^2 \bp\, \dd^2 \bq\, \dd y_1 \, \dd y_2}=& \frac{-2e^2 g^2 Q^2 N_c^2 (z_1 z_2)^3}{(2\pi)^{10}}\int \frac{\dd z}{z} \int \dd^{10}x  K_0(|\bx_{12}|Q_1) K_0(|\bx_{1^\p 2^\p}|Q_1) \Delt{2}{2^\p}\nonumber \\
&[S_{1231^\p} S_{2^\p 3} - S_{1^\p 3} S_{2^\p 3} - S_{12} + 1]  e^{i\bp\cdot(\bx_1^\p-\bx_1)}e^{i\bq\cdot(\bx_2^\p-\bx_2)}\delta(1-z_1-z_2). 
\end{align}

\begin{align}
\frac{\dd \sigma^{L}_5}{\dd^2 \bp \, \dd^2 \bq \, \dd y_1 \, \dd y_2} = & \frac{4e^2g^2Q^2 N_c^2(z_1 z_2)^3}{(2\pi)^{10} } \int_0^{z_1} \frac{\dd z}{z} \dd^{10} \bx \,K_0(|\bx_{12}|Q_1)K_0(|\bx_{1^\p 2^\p}|Q_1)\frac{1}{\bx_{31}^2} \nonumber \\
&[S_{322^\p 1^\p} S_{13} - S_{13}S_{23} - S_{1^\p 2^\p} + 1] e^{i\bp\cdot(\bx_1^\p-\bx_1)}e^{i\bq\cdot(\bx_2^\p-\bx_2)} \delta(1-z_1-z_2). \\
\frac{\dd \sigma^{L}_6}{\dd^2 \bp \, \dd^2 \bq \, \dd y_1 \, \dd y_2} = &\frac{4e^2g^2Q^2N_c^2(z_1 z_2)^3}{(2\pi)^{10}} \int_0^{z_2} \frac{\dd z}{z} \dd^{10} \bx\, K_0(|\bx_{12}|Q_1)K_0(|\bx_{1^\p 2^\p}|Q_1)\frac{1}{\bx_{32}^2}\nonumber \\
& [S_{132^\p 1^\p}S_{23} - S_{13}S_{23} - S_{1^\p 2^\p}+1]e^{i\bp\cdot(\bx_1^\p-\bx_1)}e^{i\bq\cdot(\bx_2^\p-\bx_2)}\delta(1-z_1-z_2). \\
\frac{\dd \sigma^{L}_{7}}{\dd^2 \bp \, \dd^2 \bq \, \dd y_1 \, \dd y_2} = & \frac{-2e^2g^2Q^2 N_c^2(z_1 z_2)^3}{(2\pi)^{10}} \int_0^{z_1}\frac{\dd z }{z} \dd^{10} \bx\,  K_0(|\bx_{12}|Q_1)  K_0(|\bx_{1^\p2^\p}|Q_1) \Delt{1}{2}\nonumber \\
&[S_{322^\p 1^\p} S_{13} - S_{13}S_{23} - S_{1^\p 2^\p} + 1] e^{i\bp\cdot(\bx_1^\p-\bx_1)}e^{i\bq\cdot(\bx_2^\p-\bx_2)}\delta(1-z_1-z_2).  \\
\frac{\dd \sigma^{L}_{8}}{\dd^2 \bp \, \dd^2 \bq \, \dd y_1 \, \dd y_2} = & \frac{-2e^2g^2 Q^2 N_c^2(z_1 z_2)^3}{(2\pi)^{10}} \int_0^{z_2} \frac{\dd z }{z} \dd^{10}\bx\,K_0(|\bx_{12}|Q_1)K_0(|\bx_{1^\p2^\p}|Q_1) \Delt{1}{2} \nonumber\\
& [S_{132^\p 1^\p}S_{23} - S_{13}S_{23} - S_{1^\p 2^\p}+1] e^{i\bp\cdot(\bx_1^\p-\bx_1)}e^{i\bq\cdot(\bx_2^\p-\bx_2)}\delta(1-z_1-z_2).
\end{align}

\begin{align}
\frac{\dd \sigma_9^{L}}{\dd^2 \bp \, \dd^2 \bq\, \dd y_1\, \dd y_2} = &\frac{-2e^2 g^2 Q^2 N_c^2 (z_1 z_2)^3}{(2\pi)^{10}}  \int_0^{z_1} \frac{\dd z}{z} \dd^{10} \bx \, K_0(|\bx_{12}|Q_1) K_0(|\bx_{1^\p 2^\p}|Q_1)\frac{1}{\bx_3^2} \nonumber \\
& \big[S_{122^\p 1^\p} - S_{12} - S_{1^\p 2^\p} + 1\big] e^{i\bp\cdot(\bx_1^\p-\bx_1)}e^{i\bq\cdot(\bx_2^\p-\bx_2)}\delta(1-z_1-z_2).\\
\frac{\dd \sigma_{10}^{L}}{\dd^2 \bp \, \dd^2 \bq\, \dd y_1\, \dd y_2} = &\frac{-2e^2 g^2 Q^2 N_c^2 (z_1 z_2)^3}{(2\pi)^{10}}   \int_0^{z_2} \frac{\dd z}{z} \dd^{10} \bx\,K_0(|\bx_{12}|Q_1) K_0(|\bx_{1^\p 2^\p}|Q_1)\frac{1}{\bx_3^2} \nonumber \\
&\big[S_{122^\p 1^\p} - S_{12} - S_{1^\p 2^\p} + 1\big]   e^{i\bp\cdot(\bx_1^\p-\bx_1)}e^{i\bq\cdot(\bx_2^\p-\bx_2)} \delta(1-z_1-z_2).\\
\frac{\dd \sigma_{11}^{L}}{\dd^2 \bp\, \dd^2 \bq\, \dd y_1 \, \dd y_2} = & \frac{-4e^2g^2Q^2N_c^2(z_1 z_2)^3}{(2\pi)^{10}} \int_0^{z_1} \frac{\dd z}{z}\dd^{10} \bx\,K_0(|\bx_{12}|Q_1)K_0(|\bx_{1^\p 2^\p}|Q_1) \frac{1}{\bx_3^2} \nonumber\\
&\big[S_{122^\p1^\p}-S_{12}-S_{1^\p 2^\p}+1\big] e^{i\bp\cdot(\bx_1^\p-\bx_1)}e^{i\bq\cdot(\bx_2^\p-\bx_2)}\delta(1-z_1-z_2).\\
\frac{\dd \sigma_{12}^{L}}{\dd^2 \bp\, \dd^2 \bq\, \dd y_1 \, \dd y_2} = &\frac{-4e^2g^2Q^2N_c^2(z_1 z_2)^3}{(2\pi)^{10}}  \int_0^{z_2} \frac{\dd z}{z}  \dd^{10} \bx\,K_0(|\bx_{12}|Q_1)K_0(|\bx_{1^\p 2^\p}|Q_1) \frac{1}{\bx_3^2} \nonumber\\
&\big[S_{122^\p1^\p}-S_{12}-S_{1^\p 2^\p}+1\big] e^{i\bp\cdot(\bx_1^\p-\bx_1)}e^{i\bq\cdot(\bx_2^\p-\bx_2)}  \delta(1-z_1-z_2).
\end{align}

\begin{align}
 \frac{\dd \sigma_{13(1)}^{L}}{\dd^2\bp\,\dd^2\bq\,\dd y_1\,\dd y_2} =&  \frac{e^2g^2 Q^2 N_c^2 (z_1z_2)^3}{(2\pi)^{10}}\int_{0}^{z_2}\frac{ \dd z}{z} \int \dd^{10}\bx \,K_0(|\bx_{12}|Q_1)K_0(|\bx_{1^\p 2^\p}|Q_1) \Delt{1}{2}  \nonumber \\
&[S_{12}S_{1^\p 2^\p}-S_{12}-S_{1^\p 2^\p} + 1]e^{i\bp\cdot(\bx_1^\p - \bx_1)} e^{i\bq\cdot(\bx_2^\p - \bx_2)} \delta(1-z_1-z_2).\\
 \frac{\dd \sigma_{13(2)}^{L}}{\dd^2\bp\,\dd^2\bq\,\dd y_1\,\dd y_2} =&  \frac{e^2g^2 Q^2 N_c^2 (z_1z_2)^3}{(2\pi)^{10}}\int_{0}^{z_1}\frac{ \dd z}{z} \int \dd^{10}\bx \,K_0(|\bx_{12}|Q_1)K_0(|\bx_{1^\p 2^\p}|Q_1)\Delt{1}{2} \nonumber \\
&[S_{12}S_{1^\p 2^\p}-S_{12}-S_{1^\p 2^\p} + 1]e^{i\bp\cdot(\bx_1^\p - \bx_1)} e^{i\bq\cdot(\bx_2^\p - \bx_2)}  \delta(1-z_1-z_2).
\end{align}

\begin{align}
\frac{\dd \sigma_{14(1)}^{L}}{\dd^2 \bp\, \dd^2 \bq \, \dd y_1 \, \dd y_1} =& \frac{e^2 g^2 Q^2 N_c^2 (z_1 z_2)^3}{(2\pi)^{10}} \int_0^{z_1}\frac{\dd z}{z} \dd^{10} \bx\, K_0(|\bx_{12}|Q_1) K_0(|\bx_{1^\p 2^\p}|Q_1)\left[ \frac{2}{\bx_3^2}+\frac{1}{\bx_{31}^2}+\frac{1}{\bx_{32}^2} - \frac{\bx_{12}^2}{\bx_{31}^2\bx_{32}^2}\right] \nonumber \\
&[ S_{122^\p 1^\p} - S_{12} - S_{1^\p 2^\p} + 1]  e^{i\bp\cdot(\bx_1^\p - \bx_1)} e^{i\bq\cdot(\bx_2^\p - \bx_2)} \delta(1-z_1-z_2).\\
\frac{\dd \sigma_{14(2)}^{ L}}{\dd^2 \bp\, \dd^2 \bq \, \dd y_1 \, \dd y_1} =& \frac{e^2 g^2 Q^2 N_c^2 (z_1 z_2)^3}{(2\pi)^{10}} \int_0^{z_2}\frac{\dd z}{z} \dd^{10} \bx\, K_0(|\bx_{12}|Q_1) K_0(|\bx_{1^\p 2^\p}|Q_1) \left[ \frac{2}{\bx_3^2}+\frac{1}{\bx_{31}^2}+\frac{1}{\bx_{32}^2} - \frac{\bx_{12}^2}{\bx_{31}^2\bx_{32}^2}\right]  \nonumber \\
&[ S_{122^\p 1^\p} - S_{12} - S_{1^\p 2^\p} + 1]  e^{i\bp\cdot(\bx_1^\p - \bx_1)} e^{i\bq\cdot(\bx_2^\p - \bx_2)}\delta(1-z_1-z_2).
\end{align}


\section{Dimensional Regularization}\label{dimreg}

The results in section \ref{Production cross section} include divergent integrals over the transverse momentum $\bk$ or in coordinate space as divergent integrals over $\bx_3$. Ideally, one would like to regulate these divergences using dimensional regularization. In many of the results, we have not found a way to evaluate the regularized integral, but in several it is possible. Here we show the ones that can be done.

\begin{align}
\frac{\dd \sigma^{L}_{1\times 1}}{\dd^2 \bp\, \dd^2 \bq\, \dd y_1 \, \dd y_2} = &\frac{2 e^2 g^2  Q^2 N_c^2 z_2^3 (1-z_2)^2(z_1^2+(1-z_2)^2)}{(2\pi)^{9} z_1} \int \dd^{8} x \,K_0(|\bx_{12}|Q_2)K_0(|\bx_{1^\prime 2^\prime}|Q_2)\left[ S_{122^\p 1^\p} - S_{12} - S_{1^\p 2^\p} + 1\right] \nonumber \\
&  e^{i\bp\cdot(\bx_1^\prime - \bx_1)}e^{i\bq\cdot(\bx_2^\prime - \bx_2)}\int \frac{\dd z}{z}  e^{i\frac{z}{z_1}\bp\cdot(\bx_1^\p-\bx_1)}\left[\frac{1}{\epsilon} -  \log(e^{\gamma_E}\pi \mu |\bx_1^\p-\bx_1|)\right].  \\
\frac{\dd \sigma^{L}_{2\times 2}}{\dd^2 \bp\, \dd^2 \bq\, \dd y_1 \, \dd y_2} = &\frac{2 e^2 g^2  Q^2 N_c^2 z_2^3 (1-z_1)^2(z_2^2+(1-z_1)^2)}{(2\pi)^{9} z_2} \int \dd^{8} x \,K_0(|\bx_{12}|Q_2)K_0(|\bx_{1^\prime 2^\prime}|Q_2)\left[ S_{122^\p 1^\p} - S_{12} - S_{1^\p 2^\p} + 1\right] \nonumber \\
&  e^{i\bp\cdot(\bx_1^\prime - \bx_1)}e^{i\bq\cdot(\bx_2^\prime - \bx_2)}\int \frac{\dd z}{z}e^{i\frac{z}{z_2}\bq\cdot(\bx_2^\p-\bx_2)} \left[\frac{1}{\epsilon} -  \log(e^{\gamma_E}\pi \mu |\bx_2^\p-\bx_2|)\right] .  \\
\frac{\dd \sigma^{L}_{1\times 2}}{\dd^2 \bp\, \dd^2 \bq\, \dd y_1 \, \dd y_2} = & \frac{-e^2 g^2  Q^2 N_c^2z_1 z_2 (1-z_1) (1-z_2)(z_1(1-z_1)+z_2(1-z_2))}{(2\pi)^{9} }\int \frac{\dd z}{z}\int \dd^{8} \bx \,K_0(|\bx_{12}|Q_2)K_0(|\bx_{1^\p2^\p}| Q_1)\nonumber \\ 
&[S_{12}S_{1^\p2^\p}-S_{12}-S_{1^\p2^\p}+1] e^{i\bp\cdot(\bx_1^\p-\bx_1)}e^{i\bq\cdot(\bx_2^\p-\bx_2)} \Bigg[ \left(e^{i\frac{z}{z_1}\bp\cdot\bx_{2^\p 1}} + e^{i\frac{z}{z_2}\bq\cdot\bx_{2^\p 1}}\right)\left(\frac{1}{\epsilon} -\log\left(e^{\gamma_e}\pi\mu|\bx_1-\bx_2^\p|\right)\right) \nonumber \\
&- \left(\frac{z}{z_1z_2}\right)^2 (z_2\bp-z_1\bq)^2 \int_0^1 \dd x\, e^{i z\left(\frac{x}{z_1}\bp + \frac{(1-x)}{z_2}\bq\right)\cdot\bx_{2^\p 1}} K_0\left(|\bx_{12^\p}| \frac{z}{z_1 z_2}|z_2\bp-z_1\bq|\sqrt{x(1-x)}\right)\Bigg]. \\
\frac{\dd \sigma_{11}^{L}}{\dd^2 \bp\, \dd^2 \bq\, \dd y_1 \, \dd y_2} =& \frac{-2e^2g^2Q^2N_c^2(z_1 z_2)^3}{(2\pi)^9} \int \dd^8 \bx\big[S_{122^\p1^\p}-S_{12}-S_{1^\p 2^\p}+1\big] K_0(|\bx_{12}|Q_1)K_0(|\bx_{1^\p 2^\p}|Q_1) \nonumber\\
& e^{i\bp\cdot(\bx_1^\p-\bx_1)}e^{i\bq\cdot(\bx_2^\p-\bx_2)}\int_0^{z_1} \frac{\dd z}{z} \frac{[z_1^2+(z_1-z)^2]}{z_1^2}  \left(\frac{1}{\epsilon} +\frac12 \log\left(\frac{2\pi \, z_1^2 z_2 |\bx_{12}| \mu^2}{z(z_1-z) Q_1}\right)\right). \\
\frac{\dd \sigma_{12}^{L}}{\dd^2 \bp\, \dd^2 \bq\, \dd y_1 \, \dd y_2} =& \frac{-2e^2g^2Q^2N_c^2(z_1 z_2)^3}{(2\pi)^9} \int \dd^8 \bx\big[S_{122^\p1^\p}-S_{12}-S_{1^\p 2^\p}+1\big] K_0(|\bx_{12}|Q_1)K_0(|\bx_{1^\p 2^\p}|Q_1) \nonumber\\
&e^{i\bp\cdot(\bx_1^\p-\bx_1)}e^{i\bq\cdot(\bx_2^\p-\bx_2)} \int_0^{z_2} \frac{\dd z}{z} \frac{[z_2^2+(z_2-z)^2]}{z_2^2}  \left(\frac{1}{\epsilon} +\frac12 \log\left(\frac{2\pi \, z_2^2 z_1 |\bx_{12}| \mu^2}{z(z_2-z) Q_1}\right)\right). \\
\frac{\dd \sigma_{14(1)}^{ L}}{\dd^2 \bp\, \dd^2 \bq \, \dd y_1 \, \dd y_1} =& \frac{e^2 g^2 Q^2 N_c^2 (z_1 z_2)^2}{(2\pi)^9} \int \dd^8 \bx [ S_{122^\p 1^\p} - S_{12} - S_{1^\p 2^\p} + 1]K_0(|\bx_{12}|Q_1)  K_0(|\bx_{1^\p 2^\p}|Q_1)e^{i\bp\cdot(\bx_1^\p - \bx_1)} e^{i\bq\cdot(\bx_2^\p - \bx_2)} \nonumber \\
&\int_0^{z_1} \frac{\dd z}{z} \Bigg[ [z_1 z_2 + (z_1-z)(z_2+z)] \left( \frac{1}{\epsilon} -\frac{\gamma_E}{2} + \frac12 \log\left[\frac{4\pi \mu^2 z_1^2 (z_2+z)}{z^2 (z_1-z) Q^2}\right]\right) \nonumber \\
&-z[z_1(z_1-z)+z_2(z_2+z)]\left(\frac{1}{\epsilon} - \frac{\gamma_E}{2}+ \frac12 \log\left[ \frac{4\pi \mu^2}{(z_1-z)(z_2+z)Q^2}\right]\right)\Bigg].\\
\frac{\dd \sigma_{14(2)}^{ L}}{\dd^2 \bp\, \dd^2 \bq \, \dd y_1 \, \dd y_1} =& \frac{e^2 g^2 Q^2 N_c^2 (z_1 z_2)^2}{(2\pi)^9} \int \dd^8 \bx [ S_{122^\p 1^\p} - S_{12} - S_{1^\p 2^\p} + 1] K_0(|\bx_{12}|Q_1)  K_0(|\bx_{1^\p 2^\p}|Q_1)e^{i\bp\cdot(\bx_1^\p - \bx_1)} e^{i\bq\cdot(\bx_2^\p - \bx_2)} \nonumber \\
&\int_0^{z_2} \frac{\dd z}{z} \Bigg[ [z_1 z_2 + (z_2-z)(z_1+z)] \left( \frac{1}{\epsilon} -\frac{\gamma_E}{2} + \frac12 \log\left[\frac{4\pi \mu^2 z_2^2 (z_1+z)}{z^2 (z_2-z) Q^2}\right]\right) \nonumber \\
&-z[z_1(z_1+z)+z_2(z_2-z)]\left(\frac{1}{\epsilon} - \frac{\gamma_E}{2}+ \frac12 \log\left[ \frac{4\pi \mu^2}{(z_2-z)(z_1+z)Q^2}\right]\right)\Bigg].
\end{align}

\noindent Note: These last two expressions are only valid for $z \neq 0$. We have defined $\epsilon = 2-d$ where $d$ is the number of dimensions for the transverse integral. Another type of divergence becomes apparent when looking at the results in this way, there is a $\frac{\log z}{z}$ divergence as $z \to 0$ in the virtual corrections $\sigma_{11}, \sigma_{12}, \sigma_{14(1)},$ and $\sigma_{14(2)}$. However, we find that taking the limit $z \to 0$ inside the $z$ integrals shows that the $\frac{\log z}{z}$ cancels between $\sigma_{11}$ and $\sigma_{14(1)}$, and similarly between $\sigma_{12}$ and $\sigma_{14(2)}$. 

\begin{align}
& \frac{\dd \sigma_{13(1)}^{L}}{\dd^2\bp\,\dd^2\bq\,\dd y_1\,\dd y_2} =  \frac{-e^2g^2 Q^2 N_c^2 (z_1z_2)^2}{(2\pi)^8}\int_{0}^{z_2}\frac{ \dd z}{z} (z_1+z)(z_2-z) \int \dd^8\bx \,K_0\left(|\bx_{12}|Q\sqrt{(z_1+z)(z_2-z)}\right)K_0(|\bx_{1^\p 2^\p}|Q_1)\nonumber \\
& e^{i\bp\cdot\bx_{1^\p 1}}e^{i\bq\cdot\bx_{2^\p 2}}[S_{12}S_{1^\p 2^\p}-S_{12}-S_{1^\p 2^\p} + 1]\nonumber \\
&\Bigg[-\int_0^1 \dd x  \frac{\frac{z(z_2-z)[z_1z_2+(z_1+z)(z_2-z)] (z_2\bp-z_1\bq)^2}{z_1^2z_2^3} e^{i\left[\frac{(x z_1(z_2-z)+z)}{z_2}\bq -x(z_2-z)\bp\right]\cdot\bx_{21}}i|\bx_{12}|K_1\left(-i|\bx_{12}|\sqrt{\frac{x(z_2-z)(z_2\bp-z_1\bq)^2(z+x z_1(z_2-z))}{z_1 z_2^2}}\right)}{4\pi \sqrt{\frac{x(z_2-z)(z_2\bp-z_1\bq)^2(z+x z_1(z_2-z))}{z_1 z_2^2}}}   \nonumber \\
&+\frac{z[z_1(z_1+z)+z_2(z_2-z)]}{4 z_1 z_2} e^{i[(z_1+z)\bq-(z_2-z)\bp]\cdot\bx_{21}} iH_0^{(1)}\left(\frac{(z_1+z)(z_2-z)(z_2\bp-z_1\bq)^2 |\bx_{12}|}{z_1z_2} \right) \nonumber \\
&- \frac{[z_1 z_2+(z_1+z)(z_2-z)]}{z_1z_2} \frac{e^{i\frac{z}{z_2}\bq\cdot\bx_{21}}}{2\pi} \left[\frac{1}{\epsilon}-\log\left(e^{\gamma_E}\pi \mu |\bx_{12}|\right)\right]\Bigg]\delta(1-z_1-z_2).
\end{align}

\noindent Here $H_0^{(1)}(x)$ is the Hankel function of the first kind.

\begin{align}
& \frac{\dd \sigma_{13(2)}^{L}}{\dd^2\bp\,\dd^2\bq\,\dd y_1\,\dd y_2} =  \frac{-e^2g^2 Q^2 N_c^2 (z_1z_2)^2}{(2\pi)^8}\int_{0}^{z_1}\frac{ \dd z}{z} (z_2+z)(z_1-z) \int \dd^8\bx \,K_0\left(|\bx_{12}|Q\sqrt{(z_2+z)(z_1-z)}\right)K_0(|\bx_{1^\p 2^\p}|Q_1)\nonumber \\
& e^{i\bp\cdot\bx_{1^\p 1}}e^{i\bq\cdot\bx_{2^\p 2}}[S_{12}S_{1^\p 2^\p}-S_{12}-S_{1^\p 2^\p} + 1]\nonumber \\
&\Bigg[-\int_0^1 \dd x  \frac{\frac{z(z_1-z)[z_1z_2+(z_2+z)(z_1-z)] (z_2\bp-z_1\bq)^2}{z_2^2z_1^3} e^{i\left[\frac{(x z_2(z_1-z)+z)}{z_1}\bp -x(z_1-z)\bq\right]\cdot\bx_{12}}i|\bx_{12}|K_1\left(-i|\bx_{12}|\sqrt{\frac{x(z_1-z)(z_2\bp-z_1\bq)^2(z+x z_2(z_1-z))}{z_2 z_1^2}}\right)}{4\pi \sqrt{\frac{x(z_1-z)(z_2\bp-z_1\bq)^2(z+x z_2(z_1-z))}{z_2 z_1^2}}}   \nonumber \\
&+\frac{z[z_2(z_2+z)+z_1(z_1-z)]}{4 z_1 z_2} e^{i[(z_2+z)\bp-(z_1-z)\bq]\cdot\bx_{12}} (-i)H_0^{(2)}\left(\frac{(z_2+z)(z_1-z)(z_2\bp-z_1\bq)^2 |\bx_{12}|}{z_1z_2} \right) \nonumber \\
&- \frac{[z_1 z_2+(z_2+z)(z_1-z)]}{z_1z_2} \frac{e^{i\frac{z}{z_1}\bp\cdot\bx_{12}}}{2\pi} \left[\frac{1}{\epsilon}-\log\left(e^{\gamma_E}\pi \mu |\bx_{12}|\right)\right]\Bigg]\delta(1-z_1-z_2).
\end{align}

\noindent Here the sign change in the $i\epsilon$ going from $\sigma_{13(1)}$ to $\sigma_{13(2)}$ results in taking $iH_0^{(1)} \to -iH_0^{(2)}$ in the middle term. 



\section{Useful Relations}

\begin{align}
\int \dtwo \bk \frac{e^{i\bk\cdot\bx}}{\bk^2 + A^2} = \frac{1}{2\pi} K_0(|A| | \bx|), \,\,\,\,\,\,\, A \in \mathbb{R}. \label{easyko}
\end{align}

{\small
\begin{align}
\int \frac{\dd^2 \bk_1}{(2\pi)^2} \frac{\dd^2 \bk_2}{(2\pi)^2} \frac{(Ak_1 + B k_2)^i e^{i\bk_1\cdot\bx} e^{i\bk_2\cdot\by}}{\Big[\bk_1^2+CQ^2\Big]\left[Q^2+\frac{\bk_1^2}{D} + \frac{\bk_2^2}{E} + \frac{(\bk_1-\bk_2)^2}{F}\right]} = \frac{-iAF}{(2\pi)^2 } \frac{ y^i}{\by^2} K_0\left(Q\sqrt{C\left(\bx - \frac{A}{B}\by\right)^2  - \frac{AF}{B}\by^2}\right).  \label{hardko}
\end{align}}

\noindent Note: This result is valid only if $C,D,E,F$ are all positive,  $\frac{C}{F}\left(\frac{D+F}{D} - \frac{E}{E+F}\right) = 1$, and $A + \frac{BE}{E+F} = 0.$ We have used these relations to eliminate $D$ and $E$ in the result. One can confirm that these two relations hold for all the cases we'll need. 

\begin{align}
\int \frac{\dd^2 \bk_3}{(2\pi)^2} \frac{(\bk_3-\bk)^i}{(\bk_3-\bk)^2} e^{i\bk_3\cdot\bx} = \frac{i}{2\pi}\frac{x^i}{\bx^2}e^{i\bk\cdot\bx}, \,\,\,\,\, \text{equivalently} \,\,\,\,\,\,\, \frac{\bk^i}{\bk^2} = \frac{i}{2\pi} \int \dd^2 \bx \frac{\bx^i}{\bx^2} e^{i\bk\cdot\bx}. \label{k3int}
\end{align}

\begin{align}
\int \dtwo{\bk_1} \frac{1}{[\bk_1^2+A][(\bk_1+\bk_2)^2+B]} = \frac{1}{4\pi} \frac{\log\left(\frac{A+B+\bk_2^2 + \sqrt{(A-B)^2+2(A+B)\bk_2^2+\bk_2^4}}{A+B+\bk_2^2 - \sqrt{(A-B)^2+2(A+B)\bk_2^2+\bk_2^4}}\right)}{\sqrt{(A-B)^2+2(A+B)\bk_2^2+\bk_2^4}} \,\,\,\,\,\,\,\, \text{for}\,\,\, A,B>0.
\end{align}

\begin{align}
\int \dtwo{\bk} \frac{e^{i\bk\cdot\bx} \, \log\left(A(\bk^2+B^2)\right)}{\bk^2+B^2} = \frac{-K_0(|\bx||B|)}{2\pi} \left(\gamma_E + \log\left(\frac{|\bx|}{2A|B|}\right)\right), \,\,\,\,\,\,\, A>0,B\in \mathbb{R}.
\end{align}

\begin{align}
\int \dtwo{\bk} \frac{1}{\bk^2+A^2} &\to \mu^{2-d} \int \frac{\dd^d \bk}{(2\pi)^d} \frac{1}{\bk^2+A^2} = \frac{\mu^{2-d} 2\pi^{d/2}}{(2\pi)^d \Gamma[d/2]} \int_0^\infty \dd k \frac{k^{d-1}}{k^2+A^2} = \frac{\mu^{2-d} \pi^{1-\frac{d}{2}}}{2^d \Gamma[d/2]} \frac{A^{d-2}}{\sin(\pi d/2)} \nonumber \\
&= \frac{1}{2\pi} \left[ \frac{1}{\epsilon} + \frac12 \left( \log\left[ \frac{4\pi \mu^2}{A^2}\right] - \gamma_E\right)\right] + \mathcal{O}(\epsilon), \,\,\,\,\,\,\, \epsilon = 2-d, \,\,\,\, 0<d<2. \\
\int \dtwo{\bk} \frac{1}{(\bk-\bp)^2} &\to \mu^{2-d} \int\frac{\dd^d \bk}{(2\pi)^d} \frac{1}{\bk^2} = \frac{\mu^{2-d}2\pi^{d/2}}{(2\pi)^d \Gamma[d/2]} \int_0^\infty \dd k\, k^{d-3} = \frac{\mu^{2-d}2\pi^{d/2}}{(2\pi)^d \Gamma[d/2]}  \left[\frac{k^{d-2}}{d-2}\right]^{\infty}_0.  \\
\int \frac{\dd^2 \bk}{(2\pi)^2} \frac{e^{i\bk\cdot(\bx_i-\bx_j)}}{\bk^2} &= \int\dtwo{\bx_3} \Delta^{(3)}_{ij} = \int \dtwo{\bx_3} \frac{(\bx_3-\bx_i)\cdot(\bx_3-\bx_j)}{(\bx_3-\bx_i)^2(\bx_3-\bx_j)^2} \to \frac{1}{2\pi}\left[\frac{1}{\epsilon} -  \log(e^{\gamma_E}\pi \mu |\bx_i-\bx_j|)\right] + \mathcal{O}(\epsilon).
\end{align}

\begin{align}
z_1 = \frac{p_h^+}{\zho l^+}, \,\,\,\,\,\, z_2 = \frac{q_h^+}{\zht l^+}.
\end{align}

\end{appendices}

\bibliography{mybib}
\bibliographystyle{apsrev}

\end{document}